\documentclass[preprint,prd,aps,showpacs,preprintnumbers,amsmath,amssymb]{revtex4-1}
\usepackage{graphics,epsfig,subfigure}
\usepackage{diagbox}
\usepackage[usenames]{color}
\usepackage{color}
\usepackage{graphicx}
\usepackage{amsfonts}
\usepackage{indentfirst}
\usepackage{booktabs}
\usepackage{hyperref}
\hypersetup{hypertex=ture,backref=true,colorlinks=ture,linkcolor=blue,anchorcolor=blue,citecolor=blue}
\usepackage{float}
\usepackage{multirow}

\begin{document}
\renewcommand{\baselinestretch}{1.3}
\newcommand\beq{\begin{equation}}
\newcommand\eeq{\end{equation}}
\newcommand\beqn{\begin{eqnarray}}
\newcommand\eeqn{\end{eqnarray}}
\newcommand\nn{\nonumber}
\newcommand\fc{\frac}
\newcommand\lt{\left}
\newcommand\rt{\right}
\newcommand\pt{\partial}

\title{\Large \bf Multi-state Dirac stars}
\author{Chen Liang, Shi-Xian Sun, Ji-Rong Ren\footnote{renjr@lzu.edu.cn, corresponding author}, and Yong-Qiang Wang\footnote{yqwang@lzu.edu.cn, corresponding author
}
}

\affiliation{ $^{1}$Key Laboratory of Quantum Theory and Applications of MoE, Lanzhou Center for Theoretical Physics, Lanzhou University, Lanzhou 730000, China\\
	$^{2}$Key Laboratory of Theoretical Physics of Gansu Province, Institute of Theoretical Physics $\&$ Research Center of Gravitation, Lanzhou University, Lanzhou 730000, China\\
    $^{3}$School of Physical Science and Technology, Lanzhou University, Lanzhou 730000, China}

\begin{abstract}
	In this paper, we construct the multi-state Dirac stars (MSDSs) consisting of two pairs of Dirac fields. The two pairs of Dirac fields are in the ground state and the first excited state, respectively. Each pair consists of two fields with opposite spins, ensuring spherical symmetry of the system. We discuss the solutions of the MSDSs under synchronized and nonsynchronized frequencies. By varying the mass $\tilde{\mu}_1$ of the excited state Dirac field and the frequency $\tilde{\omega}_0$ of the ground state Dirac field, we obtain different types of solutions, including single-branch and double-branch solutions. These two types of solutions do not smoothly transition into each other as the parameters $\tilde{\mu}_1$ and $\tilde{\omega}_0$ continuously change, but undergo a sudden transition when $\tilde{\mu}_1$ ($\tilde{\omega}_0$) is greater than or less than the threshold value of $0.7694$ ($0.733$). Furthermore, we analyze the characteristics of the various MSDSs solutions and analyze the relationship between the ADM mass $M$ of the MSDSs and the synchronized and nonsynchronized frequencies. Subsequently, we calculate the binding energy $E_B$ of the MSDSs and discuss the stability of the solutions. Finally, we discuss the feasibility of simulating the dark matter halos using MSDSs.
\end{abstract}

\maketitle

\section{Introduction}\label{Sec1}

Recently, there has been rapid development in the field of gravitational wave astronomy, which has provided us with new insights into compact objects such as the black holes (BHs) and the neutron stars (NSs)~\cite{LIGOScientific:2016aoc,LIGOScientific:2016vbw,LIGOScientific:2014pky}. The advancements in gravitational wave detection technology have also made it possible to search for exotic compact objects (ECOs) similar to the BHs. One prominent class of ECOs is the bosonic stars, which are particle-like configurations of massive scalar fields~\cite{Wheeler:1955zz,Kaup:1968zz,Ruffini:1969qy,Schunck:2003kk,Liebling:2012fv,Jetzer:1991jr} or vector fields~\cite{Brito:2015pxa,SalazarLandea:2016bys,Duarte:2016lig,Sanchis-Gual:2017bhw,Gorghetto:2022sue,Herdeiro:2023lze} that form under their own gravitational attraction. The repulsive force balancing gravity is provided by the Heisenberg uncertainty principle. The bosonic stars provide a promising framework for studying compact objects, and certain models of the bosonic stars can mimic the BHs~\cite{Guzman:2009zz,Herdeiro:2021lwl,Rosa:2022tfv,Cardoso:2016oxy,Barranco:2011wq}. Additionally, the bosonic stars are also considered candidates for dark matter~\cite{Lee:1995af,Arvanitaki:2009fg,Suarez:2013iw,Eby:2015hsq,Chen:2020cef,Freitas:2021cfi}.

However, particle-like configurations can also be formed by spin-1/2 fermion fields. For non-gravitational cases, attempts to construct particle-like solutions for the Dirac equation were made as early as the 1930s by  Ivanenko~\cite{Ivanenko:1938zz}. Subsequent studies have also been conducted in this regard~\cite{weyl1950remark,HEISENBERG1953897,Finkelstein:1951zz,Finkelstein:1956hrg}. However, it was not until 1970 that the exact numerical solutions for such particle-like configurations were first studied by Soler~\cite{Soler:1970xp}. When gravitational interactions are considered, numerical calculations become more challenging. In 1999, Finster et al. constructed the exact numerical solutions for the Einstein-Dirac system, which couples spinor fields with Einstein's gravity, for the first time~\cite{Finster:1998ws}. These particle-like configurations, formed by spin-1/2 fermions under their own gravitational attraction, are known as the Dirac stars. Subsequently, research on the Dirac stars has been extended to include charged~\cite{Finster:1998ux} and gauge field~\cite{Finster:2000ps} additions, and the existence of the Dirac star solutions has been proven~\cite{RotaNodari:2009cfo,nodari2010perturbation}. Recently, the rotational Dirac star solutions~\cite{Herdeiro:2019mbz} and their charged counterparts~\cite{Herdeiro:2021jgc} have been provided for the first time by Herdeiro et al. Some comparative studies between the bosonic stars and 
the Dirac stars have been conducted in~\cite{Herdeiro:2017fhv,Herdeiro:2020jzx}. Additionally, various interesting studies on the Einstein-Dirac system have been carried out~\cite{Liang:2022mjo,Ma:2023vfa,Blazquez-Salcedo:2019qrz,Blazquez-Salcedo:2019uqq,Dzhunushaliev:2019kiy,Dzhunushaliev:2019uft,Dzhunushaliev:2018jhj,Daka:2019iix,Kain:2023jgu,Lee:1986tr}.

In 2010, Bernal et al. constructed the multi-state boson stars (MSBSs) composed of two complex scalar fields in their ground state and the first excited state and analyzed the stability of the solutions~\cite{Bernal:2009zy}. Subsequently, the MSBSs were extended to include rotation~\cite{Li:2019mlk} and self-interactions~\cite{Li:2020ffy}. It is possible that the Dirac field, under its own gravitational attraction, can also form multi-state configurations. In this work, we numerically solve the Einstein-Dirac system and construct spherically symmetric multi-state Dirac stars (MSDSs), where two coexisting states of the Dirac field are present.

The paper is organized as follows. In Sec.~\ref{sec2}, we introduce the Einstein-Dirac system, which couples four-dimensional Einstein's gravity with two sets of Dirac fields. In Sec.~\ref{sec3}, we investigate the boundary conditions of the MSDSs. In Sec.~\ref{sec4}, we present the numerical results and analyze the solutions of the MSDSs under synchronized and nonsynchronized frequencies. We also discuss the binding energy of the solutions and the problem of the galactic halos. We conclude in Sec.~\ref{sec5}.

\section{The model setup}\label{sec2}

We consider a system composed of multiple matter fields that are minimally coupled to Einstein's gravity. The matter fields consist of two pairs of Dirac spinor fields, with each pair containing two spinors of opposite spin. This arrangement ensures that the system exhibits spherical symmetry. One pair is in the ground state, while the other pair is in the first excited state. For such a system, the action is given by:
\begin{equation}\label{equ1}
  S=\int \sqrt{-g} d^4 x \left( \frac{R}{16\pi G} + {\cal L}_{0} + {\cal L}_{1}\right)\,,
\end{equation}
\noindent where $R$ is the Ricci scalar, $G$ is the gravitational constant,and ${\cal L}_{0}$ and ${\cal L}_{1}$ are the Lagrangians of the spinor fields in the ground state and first excited state, respectively,
\begin{equation}\label{equ2}
  {\cal L}_{0}=-i\sum\limits_{k=1}^2 \left[ \frac{1}{2}\left(\hat{D}_\mu\overline{\Psi}^{(k)}_0\gamma^\mu\Psi^{(k)}_0 - \overline{\Psi}^{(k)}_0\gamma^\mu\hat{D}_\mu\Psi^{(k)}_0\right) + \mu_0\overline{\Psi}^{(k)}_0\Psi^{(k)}_0\right]\,,
\end{equation}
\begin{equation}\label{equ3}
  {\cal L}_{1}=-i\sum\limits_{k=1}^2 \left[ \frac{1}{2}\left(\hat{D}_\mu\overline{\Psi}^{(k)}_1\gamma^\mu\Psi^{(k)}_1 - \overline{\Psi}^{(k)}_1\gamma^\mu\hat{D}_\mu\Psi^{(k)}_1\right) + \mu_1\overline{\Psi}^{(k)}_1\Psi^{(k)}_1\right]\,,
\end{equation}
 \noindent where $\Psi^{(k)}_n$ are spinors with mass $\mu_n$ and $n$ radial nodes, and the index $k=1,2$, corresponds to spinors with opposite spin. The variations of the action~(\ref{equ1}) with respect to the metric and the field functions yield the Einstein equations and the Dirac equation:
\begin{equation}\label{equ4}
  R_{\alpha\beta} - \frac{1}{2}g_{\alpha\beta}R = 8\pi G\left(T^0_{\alpha\beta} + T^1_{\alpha\beta}\right)\,,
\end{equation}
\begin{equation}\label{equ5}
 \gamma^\mu\hat{D}_\mu\Psi^{(k)}_n - \mu_n\Psi^{(k)}_n = 0\,,\quad n = 0,\,1\,,
\end{equation}
 \noindent where $T^0_{\alpha\beta}$ and $T^1_{\alpha\beta}$ are the energy-momentum tensors of the two sets of spinor fields,
 \begin{equation}
    T^0_{\alpha\beta} = \sum\limits_{k=1}^2 -\frac{i}{2}\left(\overline{\Psi}^{(k)}_0\gamma_\alpha\hat{D}_\beta\Psi^{(k)}_0 + \overline{\Psi}^{(k)}_0\gamma_\beta\hat{D}_\alpha\Psi^{(k)}_0 - \hat{D}_\alpha\overline{\Psi}^{(k)}_0\gamma_\beta\Psi^{(k)}_0 - \hat{D}_\beta\overline{\Psi}^{(k)}_0\gamma_\alpha\Psi^{(k)}_0\right)\,,
  \end{equation}
  \begin{equation}
    T^1_{\alpha\beta} = \sum\limits_{k=1}^2 -\frac{i}{2}\left(\overline{\Psi}^{(k)}_1\gamma_\alpha\hat{D}_\beta\Psi^{(k)}_1 + \overline{\Psi}^{(k)}_1\gamma_\beta\hat{D}_\alpha\Psi^{(k)}_1 - \hat{D}_\alpha\overline{\Psi}^{(k)}_1\gamma_\beta\Psi^{(k)}_1 - \hat{D}_\beta\overline{\Psi}^{(k)}_1\gamma_\alpha\Psi^{(k)}_1\right)\,.
  \end{equation}
Furthermore, it can be seen from Eq.~(\Ref{equ2}) and Eq.~(\Ref{equ3}) that the Lagrangians of the spinor fields are invariant under a global $U(1)$ transformation $\Psi^{(k)}_n\rightarrow e^{i\alpha}\Psi^{(k)}_n$, where $\alpha$ is an arbitrary constant. As a result, the system possesses a conserved current:
\begin{equation}
  J_n^{\mu} = \overline{\Psi}_n\gamma^\mu\Psi_n,\quad n = 0,\,1\,.
\end{equation}
Integrating the timelike component of the conserved current over a spacelike hypersurface $\cal{S}$ yields the Noether charge:
\begin{equation}\label{equ9}
 Q_n = \int_{\cal S}J_n^t\,,\quad n = 0,\,1\,.
\end{equation}

To construct spherically symmetric solutions, we choose the metric to be of the following form:
\begin{equation}\label{equ10}
 ds^2 = -N(r)\sigma^2(r)dt^2 + \frac{dr^2}{N(r)} + r^2\left(d\theta^2 + \sin^2\theta d\varphi^2\right)\,,
\end{equation}
where $N(r) = 1 - {2m(r)}/{r}$. The two pairs of Dirac fields are given by~\cite{Herdeiro:2017fhv}:
\begin{equation}
  \Psi^{(1)}_n = \begin{pmatrix}\cos(\frac{\theta}{2})[(1 + i)f_n(r) - (1 - i)g_n(r)]\\ i\sin(\frac{\theta}{2})[(1 - i)f_n(r) - (1 + i)g_n(r)]\\-i\cos(\frac{\theta}{2})[(1 - i)f_n(r) - (1 + i)g_n(r)]\\ -\sin(\frac{\theta}{2})[(1 + i)f_n(r) - (1 - i)g_n(r)] \end{pmatrix}e^{i\frac{\varphi}{2} - i\omega_nt}\,,
 \end{equation}
 \begin{equation}
  \Psi^{(2)}_n = \begin{pmatrix}i\sin(\frac{\theta}{2})[(1 + i)f_n(r) - (1 - i)g_n(r)]\\ \cos(\frac{\theta}{2})[(1 - i)f_n(r) - (1 + i)g_n(r)]\\ \sin(\frac{\theta}{2})[(1 - i)f_n(r) - (1 + i)g_n(r)]\\ i\cos(\frac{\theta}{2})[(1 + i)f_n(r) - (1 - i)g_n(r)] \end{pmatrix}e^{-i\frac{\varphi}{2} - i\omega_nt}\,,
\end{equation}
where the index $n$ also represents the number of radial nodes, and $\omega_n$ is the frequency of the Dirac field with $n$ radial nodes. We only consider the cases of $n=0,1$ in this paper. 

Substituting the above ansatz into the field equations~(\ref{equ4}--\ref{equ5}) yields the following system of ordinary differential equations:
\begin{equation}\label{equ13}
  f^\prime_n + \left(\frac{N^\prime}{4N} + \frac{\sigma^\prime}{2\sigma} + \frac{1}{r\sqrt{N}} + \frac{1}{r}\right)f_n + \left(\frac{\mu_n}{\sqrt{N}} - \frac{\omega_n}{N\sigma}\right)g_n = 0 \,,
 \end{equation}
 \begin{equation}\label{equ14}
  g^\prime_n + \left(\frac{N^\prime}{4N} + \frac{\sigma^\prime}{2\sigma} - \frac{1}{r\sqrt{N}} + \frac{1}{r}\right)g_n + \left(\frac{\mu_n}{\sqrt{N}} + \frac{\omega_n}{N\sigma}\right)f_n = 0\,,
 \end{equation}
 \begin{equation}\label{equ15}
  m^\prime = \frac{32\pi G r^2}{\sqrt{N}\sigma} \left[\omega_0(f^2_0 + g^2_0) + \omega_1(f^2_1 + g^2_1) \right] \,,
 \end{equation}
 \begin{equation}\label{equ16}
  \frac{\sigma^\prime}{\sigma} = \frac{32\pi Gr}{\sqrt{N}}\left[g_0f_0^\prime - f_0g_0^\prime + g_1f_1^\prime - f_1g_1^\prime + \frac{\omega_0(f_0^2 + g_0^2)}{N\sigma} + \frac{\omega_1(f_1^2 + g_1^2)}{N\sigma}\right]\,.
\end{equation}
And the Noether charges of the system are:
\begin{equation}\label{equ17}
 Q_n = 16\pi\int_0^\infty r^2\frac{f^2_n + g^2_n}{\sqrt{N}}dr\,,\quad n = 0,\,1\,.
\end{equation}

\section{Boundary conditions}\label{sec3}

To solve the system of ordinary differential equations obtained in the previous section, it is necessary to impose appropriate boundary conditions. First, for a regular, asymptotically flat spacetime, the metric function should satisfy the following boundary conditions:
\begin{equation}
m(0) = 0,\qquad \sigma(0) = \sigma_0,\qquad m(\infty) = M,\qquad \sigma(\infty) = 1,
\end{equation}
where the ADM mass $M$ and $\sigma_0$ are unknown constants. In addition, the matter field vanishes at infinity:
\begin{equation}
  f_n(\infty) = 0,\qquad g_n(\infty) = 0\,,\quad n=0\,,1\,.
\end{equation}
Expanding equations~(\ref{equ13}--\ref{equ14}) near the origin, we obtain that the field functions satisfy the following condition at the origin:
\begin{equation}
   f_n(0) = 0,\qquad \left.\frac{dg_n(r)}{dr}\right|_{r = 0} = 0.
\end{equation}

\section{Numerical results}\label{sec4}

In order to facilitate numerical calculations, we employ the following dimensionless quantities:
\begin{equation}
  r \rightarrow r/\rho,\quad f_n \rightarrow \frac{\sqrt{4\pi\rho}}{M_{Pl}}f_n,\quad g_n \rightarrow \frac{\sqrt{4\pi\rho}}{M_{Pl}}g_n,\quad \omega_n \rightarrow \omega_n\rho,\quad \mu_n \rightarrow \mu_n\rho,
\end{equation}
where $M_{Pl} = 1/\sqrt{G}$ is the Planck mass. For any physical quantity $A$, we denote the dimensionless quantity under the conditions of $\rho = 1/\mu_0$ and $\rho = 1/\mu_1$ as $\tilde{A}$ and $\overline{A}$, respectively. To facilitate computation, we define the radial coordinate $x$ as follows:
\begin{equation}
x = \frac{\tilde{r}}{1+\tilde{r}},
\end{equation}
where the radial coordinate $\tilde{r}\in[0,\infty)$, so $x\in[0,1]$. We utilize the finite element method to numerically solve the system of differential equations. The integration region $0\le x\le 1$ is discretized into 1000 grid points. The Newton-Raphson method is employed as our iterative approach. To ensure the accuracy of the computational results, we enforce a relative error criterion of less than $10^{-5}$.

To ensure the accuracy of our numerical calculations, it is crucial to verify the numerical precision by validating physical constraints~\cite{Herdeiro:2021teo,Herdeiro:2022ids}, in addition to employing the aforementioned numerical analysis methods. In this study, we examined the equivalence between the asymptotic mass and the Komar mass of the numerical solution, and the results consistently showed a discrepancy of less than $10^{-5}$ between these two quantities.

We denote the Dirac stars in the ground state and the first excited state as $D_0$ and $D_1$, respectively, and the multi-state Dirac stars as $D_0D_1$ (or MSDSs). The representation of the gamma matrices and the choice of the tetrad in the Dirac equation are the same as in~\cite{Liang:2022mjo}.

\subsection{ Synchronized frequency }

Through the analysis of the numerical calculations, we observed that the solution of the MSDSs under synchronized frequency depends on the ratio of the masses of the excited state and ground state Dirac fields: $\mu_1/\mu_0$, which is the dimensionless mass $\tilde{\mu}_1$ of the first excited state Dirac field. By varying the value of the mass $\tilde{\mu}_1$, various MSDSs solutions can be obtained. Based on the number of branches in the obtained MSDSs solutions, we categorize them into single-branch and double-branch solutions. For $0.7694 \le \tilde{\mu}_1 < 1$, the MSDSs solution corresponds to a single-branch solution, whereas for $0.7573 \le \tilde{\mu}_1 < 0.7694$, the MSDSs solution corresponds to a double-branch solution. In the following discussion, we will delve into the characteristics of these two types of solutions.

\begin{figure}[!htbp]
\begin{center}
    \includegraphics[height=.24\textheight]{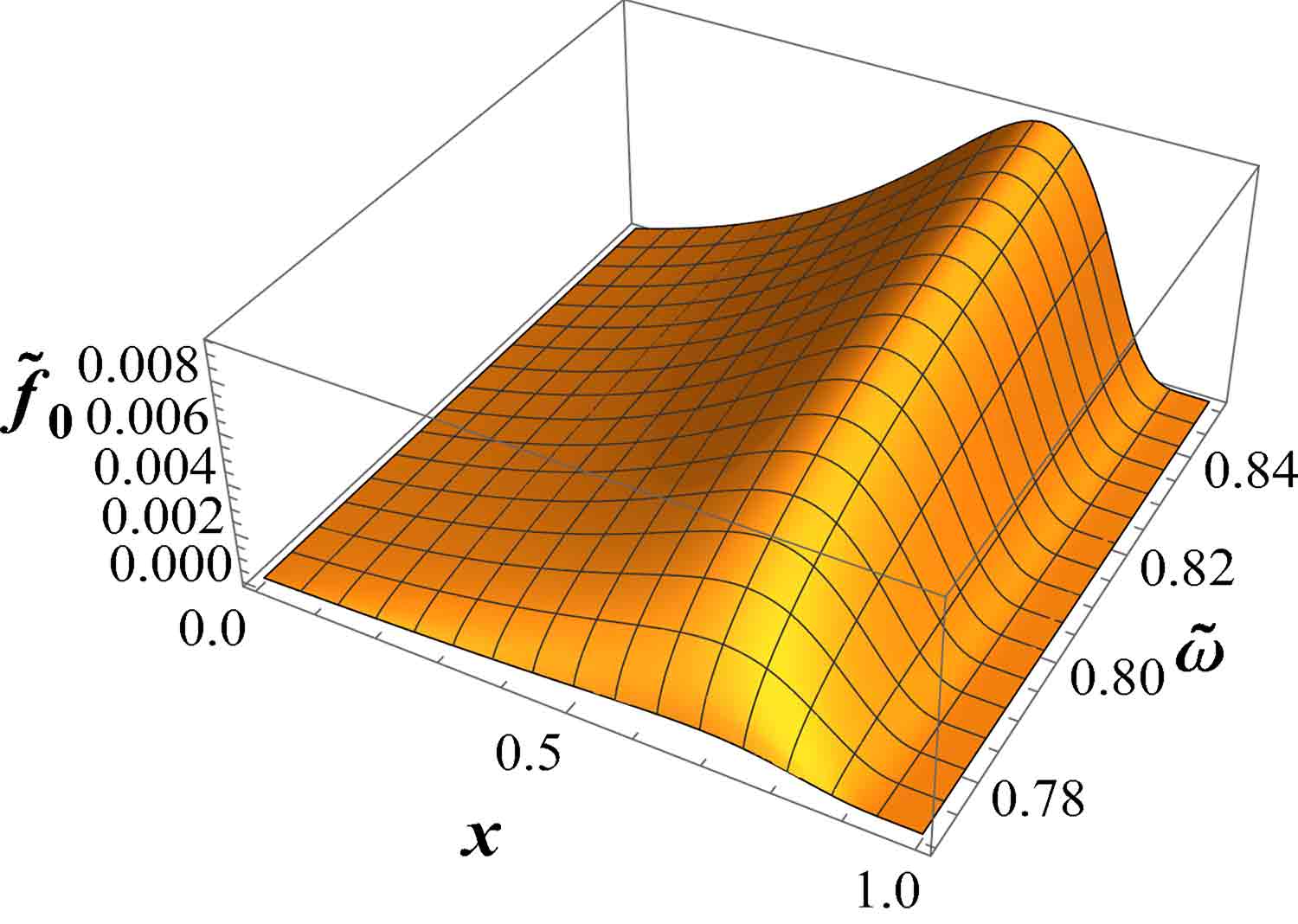}
    \includegraphics[height=.24\textheight]{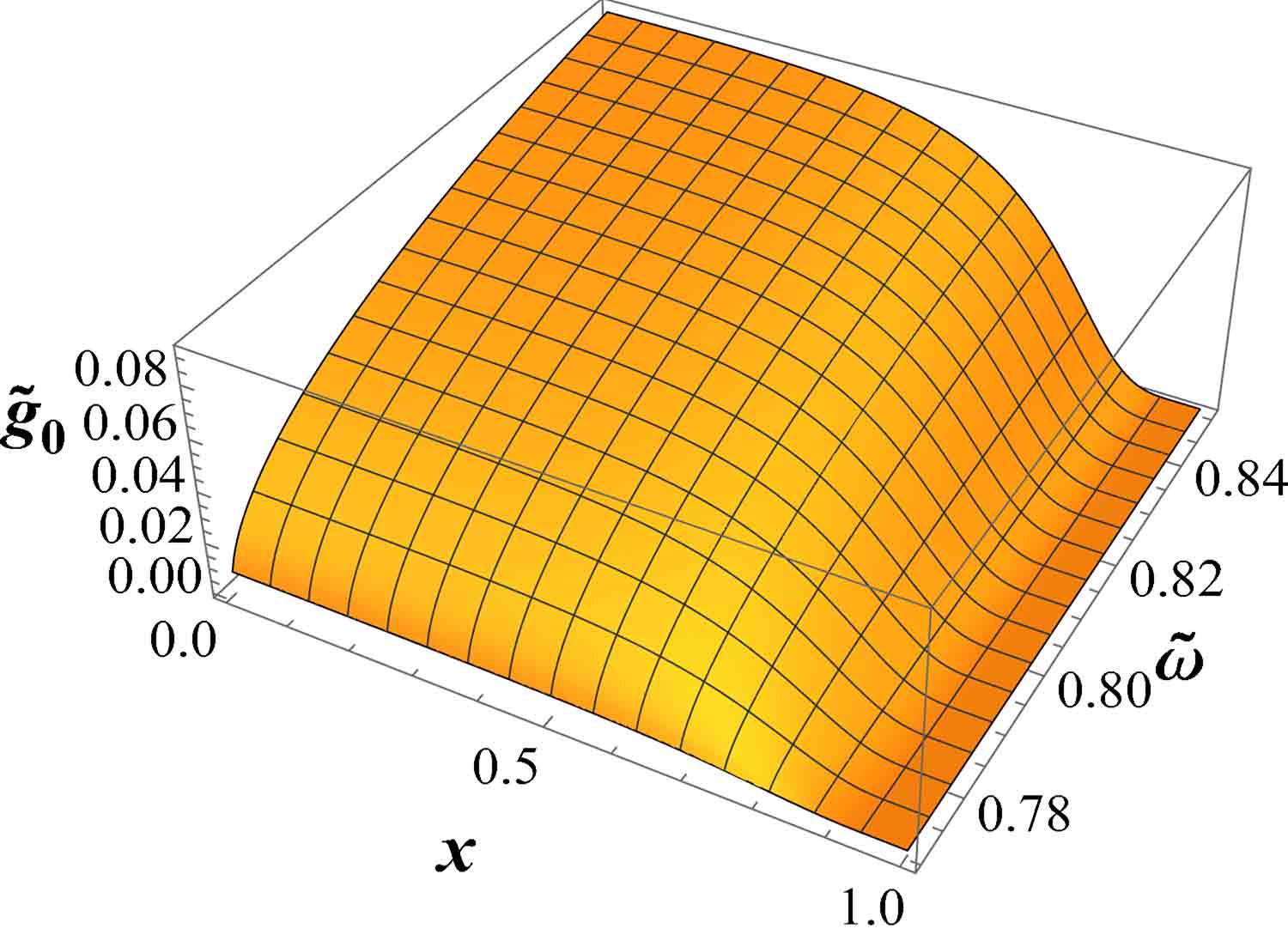}
    \includegraphics[height=.24\textheight]{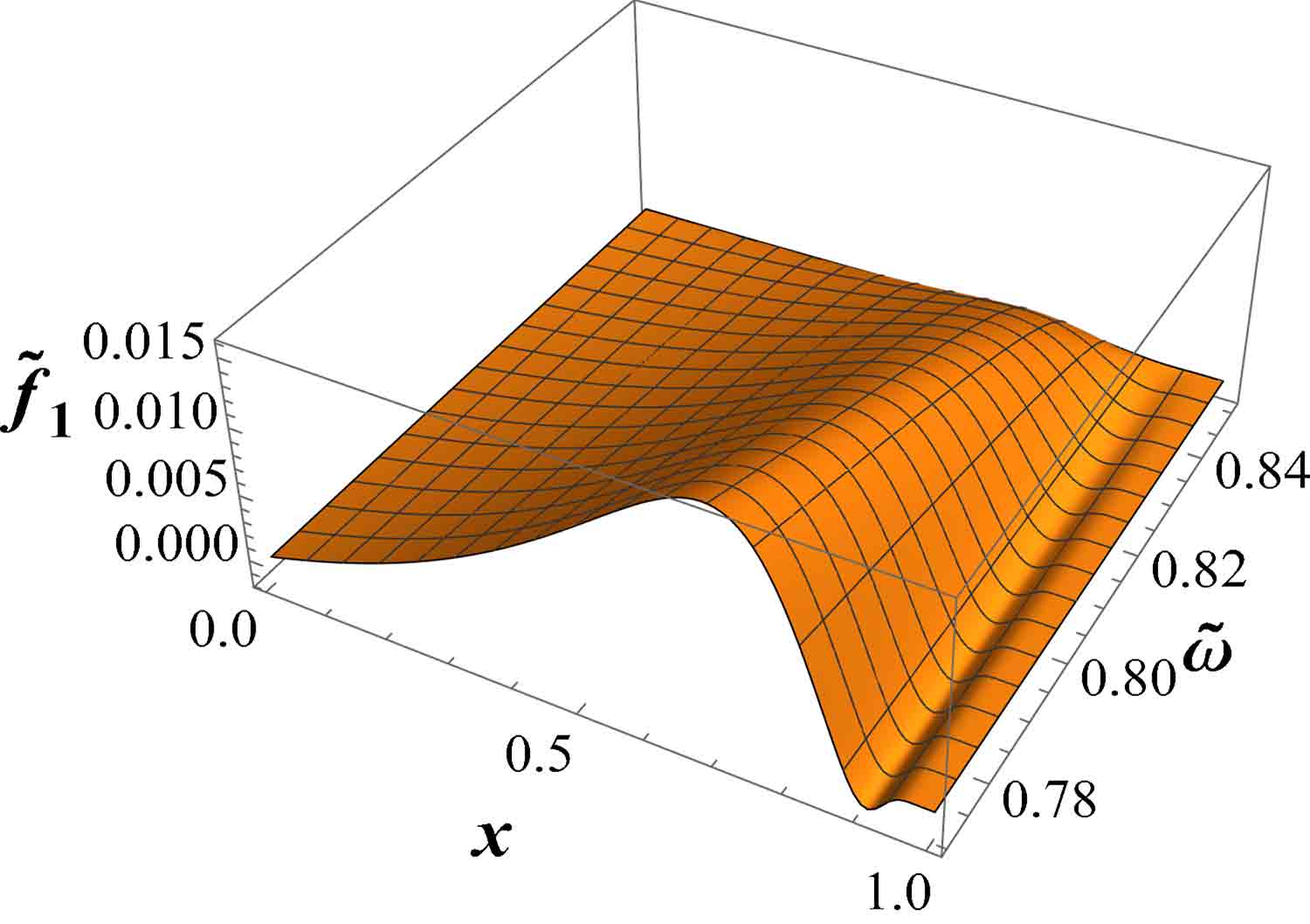}
    \includegraphics[height=.24\textheight]{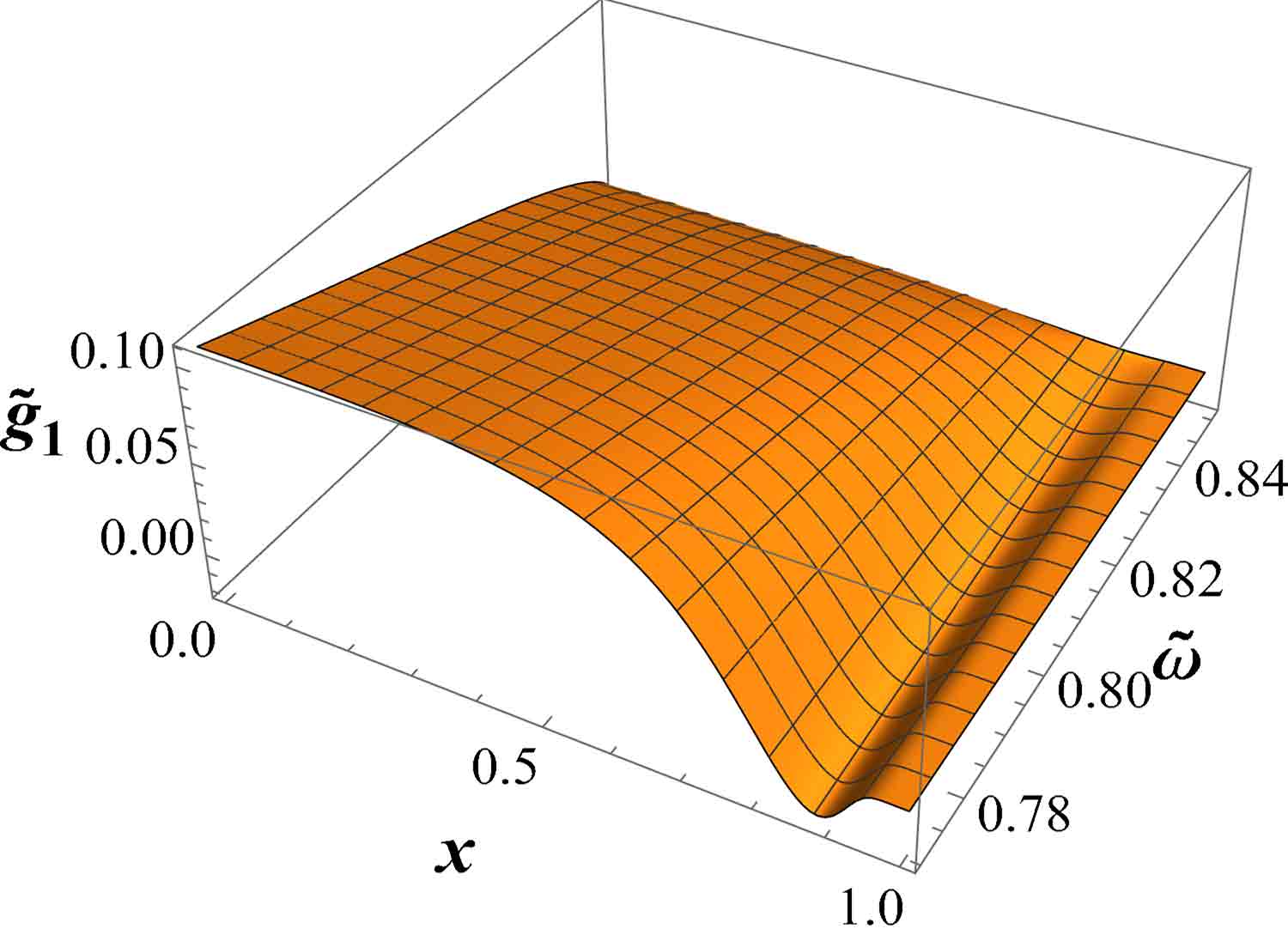}
\end{center}
\caption{The matter functions $\tilde{f}_0$, $\tilde{g}_0$, $\tilde{f}_1$ and $\tilde{g}_1$ as functions of $x$ and $\tilde{\omega}$ for $\tilde{\mu}_1 = 0.898$.}
\label{sf_single_f_g}
\end{figure}

\subsubsection{Single-branch}

\begin{figure}[!htbp]
\begin{center}
    \includegraphics[height=.23\textheight]{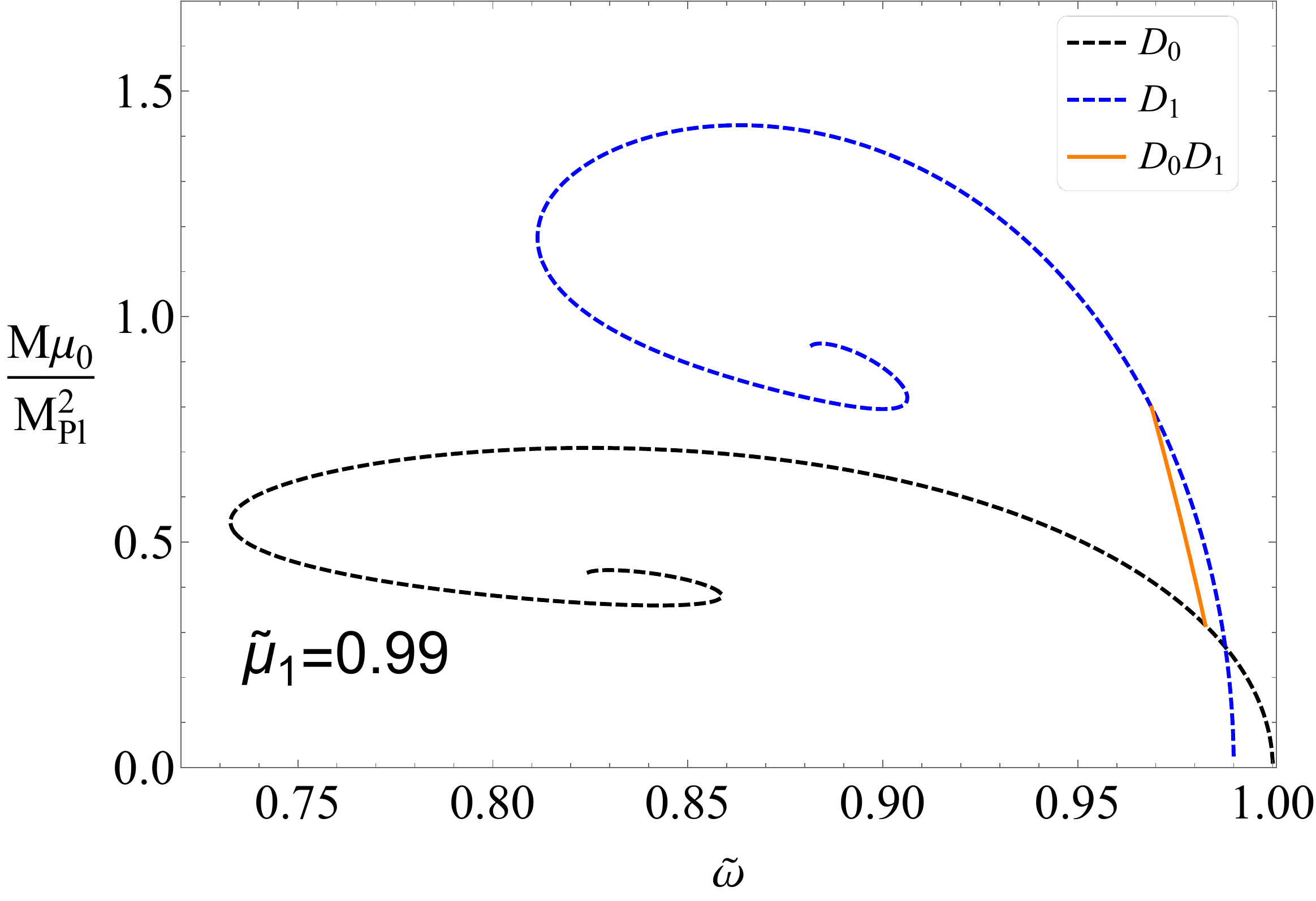}
    \includegraphics[height=.23\textheight]{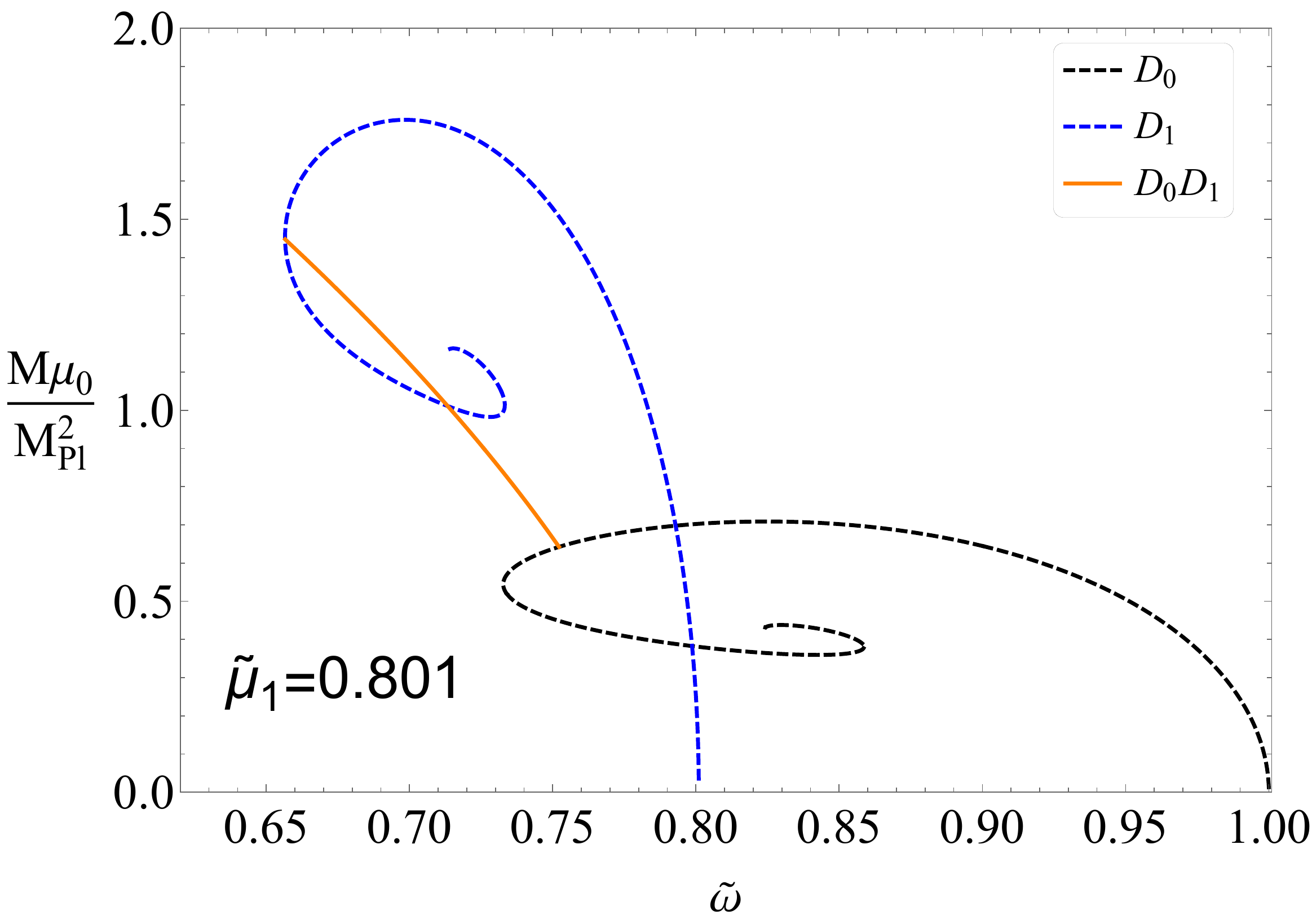}
    \includegraphics[height=.23\textheight]{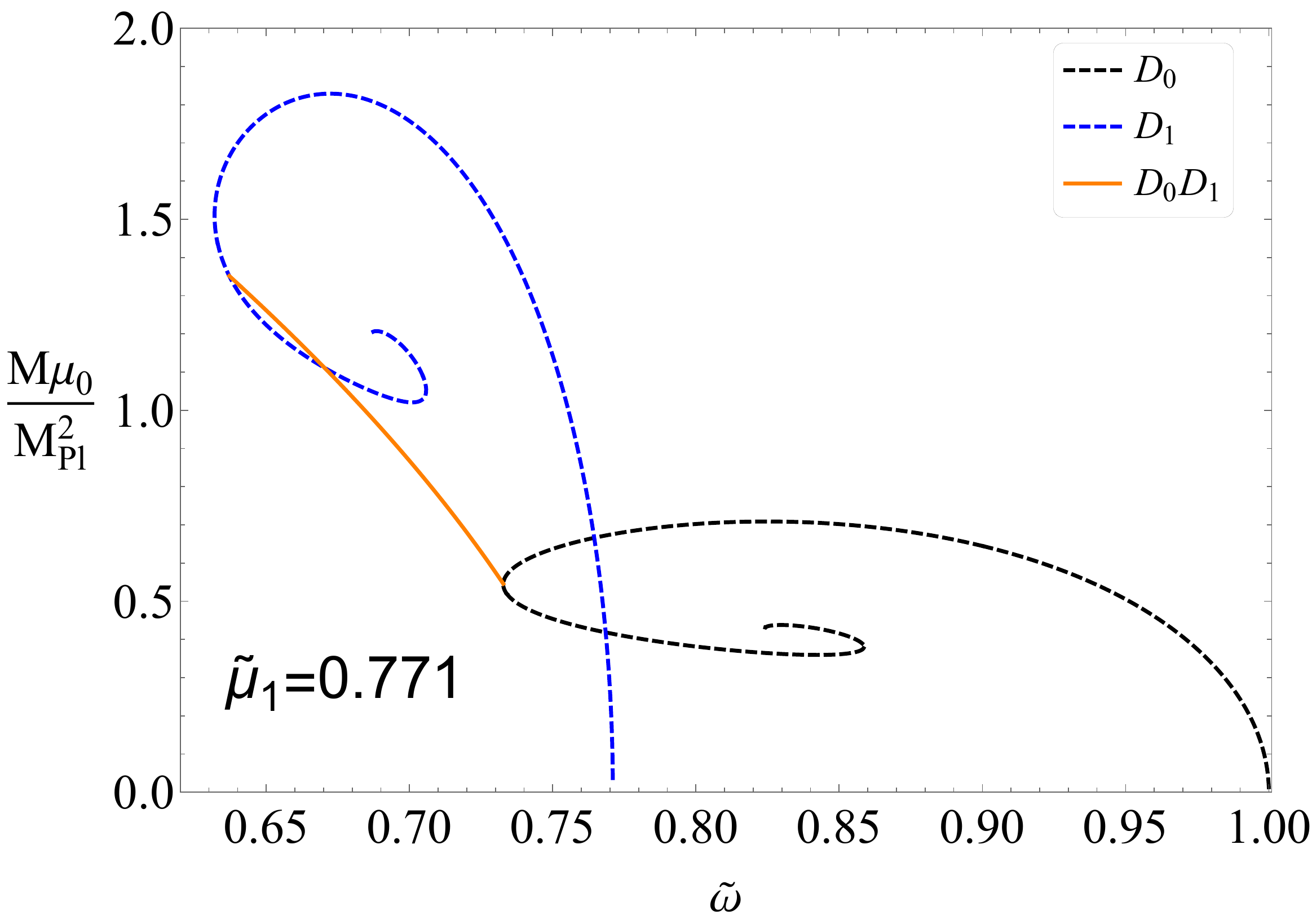}
    \includegraphics[height=.23\textheight]{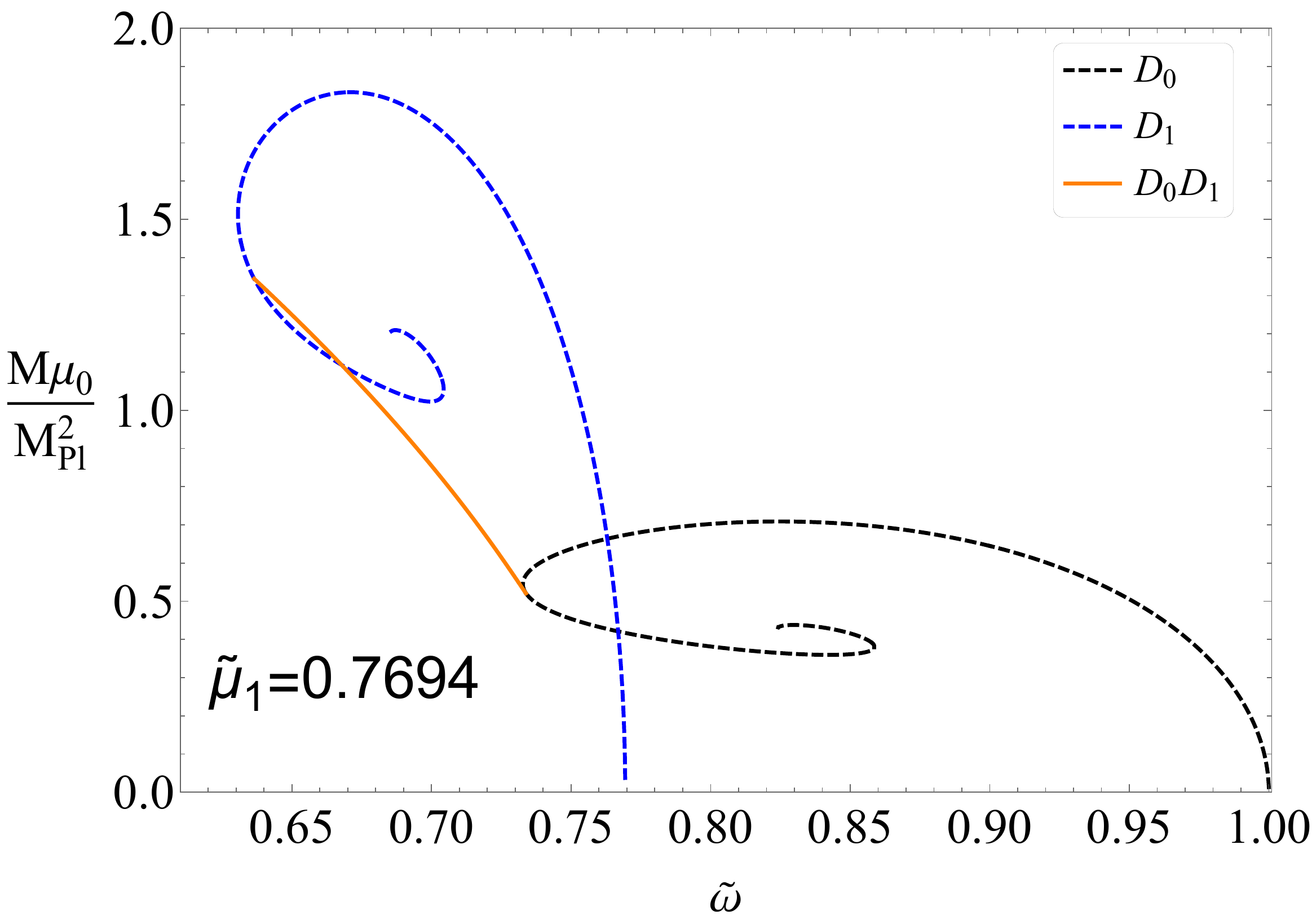}
    \includegraphics[height=.23\textheight]{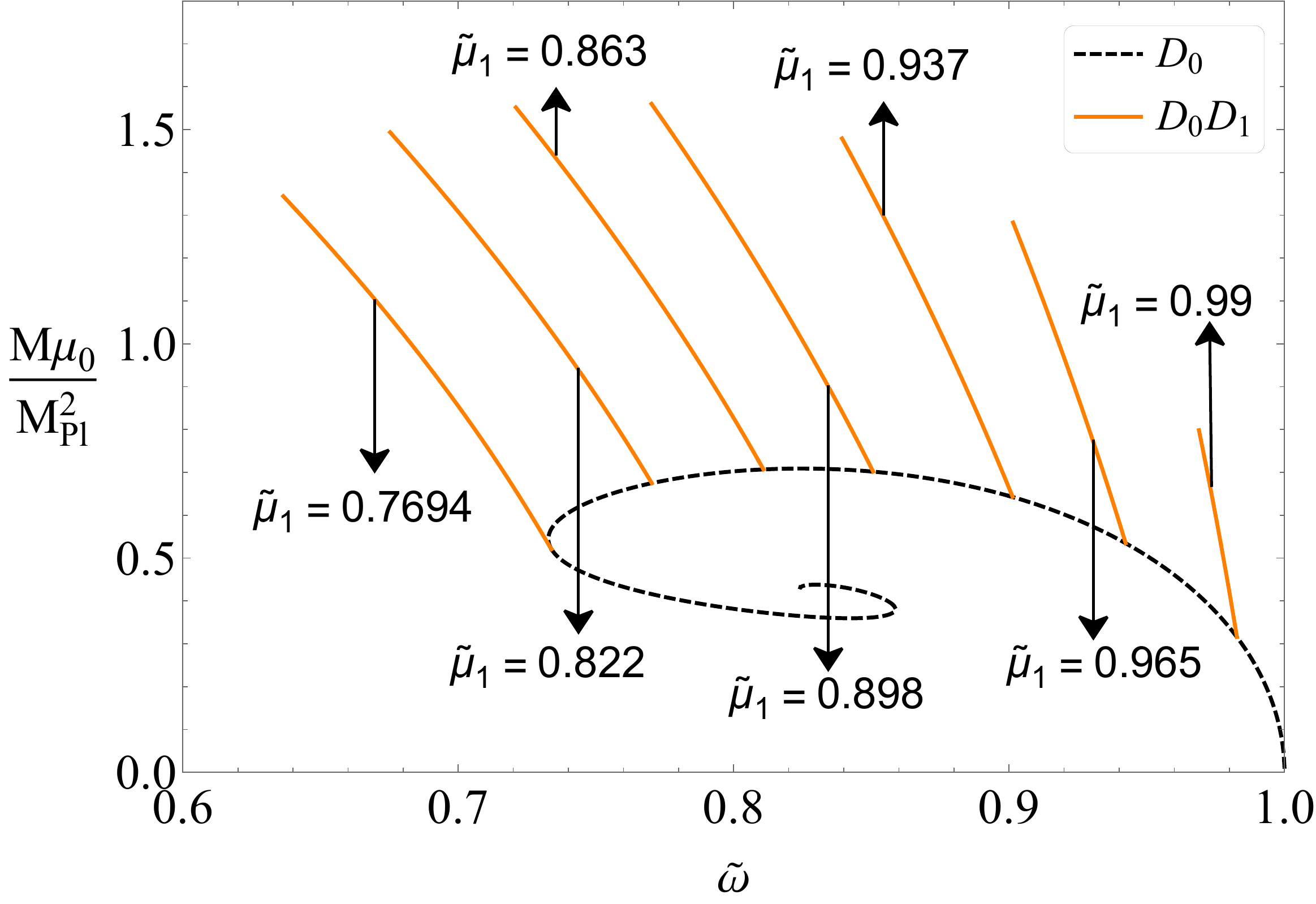}
    \includegraphics[height=.23\textheight]{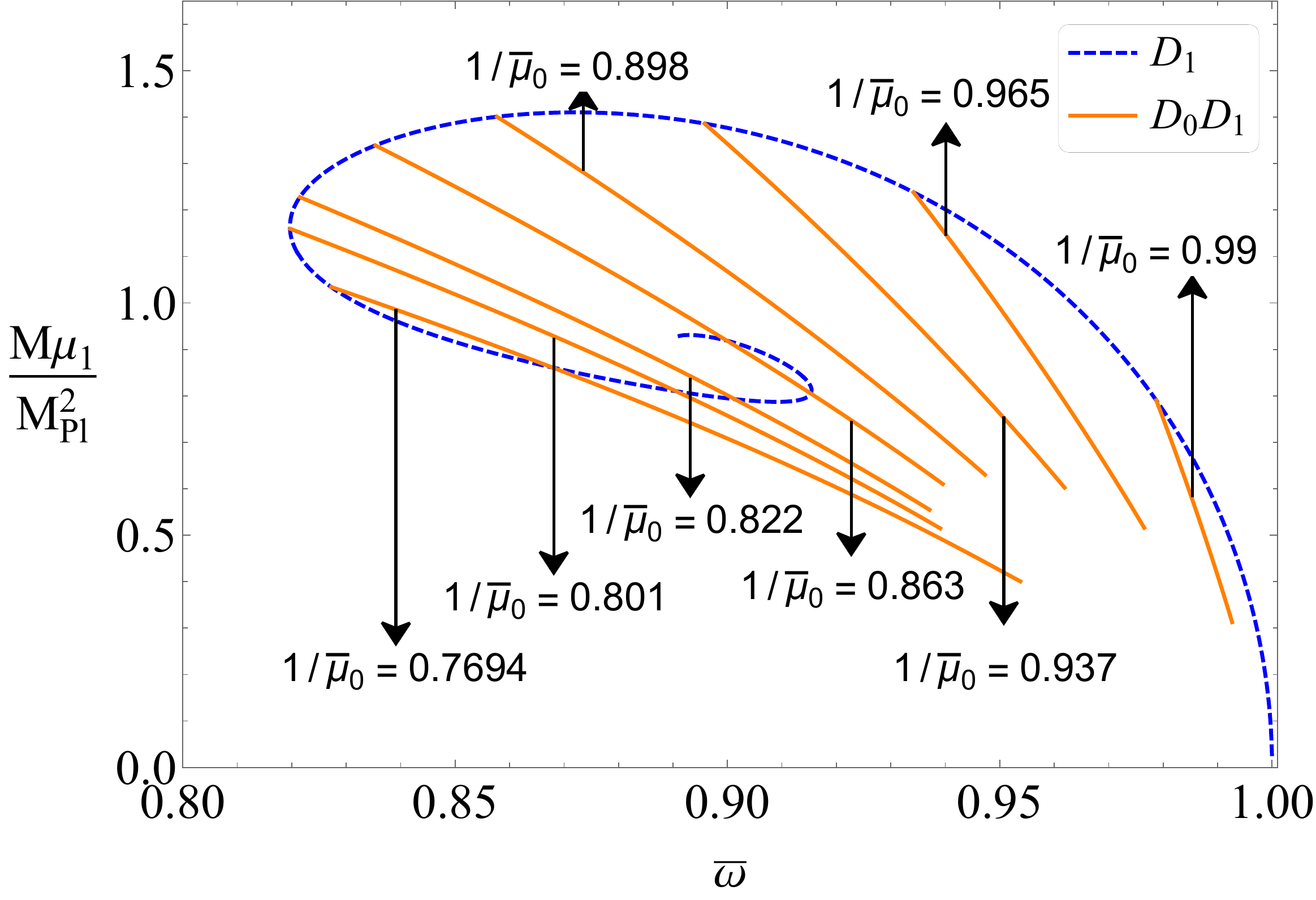}
\end{center}
\caption{ The ADM mass $M$ of the MSDSs as a function of the synchronized frequency $\tilde{\omega}$ ($\overline{\omega}$) for several values of  $\tilde{\mu}_1$ ($1/\overline{\mu}_0$).
}
\label{sf_single_k-adm}
\end{figure}

We first discuss the more general single-branch solution in the multi-field system~\cite{Liang:2022mjo,Ma:2023vfa,Zeng:2021oez,Li:2020ffy,Li:2019mlk}. The characteristic change in the radial profile of the matter field forming the MSDSs as the synchronized frequency $\tilde{\omega}$ continuously varies is shown in Fig.~\ref{sf_single_f_g}. The field functions depicted in the figure were obtained under the condition of a fixed mass $\tilde{\mu}_1 = 0.898$. The two upper plots correspond to the ground state Dirac field functions $\tilde{f}_0$ and $\tilde{g}_0$, revealing the absence of nodes in these field functions. The two lower plots represent the excited state Dirac field functions $\tilde{f}_1$ and $\tilde{g}_1$, where each of these field functions exhibits a single node, indicating that the Dirac field is in the first excited state. It is evident that as the synchronized frequency $\tilde{\omega}$ increases, the peak values of the ground state Dirac field functions $\tilde{f}_0$ and $\tilde{g}_0$ also gradually increase, while the peak values of the excited state Dirac field functions $\tilde{f}_1$ and $\tilde{g}_1$ gradually decrease. Additionally, when the synchronized frequency approaches the minimum value at which the single-branch solution can exist, the ground-state Dirac field tends to disappear, while the opposite is true for the excited state Dirac field.

Next, we analyze the characteristics of the ADM mass $M$ of the single-branch solution of the MSDSs as the synchronized frequency $\tilde{\omega}$ varies. As shown in Fig.~\ref{sf_single_k-adm}, the black dashed line represents the ground state Dirac stars ($D_0$), the blue dashed line represents the first excited state Dirac stars ($D_1$), and the orange line represents the MSDSs ($D_0D_1$). The ADM mass of the system monotonically decreases as the synchronized frequency increases. It can be observed that the upper end of the orange line intersects with the blue dashed line, where the MSDSs degenerate into the $D_1$; the lower end of the orange line intersects with the black dashed line, where the MSDSs degenerate into the $D_0$. This degeneration of the MSDSs is manifested in Fig.~\ref{sf_single_f_g} as the disappearance of the ground state or excited state field functions. 

Furthermore, when the mass $\tilde{\mu}_1$ of the excited state Dirac field is small, both endpoints of the orange line are located on the first branch of the black and blue dashed lines. As the mass $\tilde{\mu}_1$ decreases to $0.801$, the intersection point of the orange line and the blue dashed line is located at the inflection point between the first and second branches of the blue dashed line. As the mass $\tilde{\mu}_1$ further decreases to $0.771$, the intersection point of the orange line and the black dashed line is located at the inflection point between the first and second branches of the black dashed line. When $\tilde{\mu}_1$ decreases to the minimum frequency at which the single-branch solution can exist, $0.7694$, the orange line and the black dashed line exhibit a "tangent" form. The two plots at the bottom of Fig.~\ref{sf_single_k-adm} intuitively illustrate the variation of the orange line endpoints. It should be noted that the horizontal axis of the lower-right plot represents $\overline{\omega}$, not $\tilde{\omega}$, in order to show the changing trend of the intersection points of the orange line and the blue dashed line.

\begin{figure}[!htbp]
\begin{center}
    \includegraphics[height=.20\textheight]{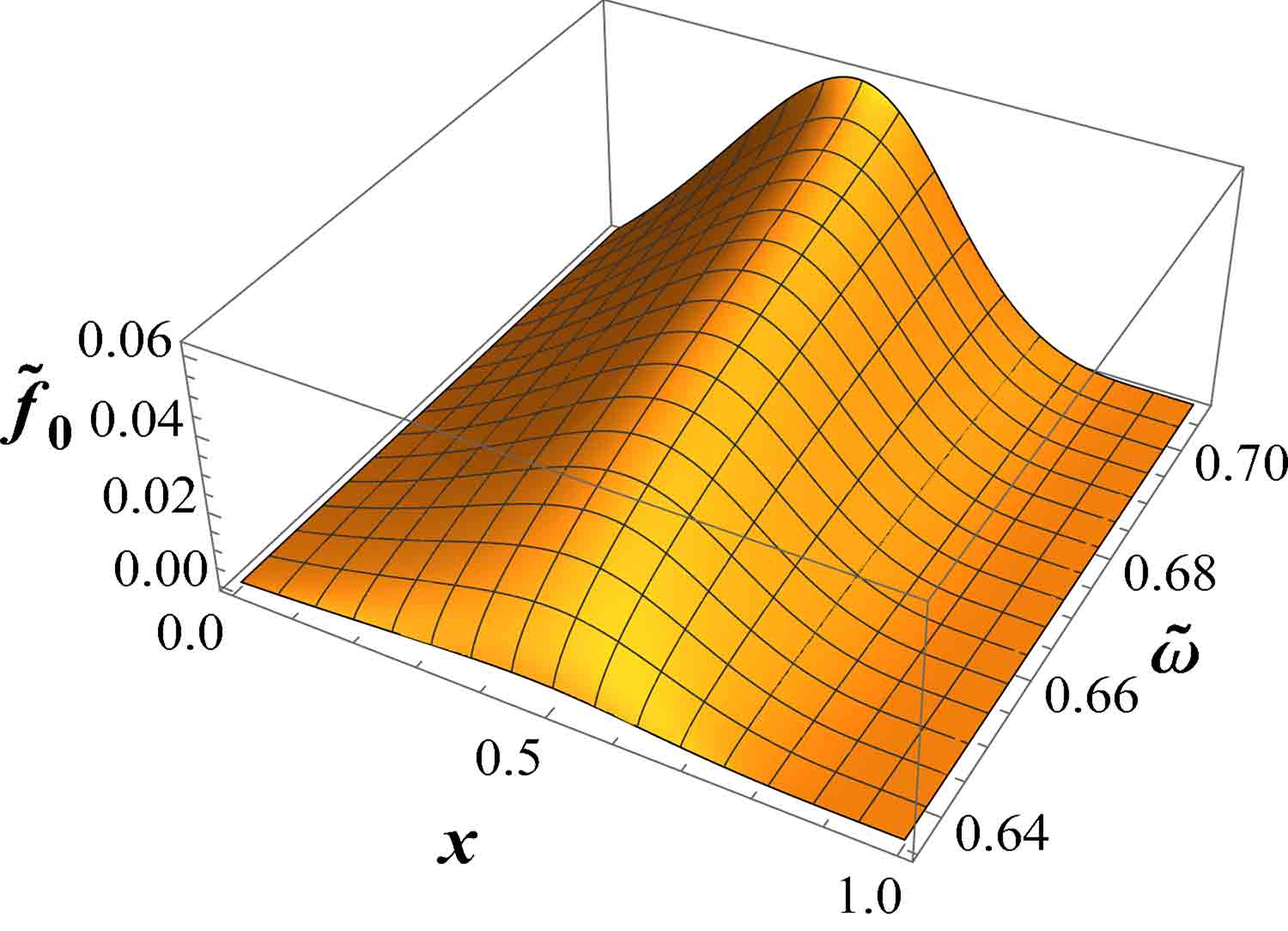}
    \includegraphics[height=.20\textheight]{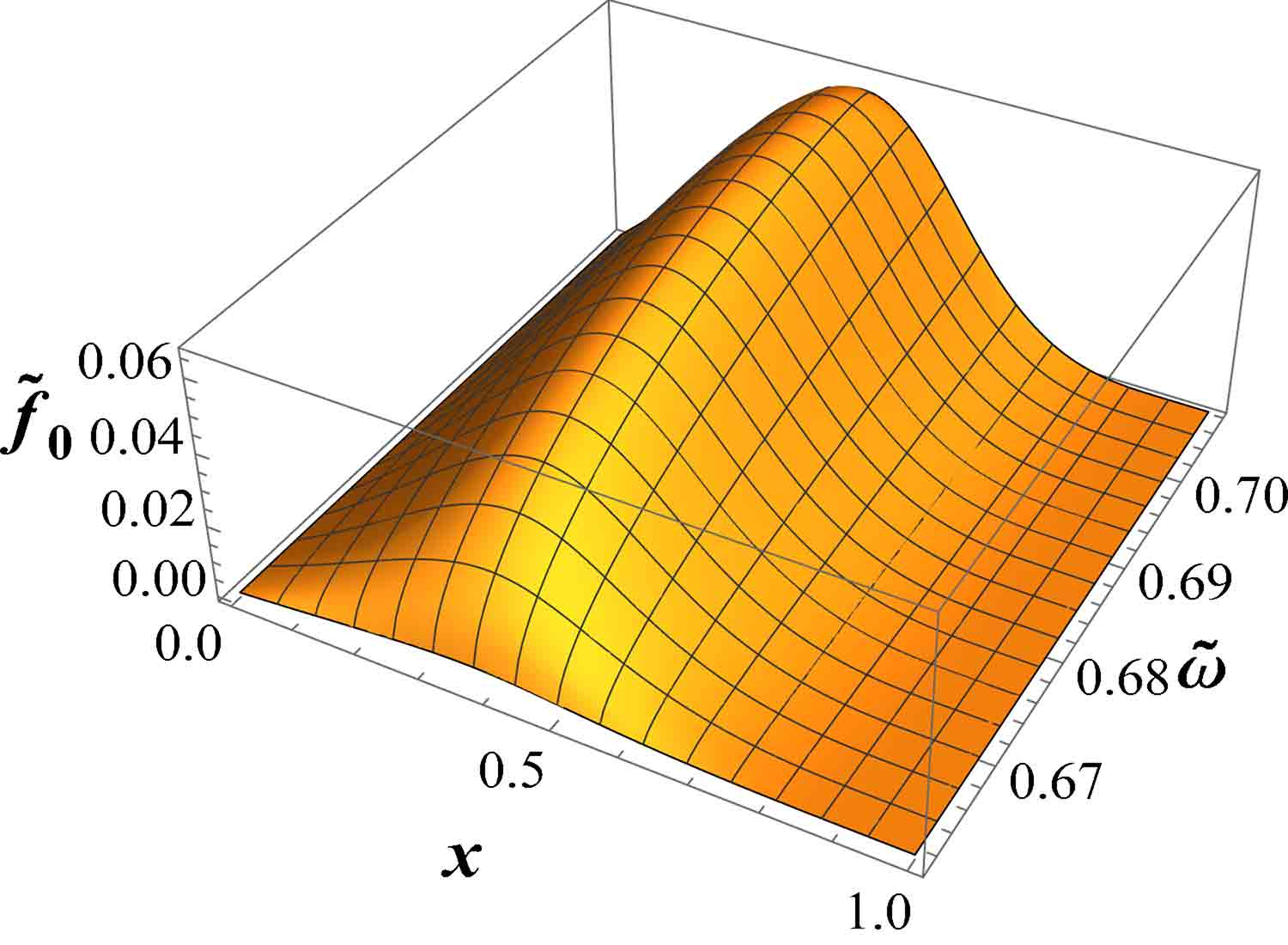}
    \includegraphics[height=.20\textheight]{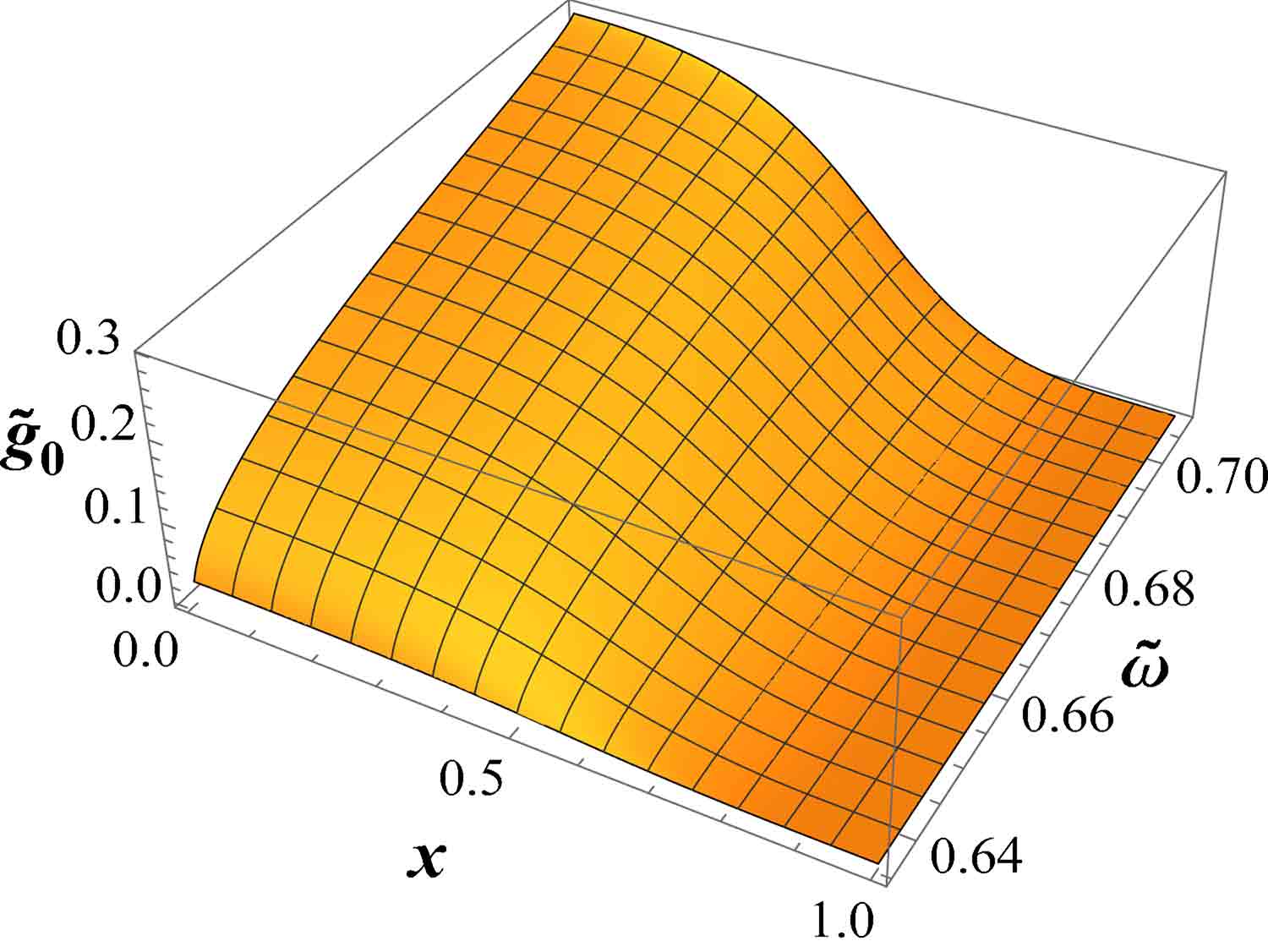}
    \includegraphics[height=.20\textheight]{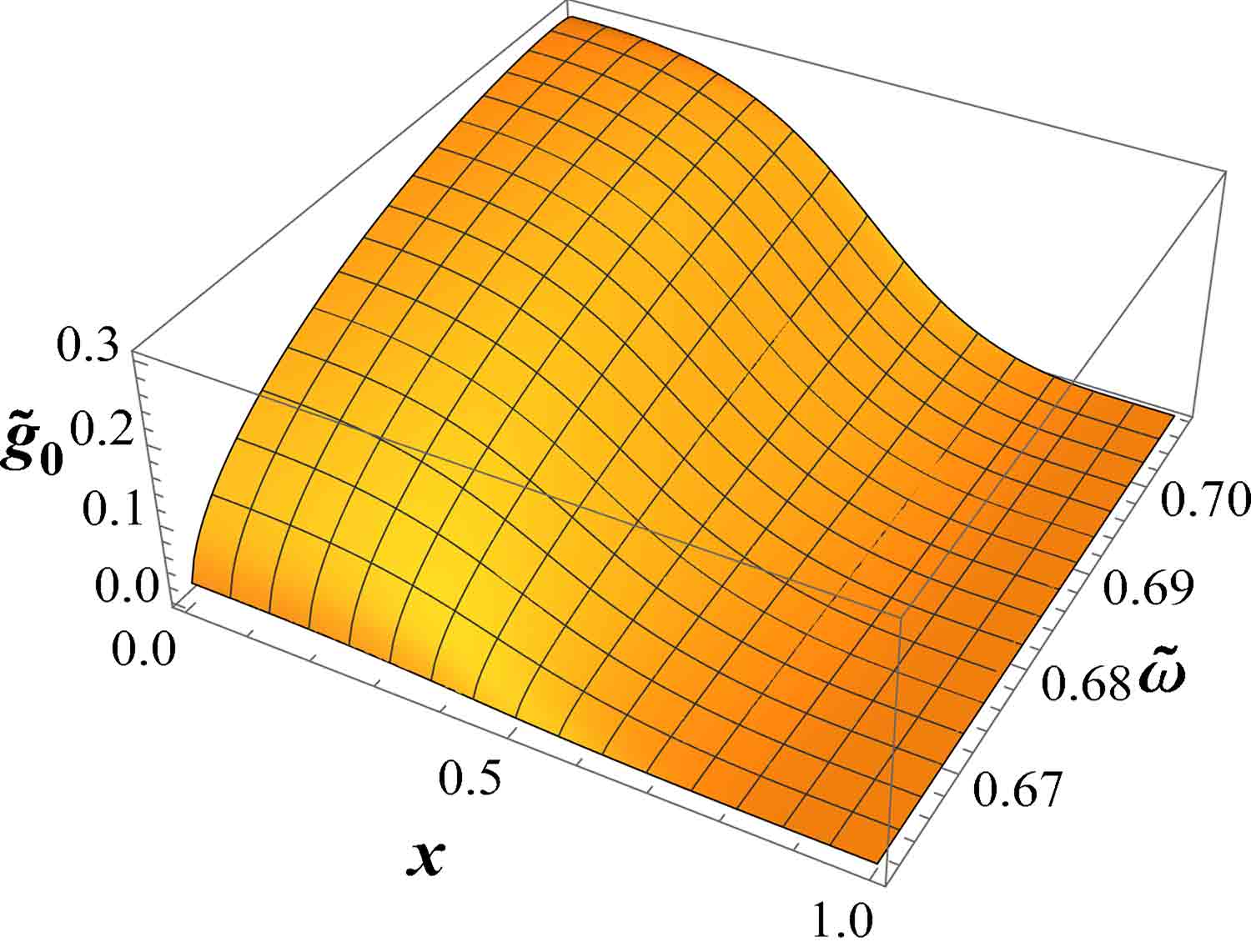}
    \includegraphics[height=.20\textheight]{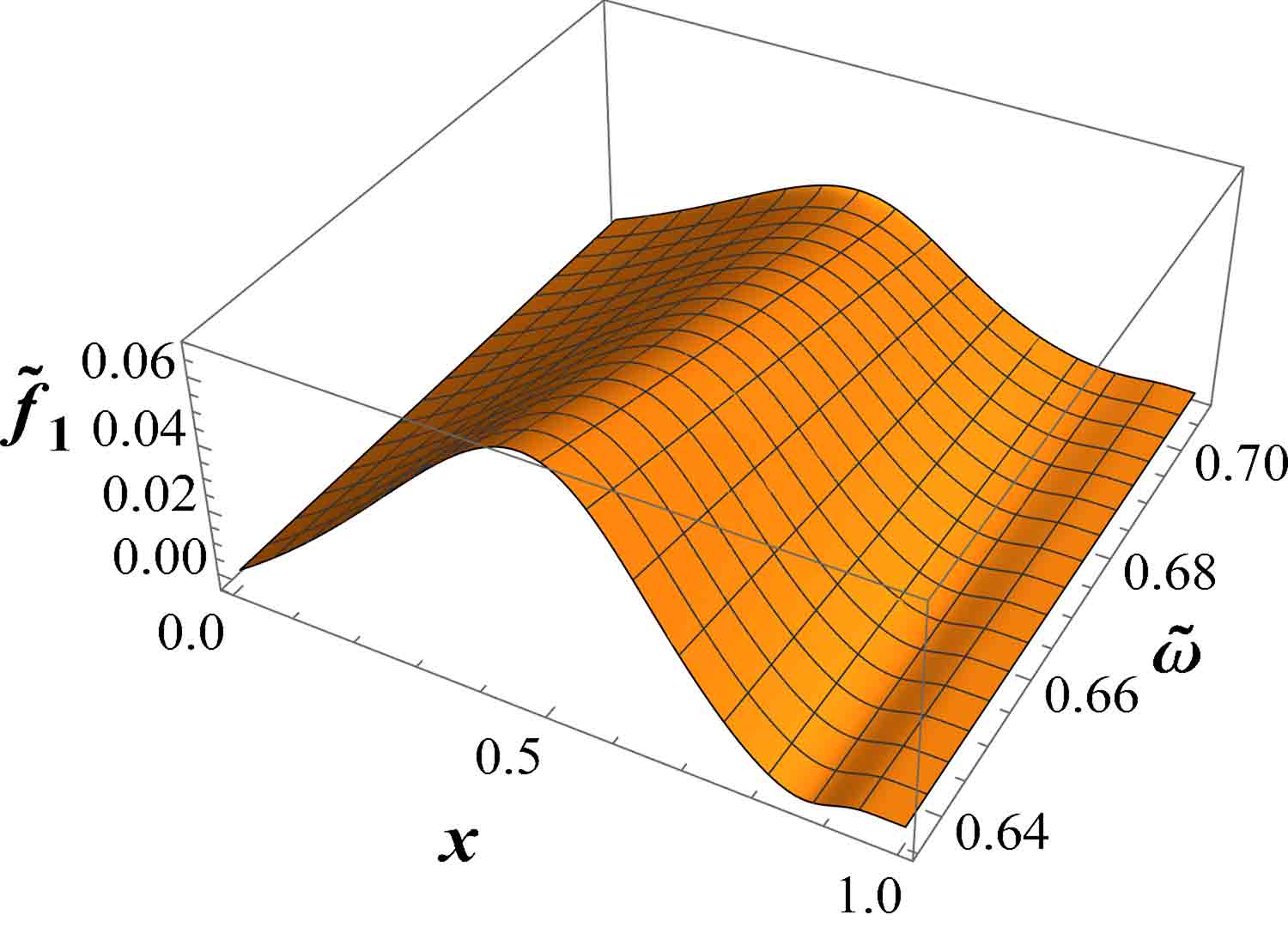}
    \includegraphics[height=.20\textheight]{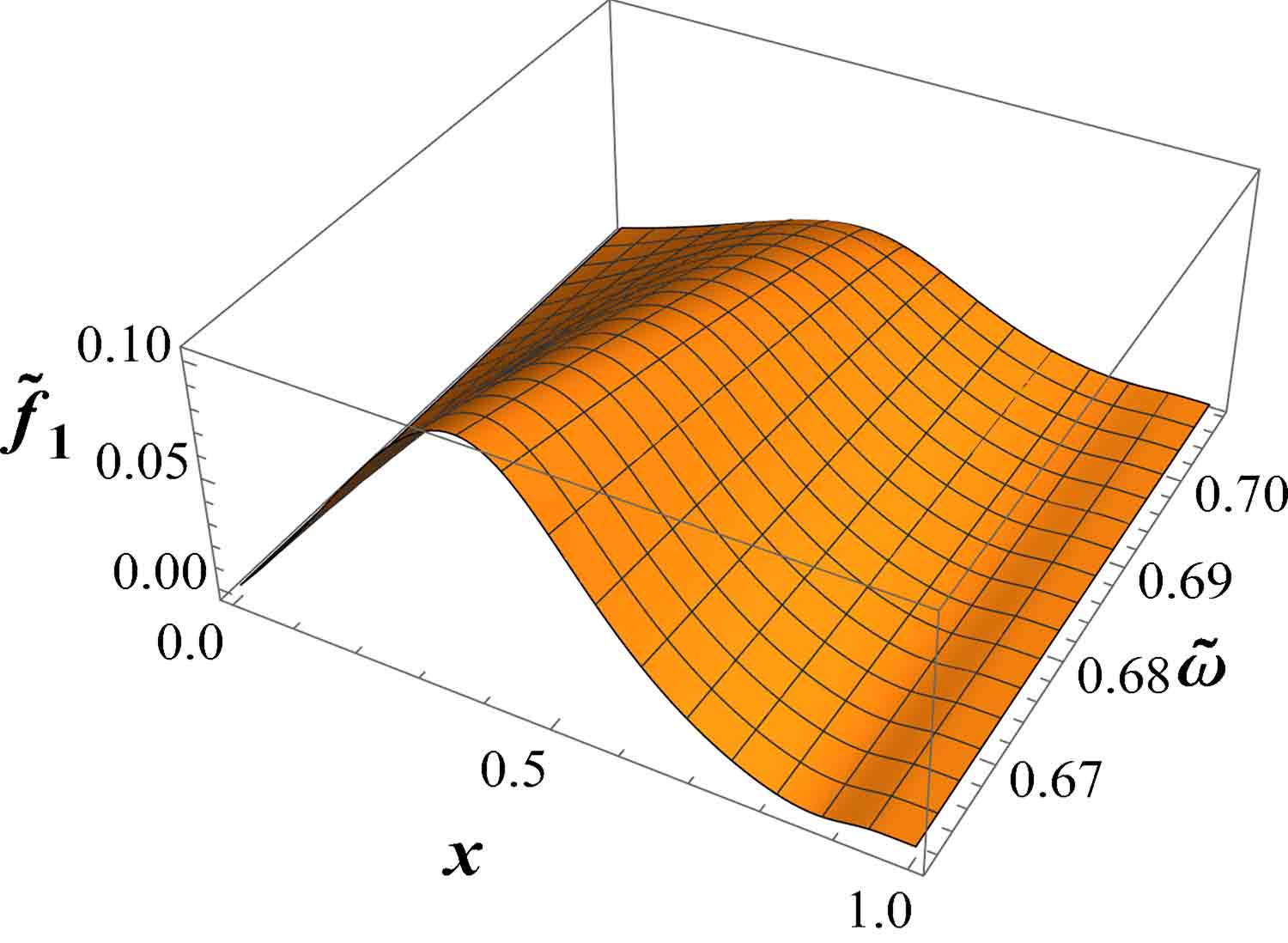}
    \includegraphics[height=.20\textheight]{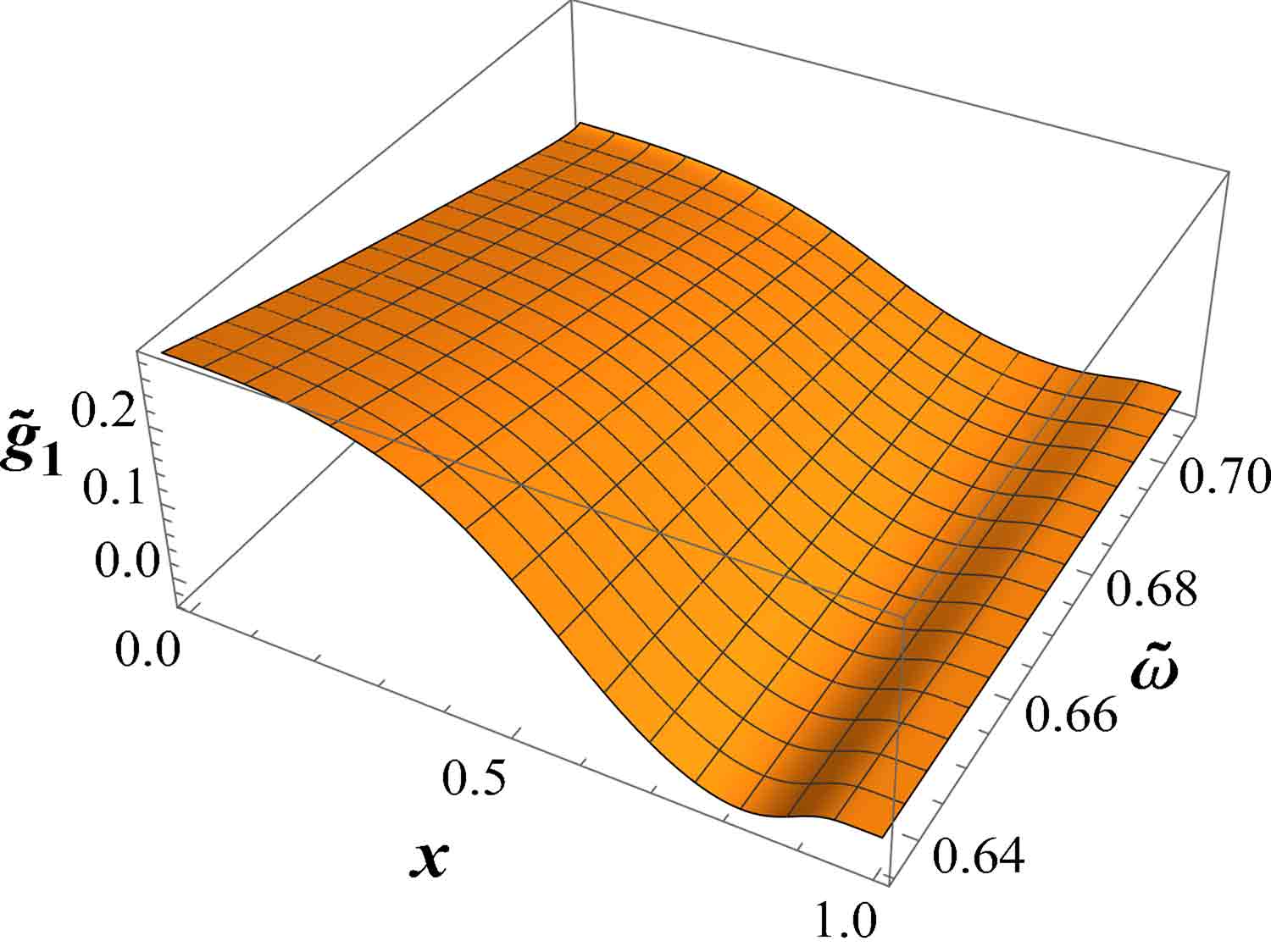}
    \includegraphics[height=.20\textheight]{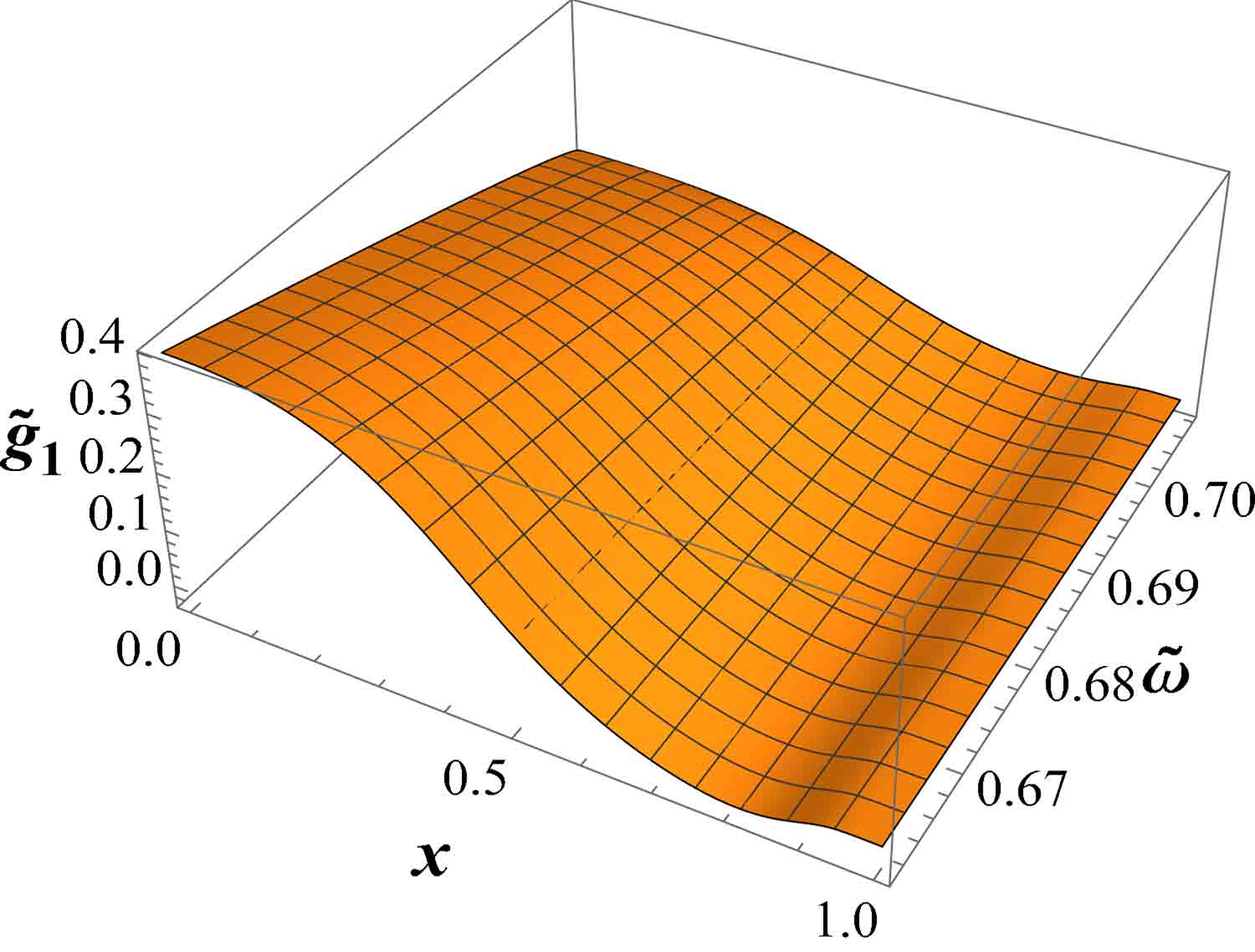}
\end{center}
\caption{The matter functions $\tilde{f}_0$, $\tilde{g}_0$, $\tilde{f}_1$ and $\tilde{g}_1$ on the first (left column) and second branches (right column) of the MSDSs solution function as functions of $x$ and $\tilde{\omega}$ for $\tilde{\mu}_1 = 0.7645$.}
\label{sf_double_f_g}
\end{figure}

\subsubsection{Double-Branch}
\begin{figure}[!htbp]
\begin{center}
    \includegraphics[height=.20\textheight]{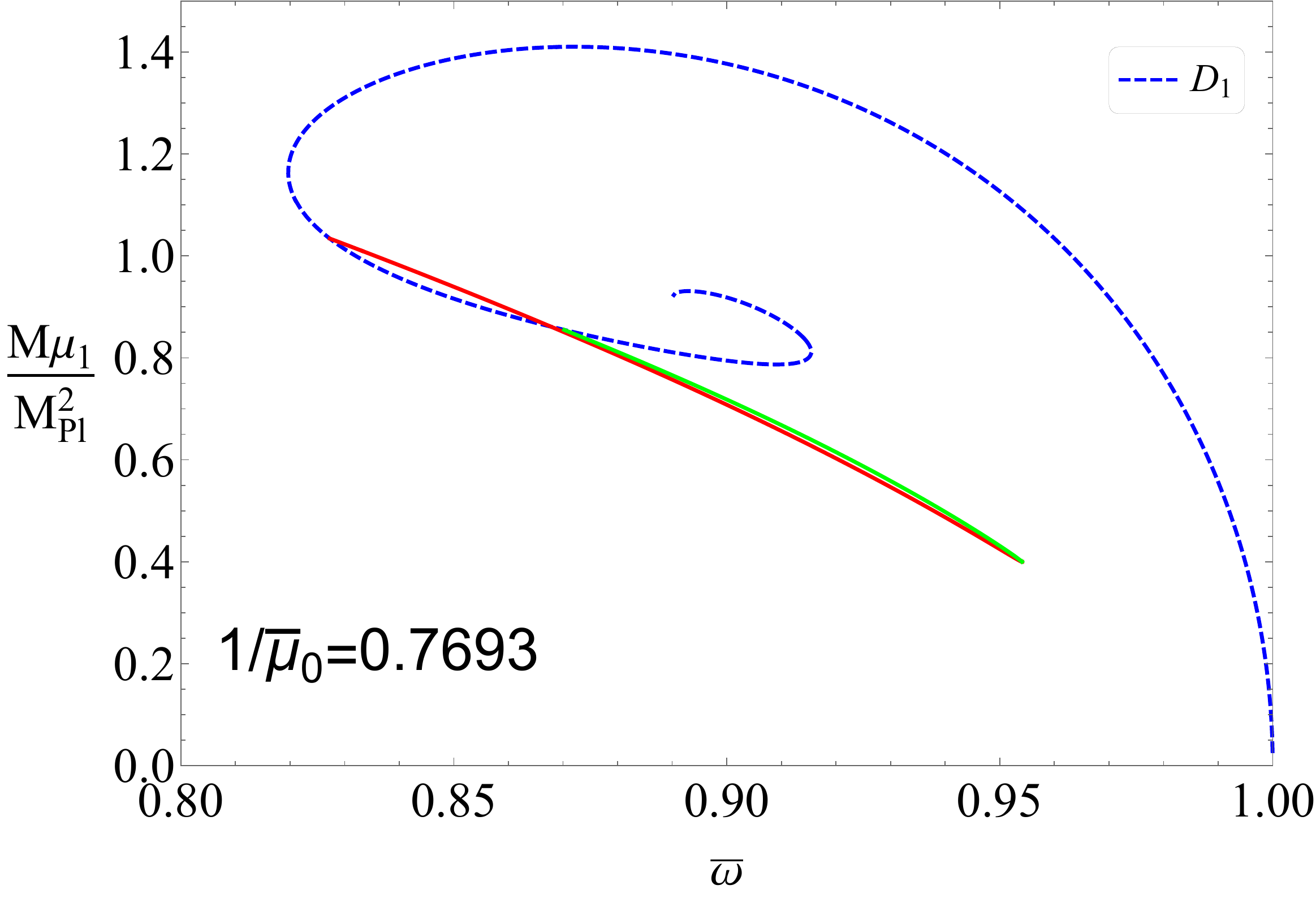}
    \includegraphics[height=.20\textheight]{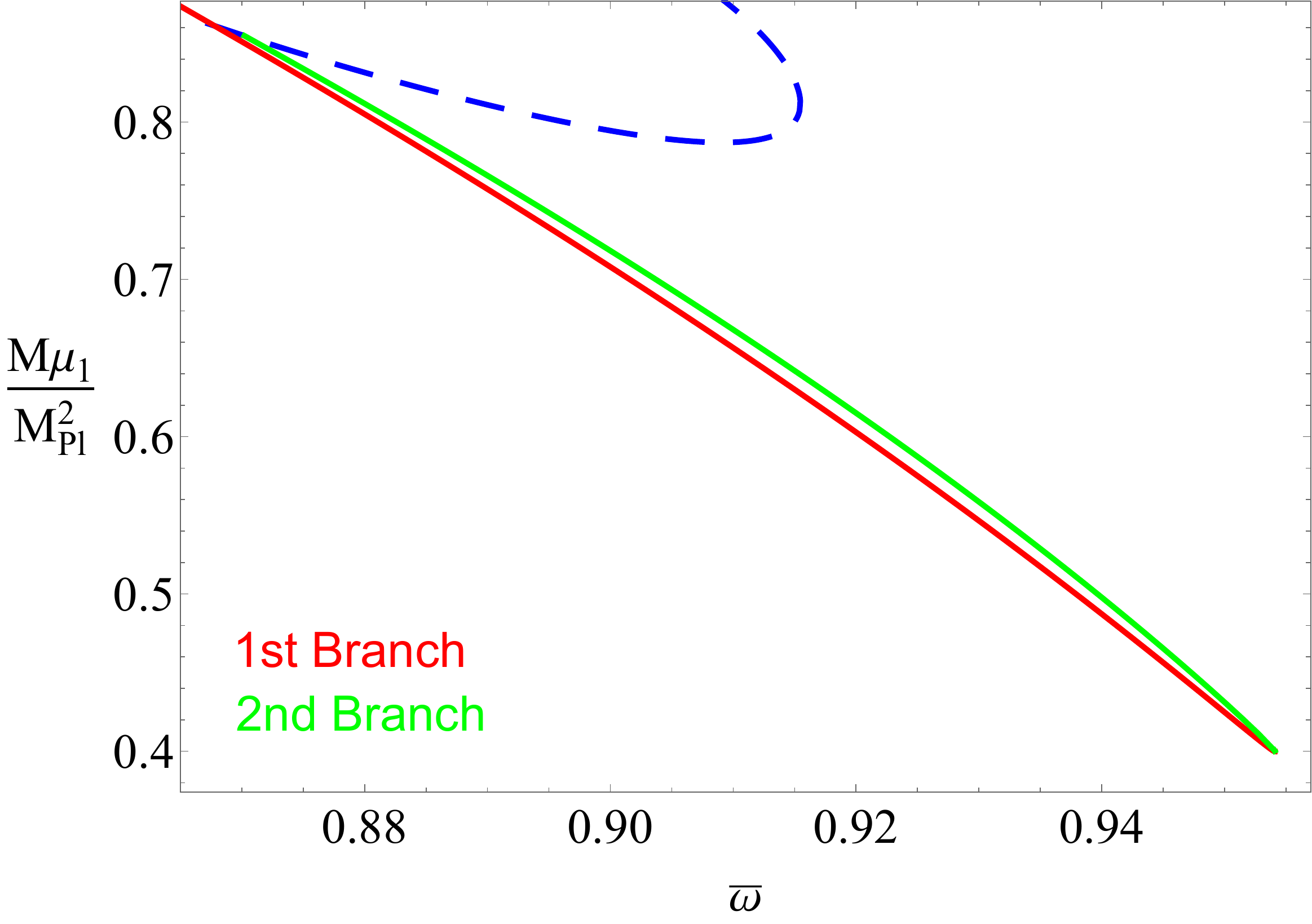}
    \includegraphics[height=.20\textheight]{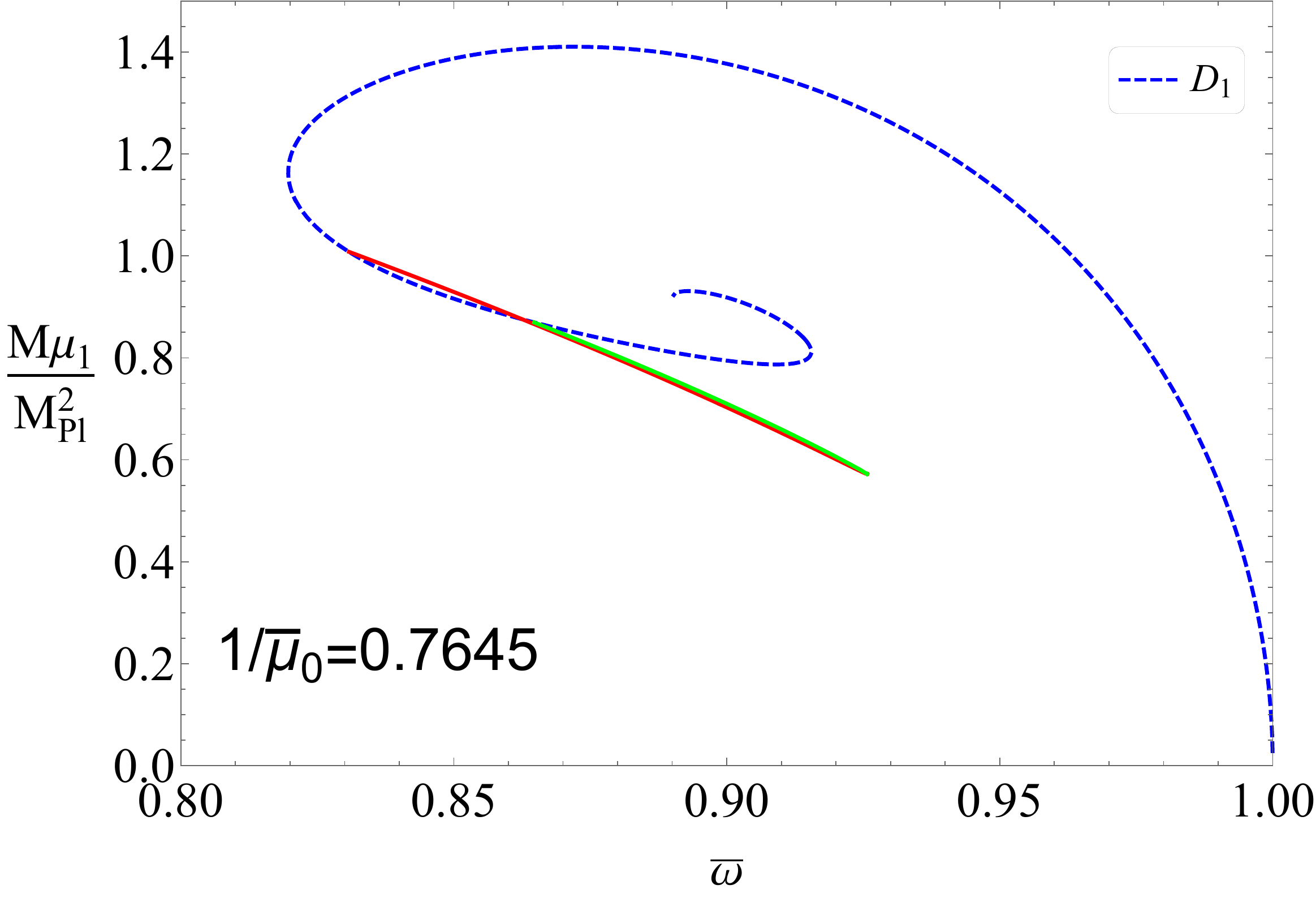}
    \includegraphics[height=.20\textheight]{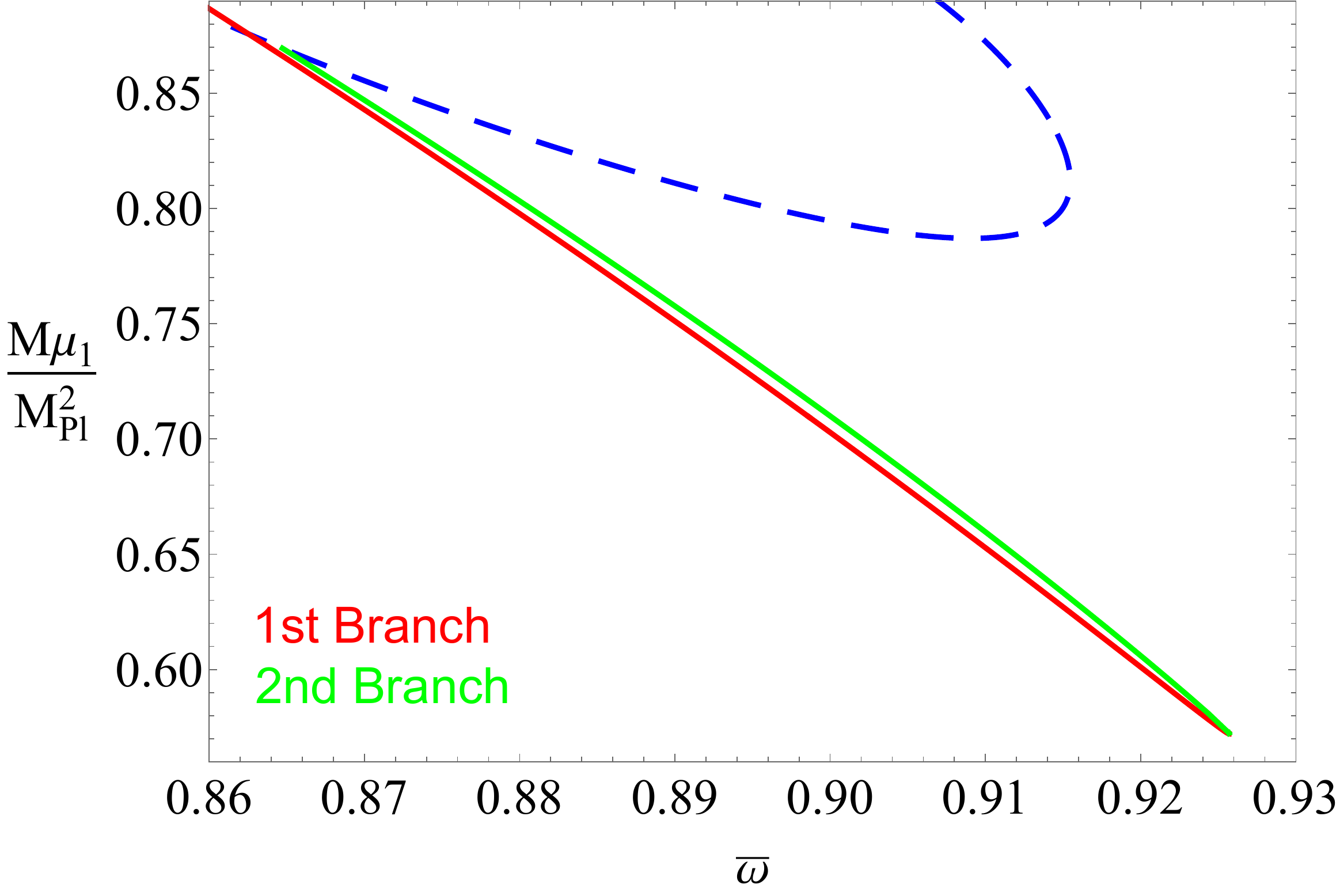}
    \includegraphics[height=.20\textheight]{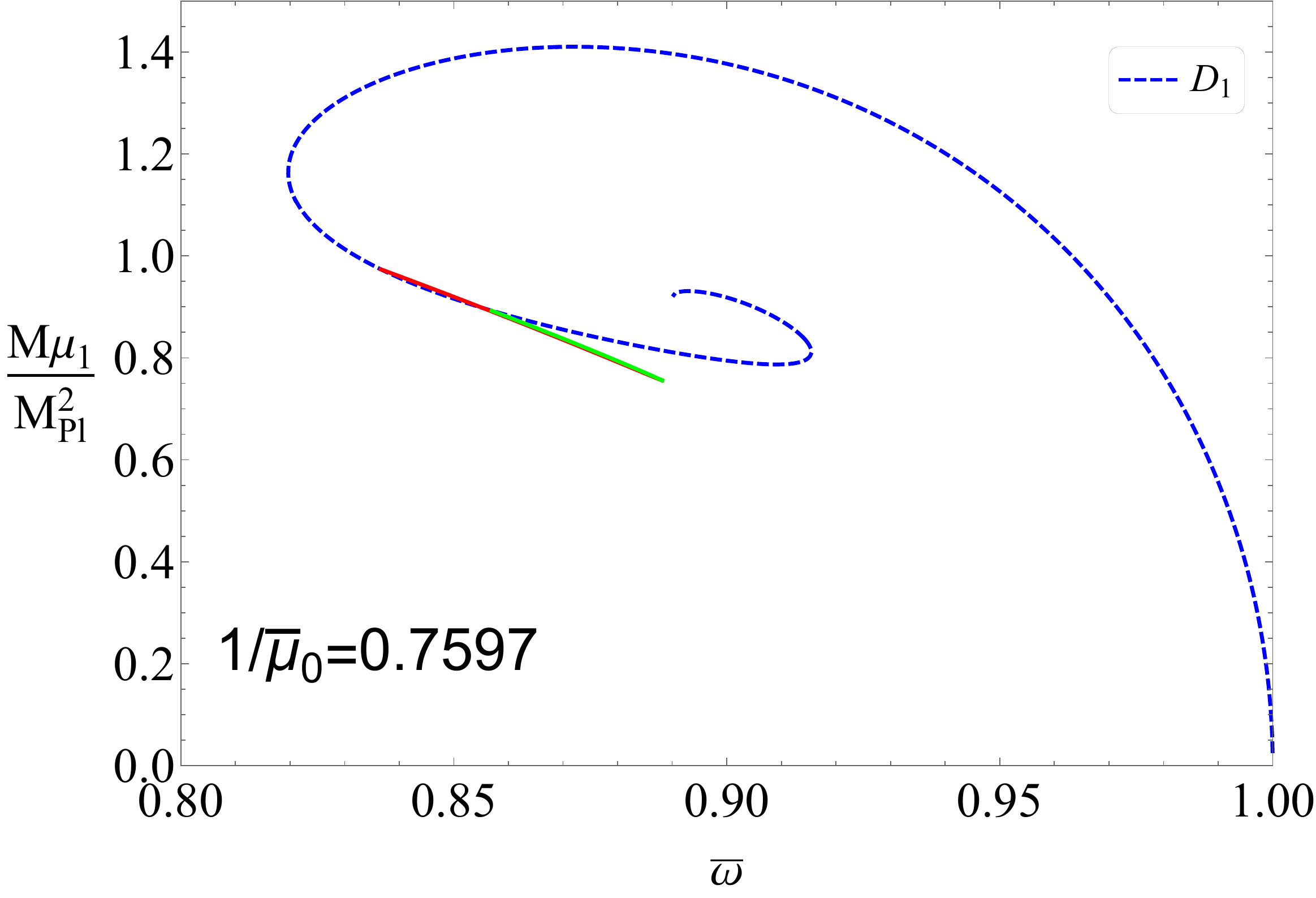}
    \includegraphics[height=.20\textheight]{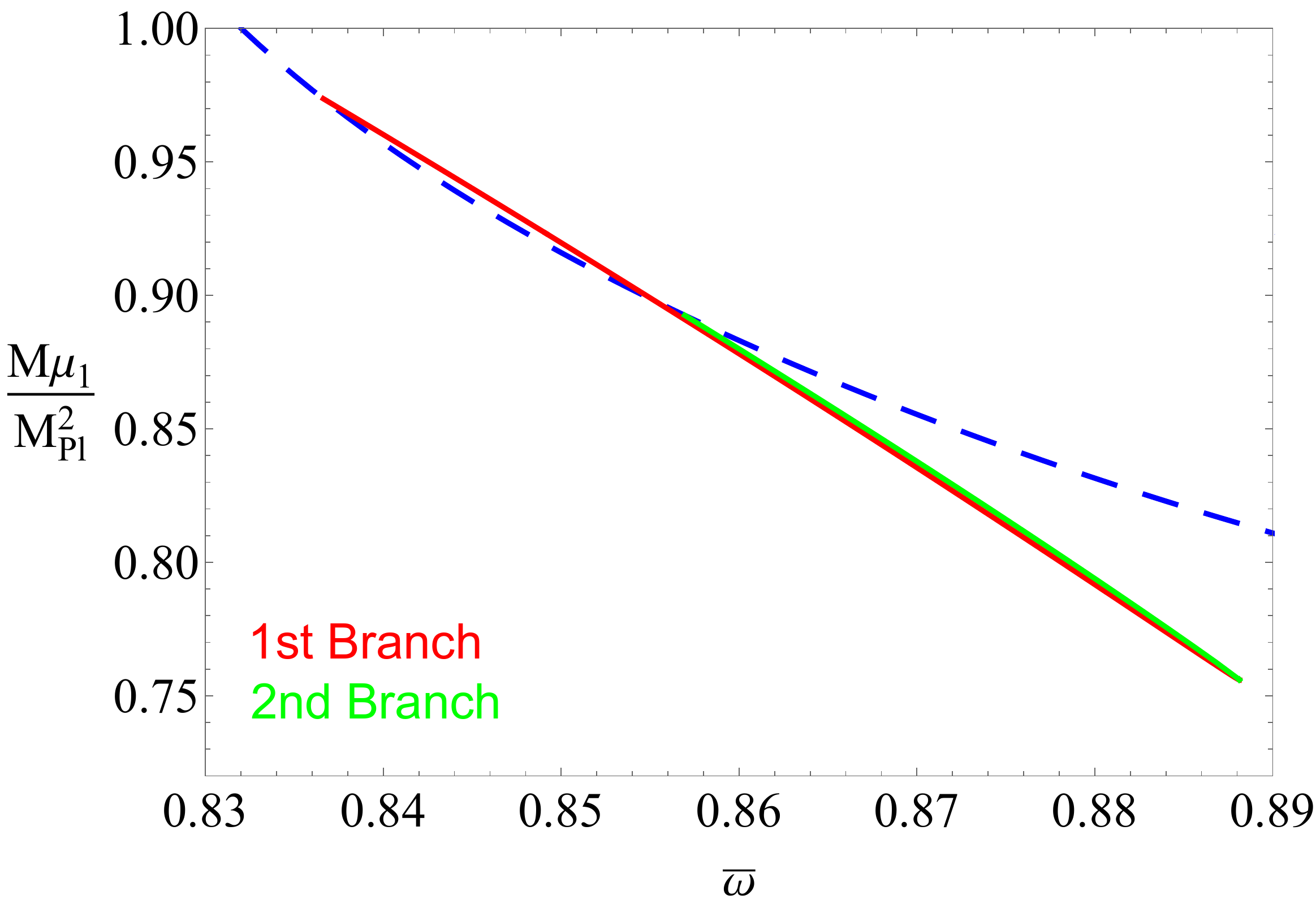}
    \includegraphics[height=.20\textheight]{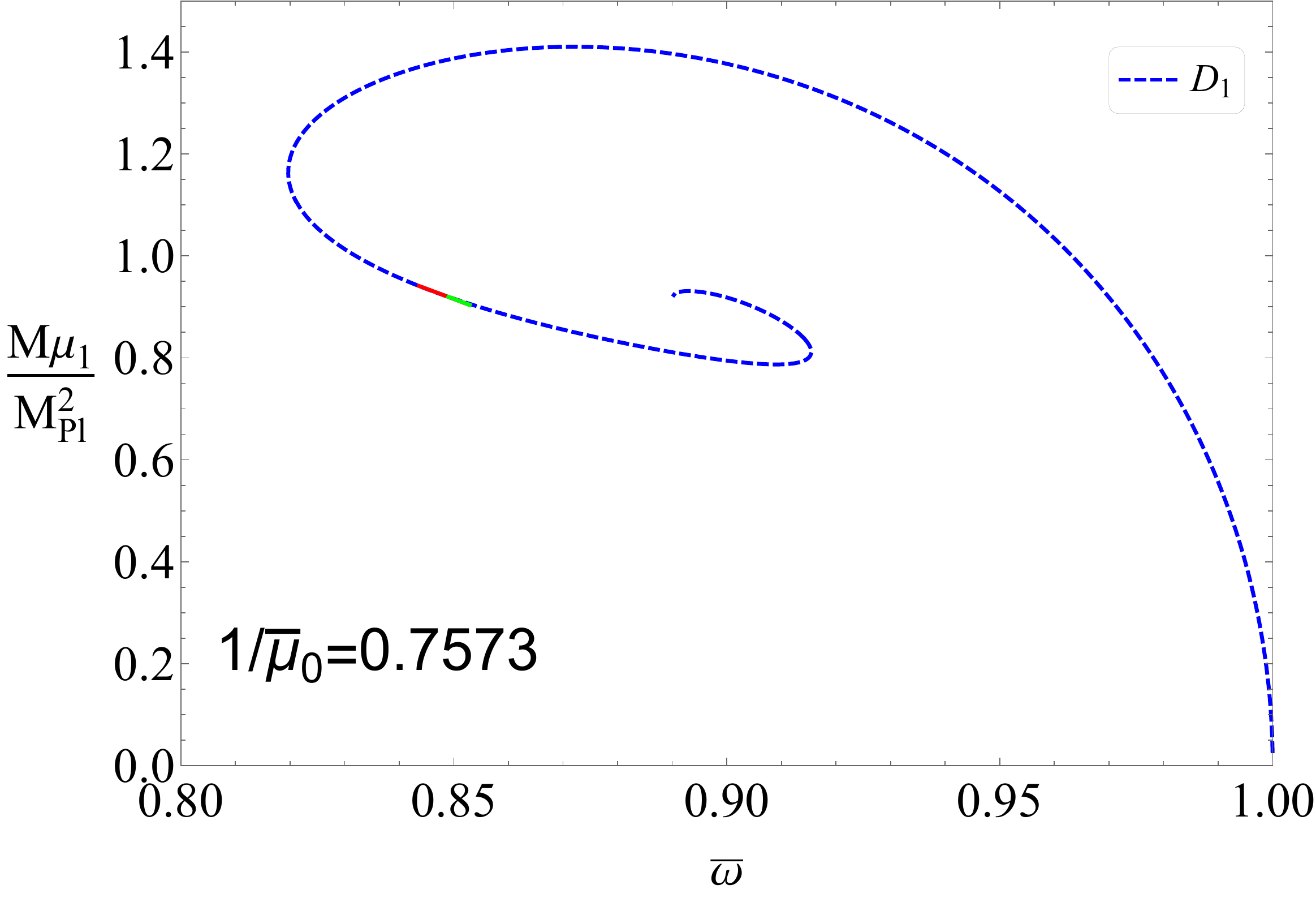}
    \includegraphics[height=.20\textheight]{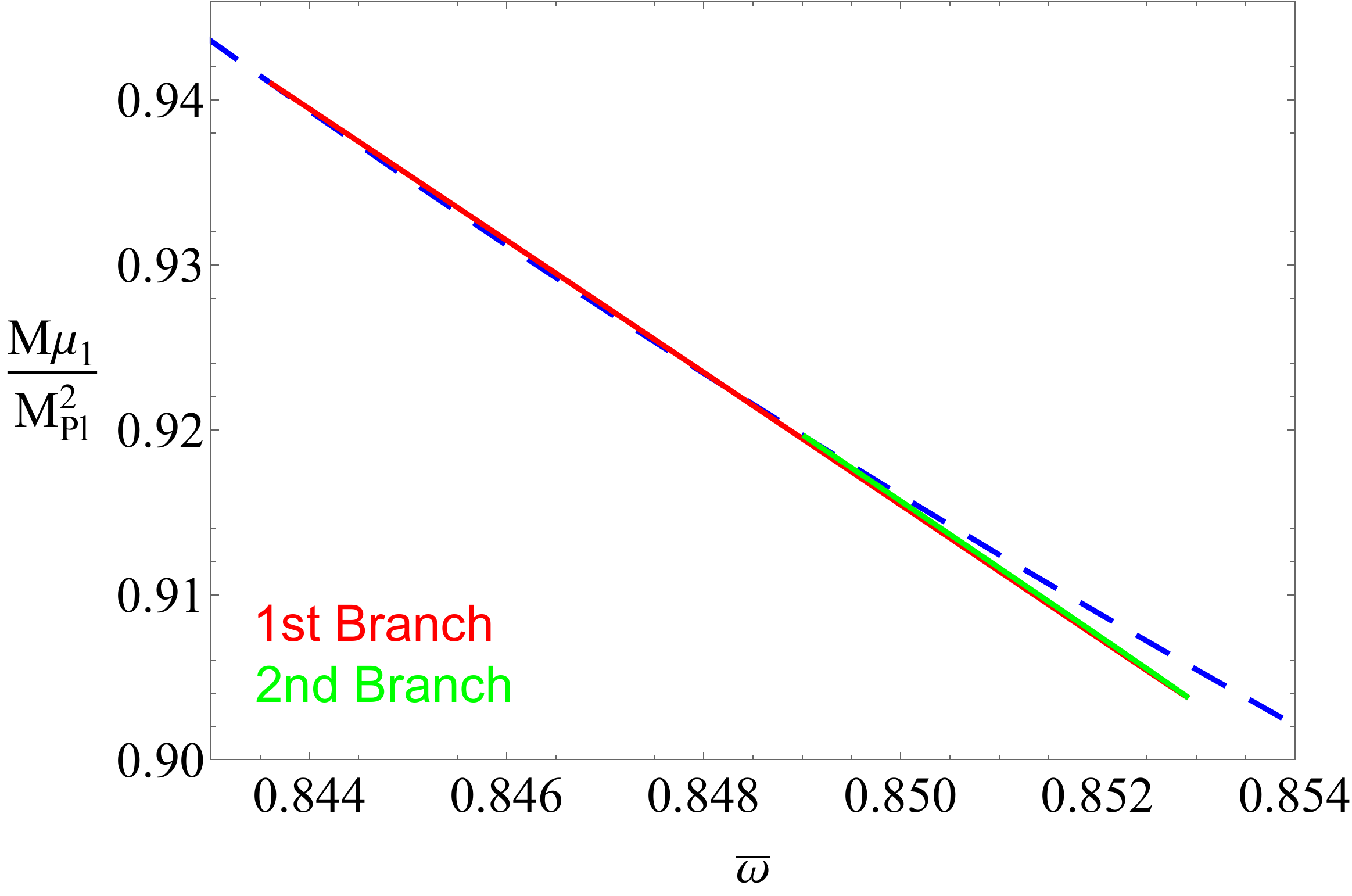}
\end{center}
\caption{The ADM mass $M$ of the MSDSs as a function of the synchronized frequency $\overline{\omega}$ for several values of $1/\overline{\mu}_0$.}
\label{sf_double_k-adm}
\end{figure}

In addition to the single-branch solution described in the preceding section, the solutions of MSDSs exhibit two branches when the synchronized frequency $\tilde{\omega}$ is sufficiently low. In the following, we will first discuss the variation of the matter field functions with respect to the synchronized frequency $\tilde{\omega}$ for the double-branch solution. As shown in Fig.~\ref{sf_double_f_g}, we demonstrate the relationship between the radial profile of the matter fields that constitute the MSDSs and the synchronized frequency $\tilde{\omega}$ under the condition of a fixed mass $\tilde{\mu}_1 = 0.898$. For the first branch field functions of the MSDSs displayed in the left column of the figure, as the synchronized frequency increases, the peak values of the ground state Dirac field functions $\tilde{f}_0$ and $\tilde{g}_0$ also increase, while the peak values of the excited state Dirac field functions $\tilde{f}_1$ and $\tilde{g}_1$ gradually decrease. For the second branch field functions displayed in the right column, as the synchronized frequency decreases, the peak values of the ground state field functions $\tilde{f}_0$ and $\tilde{g}_0$ gradually decrease, while the peak values of the excited state field functions $\tilde{f}_1$ and $\tilde{g}_1$ gradually increase. It is worth noting that the ground state Dirac field disappears at the minimum synchronized frequency of the first and second branches, while the excited state field always exists.

Next, we analyze the characteristics of the ADM mass $M$ of the double-branch solutions of the MSDSs as a function of the synchronized frequency. Fig.~\ref{sf_double_k-adm} depicts the relationship between the ADM mass $M$ and the synchronized frequency $\overline{\omega}$ for different masses $\overline{\mu}_0$ of the MSDSs. The blue dashed line represents the $D_1$ solutions, while the red and green lines represent the first and second branch solutions of the MSDSs, respectively.

In Fig.~\ref{sf_double_k-adm}, the MSDSs solutions exhibit a double-branch structure when $1/\overline{\mu}_0$ ($\tilde{\mu}_1$) is less than the threshold value of $0.7964$. This change in solutions occurs abruptly, and the single-branch solutions do not gradually extend into a second branch as $1/\overline{\mu}_0$ decreases. Extensive numerical calculations were conducted, and even when $1/\overline{\mu}_0$ is gradually reduced by an order of magnitude of $10^{-5}$, the single-branch solutions fail to transition continuously to the double-branch solutions. We refer to this phenomenon as \textit{bifurcation}. It is worth noting that this phenomenon has not been observed in previous studies of multi-field soliton models~\cite{Liang:2022mjo, Ma:2023vfa}.

The two branches of the double-branch solution are very close to each other, but their intersections with the blue dashed line are clearly different. As $1/\overline{\mu}_0$ decreases, the synchronized frequency range of the two branches of the MSDSs gradually decreases. For all double-branch solutions, the intersections of the two branches with the blue dashed line are always in the second branch of the blue dashed line. When $1/\overline{\mu}_0$ decreases to $0.7573$, the second branch solution is very close to some of the solutions in the first branch, to the extent that the two branches almost coincide. It is noteworthy that at this point, the first branch of the MSDSs appears to be "tangential" to the blue dashed line, which is a similar feature to the single-branch solution of the MSDSs mentioned earlier.

\subsection{ Nonsynchronized frequency }
In this section, we discuss the nonsynchronized frequency solutions of the MSDSs. To analyze the influence of the parameters on the numerical solutions, we set the masses of the ground state and excited state Dirac fields to be the same ($\tilde{\mu}_0 = \tilde{\mu}_1 = 1$) and investigate how the MSDSs change with the frequency of the excited state Dirac field $\tilde{\omega}_1$ under a fixed frequency of the ground state Dirac field $\tilde{\omega}_0$. Through a series of numerical calculations, we find that the nonsynchronized frequency solutions of the system can be divided into two types. When the frequency of the ground state Dirac field is in the range of $0.733 \le \tilde{\omega}_0 < 1$, the solution of the MSDSs is a single-branch solution. When the frequency of the ground state Dirac field is in the range of $0.6971 \le \tilde{\omega}_0 < 0.733$, the solution of the MSDSs is a double-branch solution.

\subsubsection{Single-branch}
\begin{figure}[!htbp]
\begin{center}
    \includegraphics[height=.24\textheight]{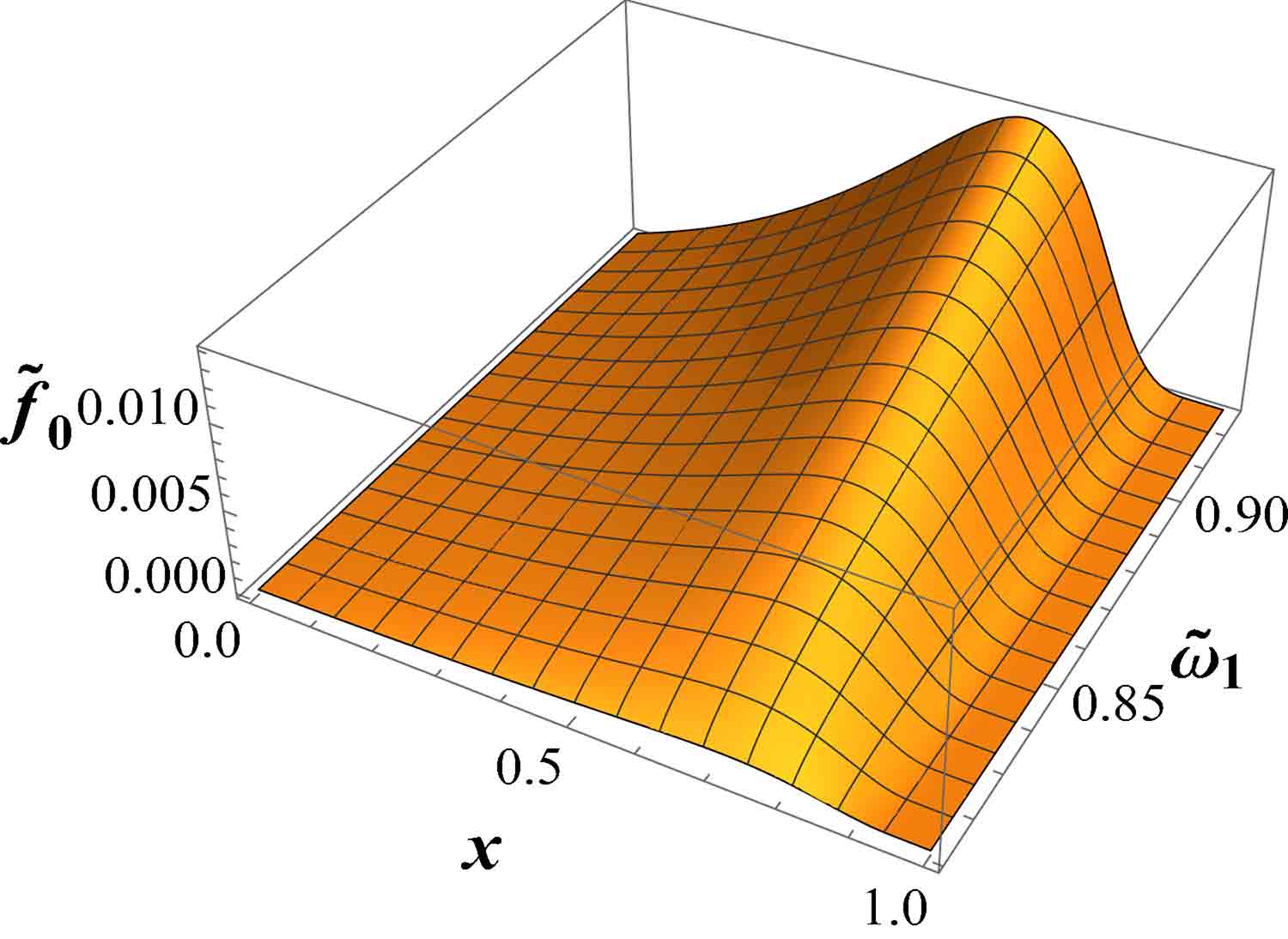}
    \includegraphics[height=.24\textheight]{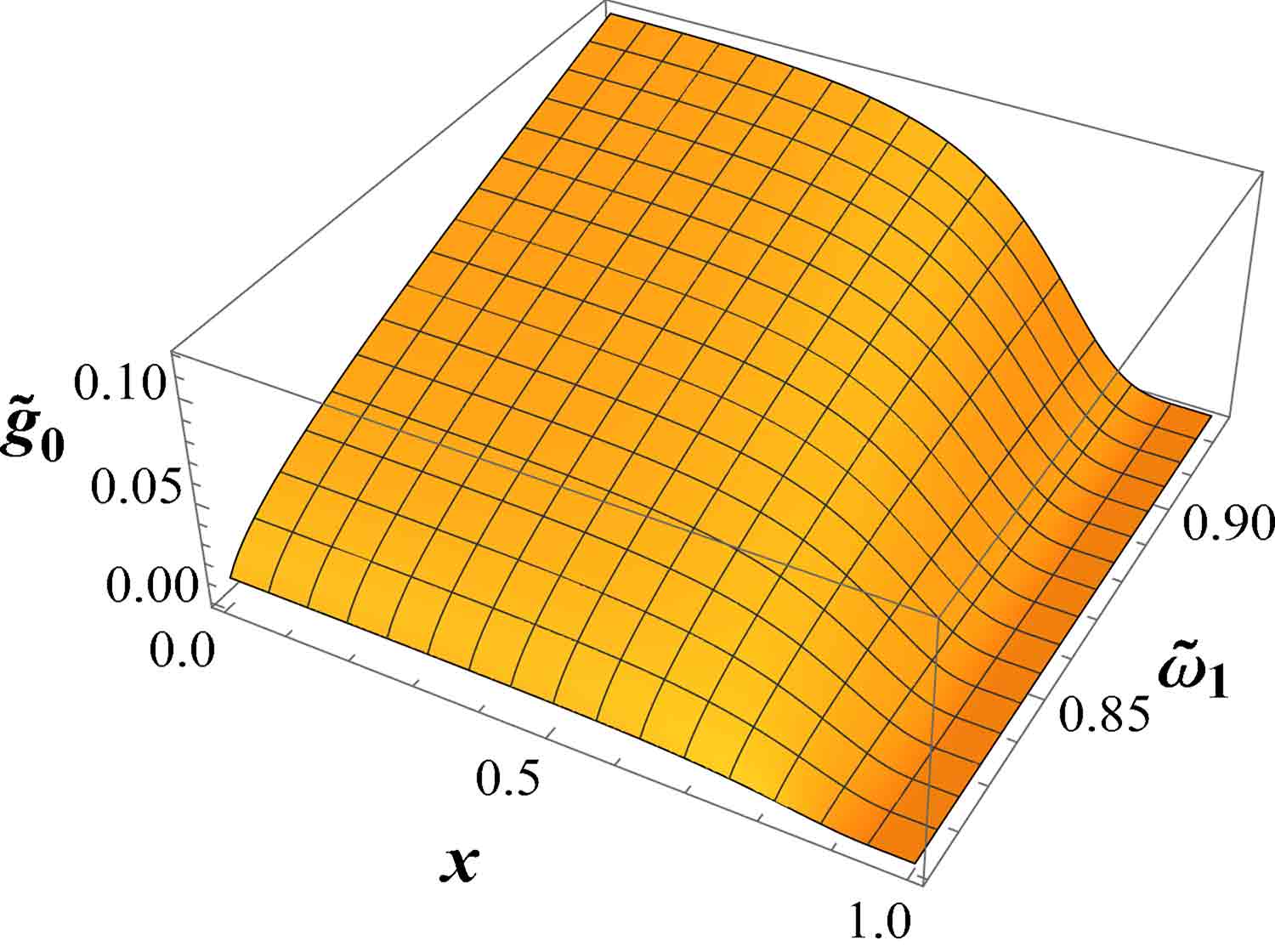}
    \includegraphics[height=.24\textheight]{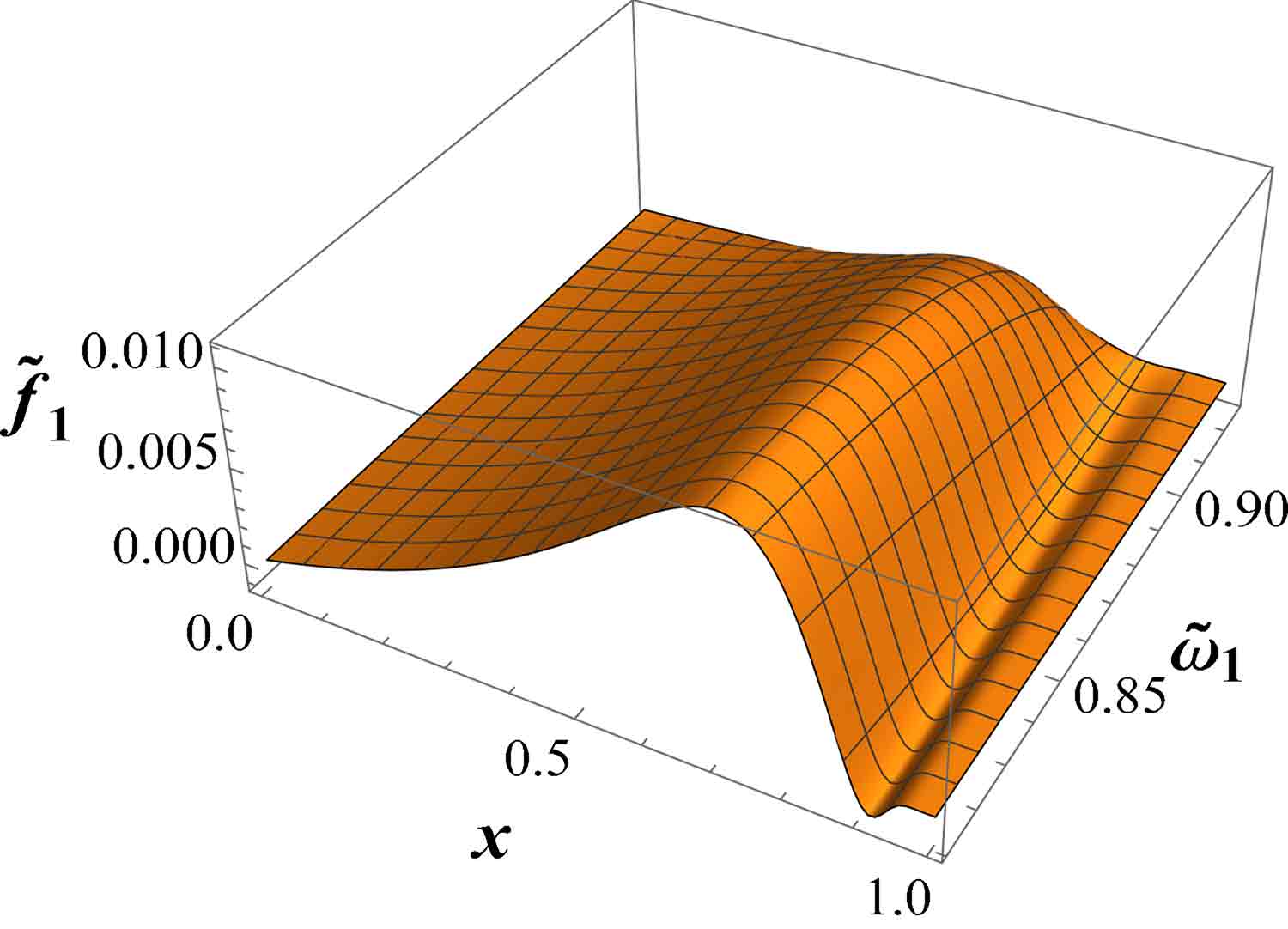}
    \includegraphics[height=.24\textheight]{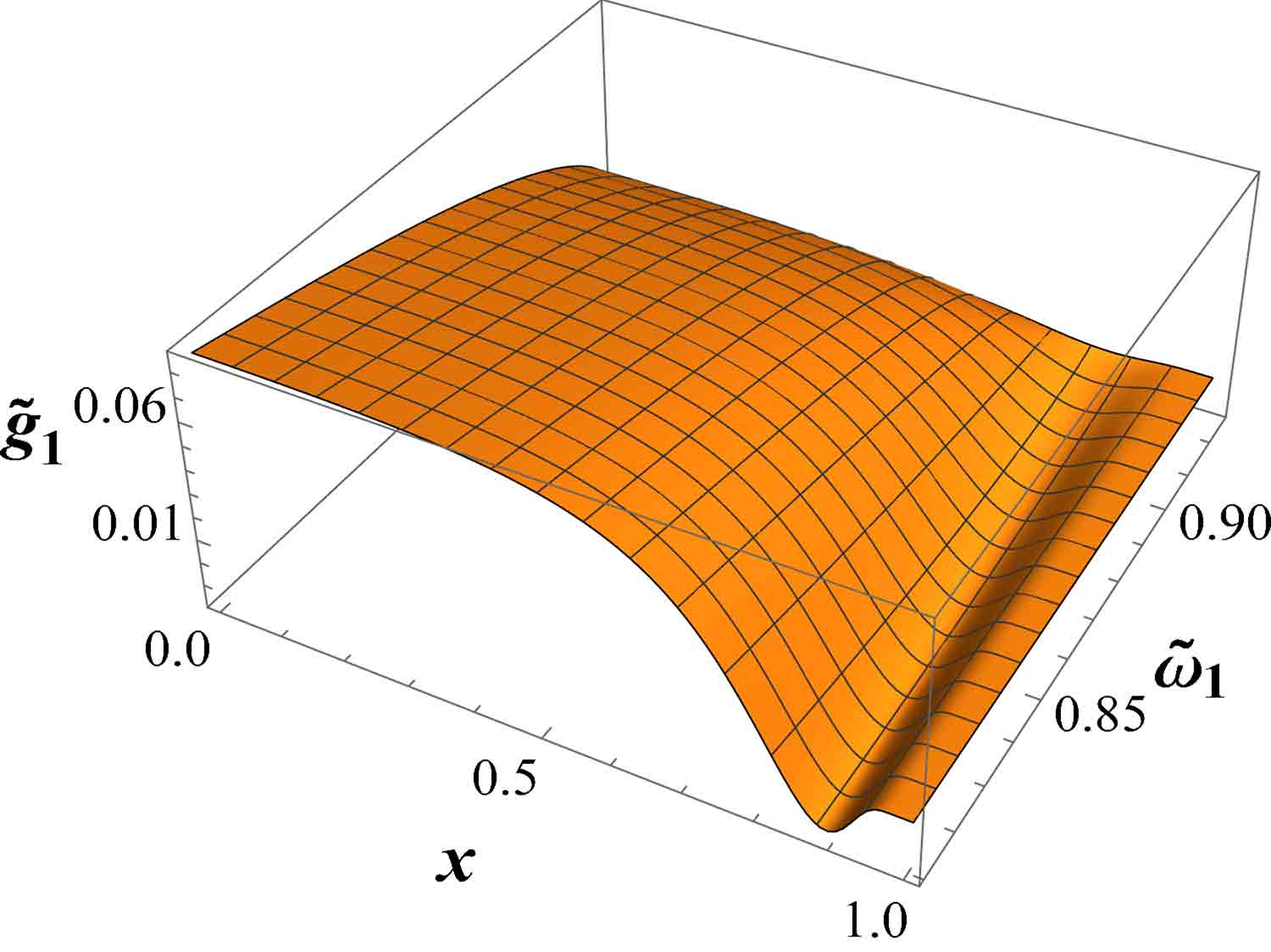}
\end{center}
\caption{The matter functions $\tilde{f}_0$, $\tilde{g}_0$, $\tilde{f}_1$ and $\tilde{g}_1$ as functions of $x$ and $\tilde{\omega}_1$ for $\tilde{\omega}_0 = 0.819$.}
\label{nsf_single_f_g}
\end{figure}

We first discuss the single-branch solution. The radial profile of the ground-state and excited state Dirac field functions as a function of the excited state frequency $\tilde{\omega}_1$ is shown in Fig.~\ref{nsf_single_f_g}. It can be observed that as the frequency $\tilde{\omega}_1$ increases, the maximum values of the ground state Dirac field functions $\tilde{f}_0$ and $\tilde{g}_0$ gradually increase, while the maximum and minimum absolute values of the first excited-state Dirac field functions $\tilde{f}_1$ and $\tilde{g}_1$ decrease. Similar to the single-branch solution at synchronized frequency, for the single-branch solution at nonsynchronized frequency, the ground state fields almost disappear at the minimum value of the frequency $\tilde{\omega}_1$, and the excited state fields tend to disappear at the maximum value of the frequency $\tilde{\omega}_1$.

\begin{figure}[!htbp]
\begin{center}
    \includegraphics[height=.23\textheight]{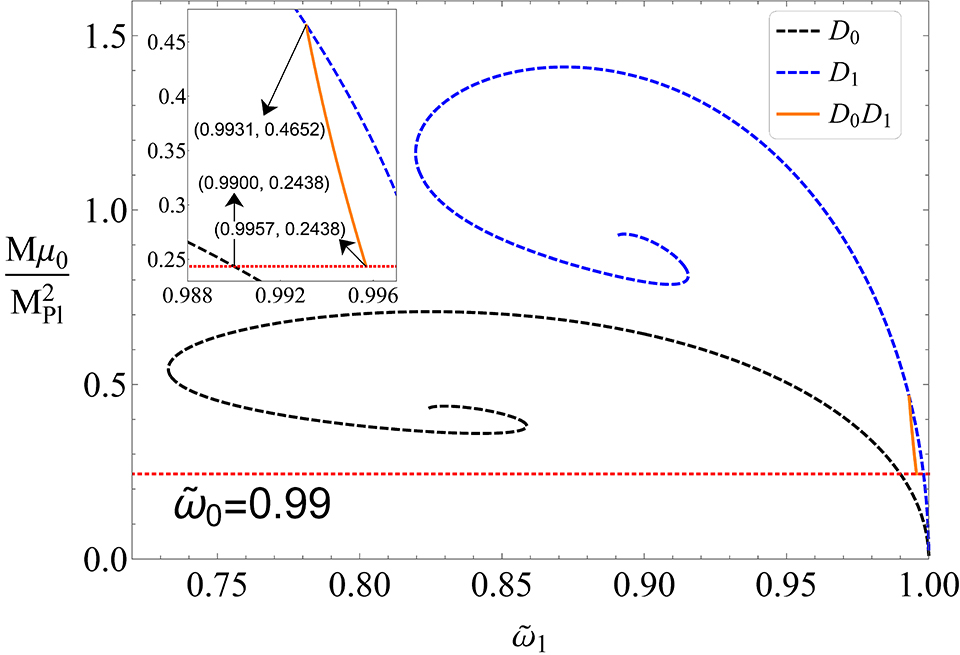}
    \includegraphics[height=.23\textheight]{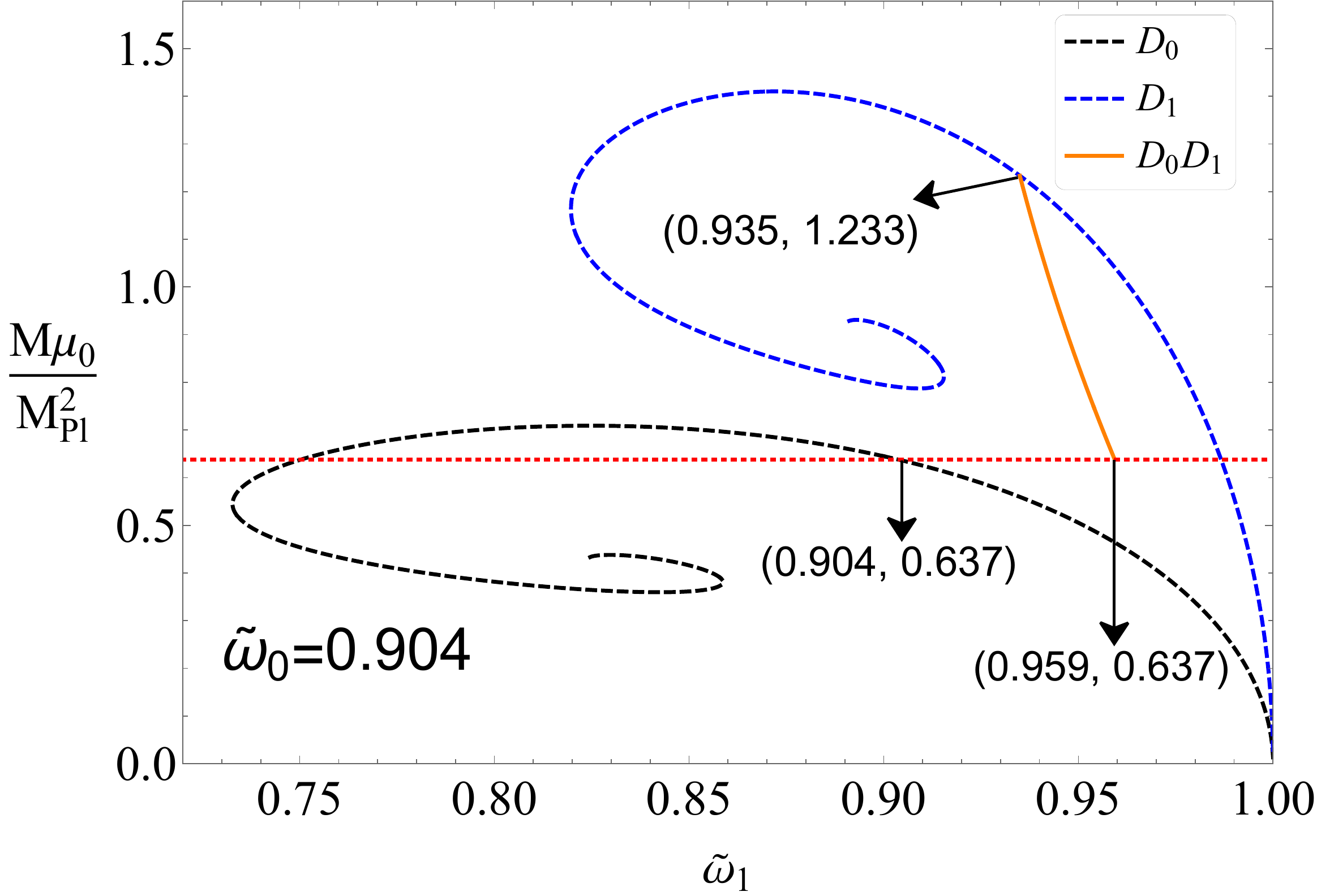}
    \includegraphics[height=.23\textheight]{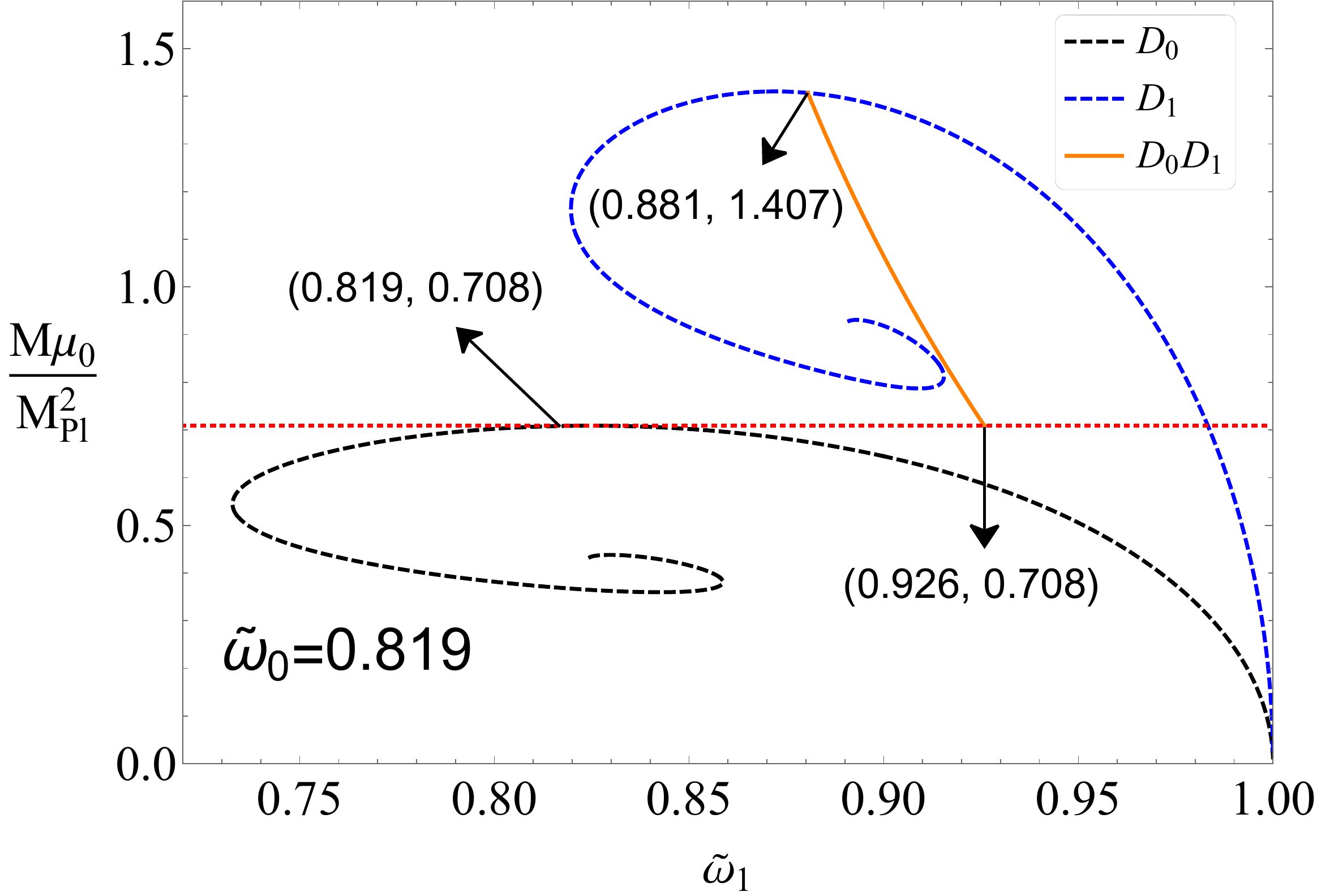}
    \includegraphics[height=.23\textheight]{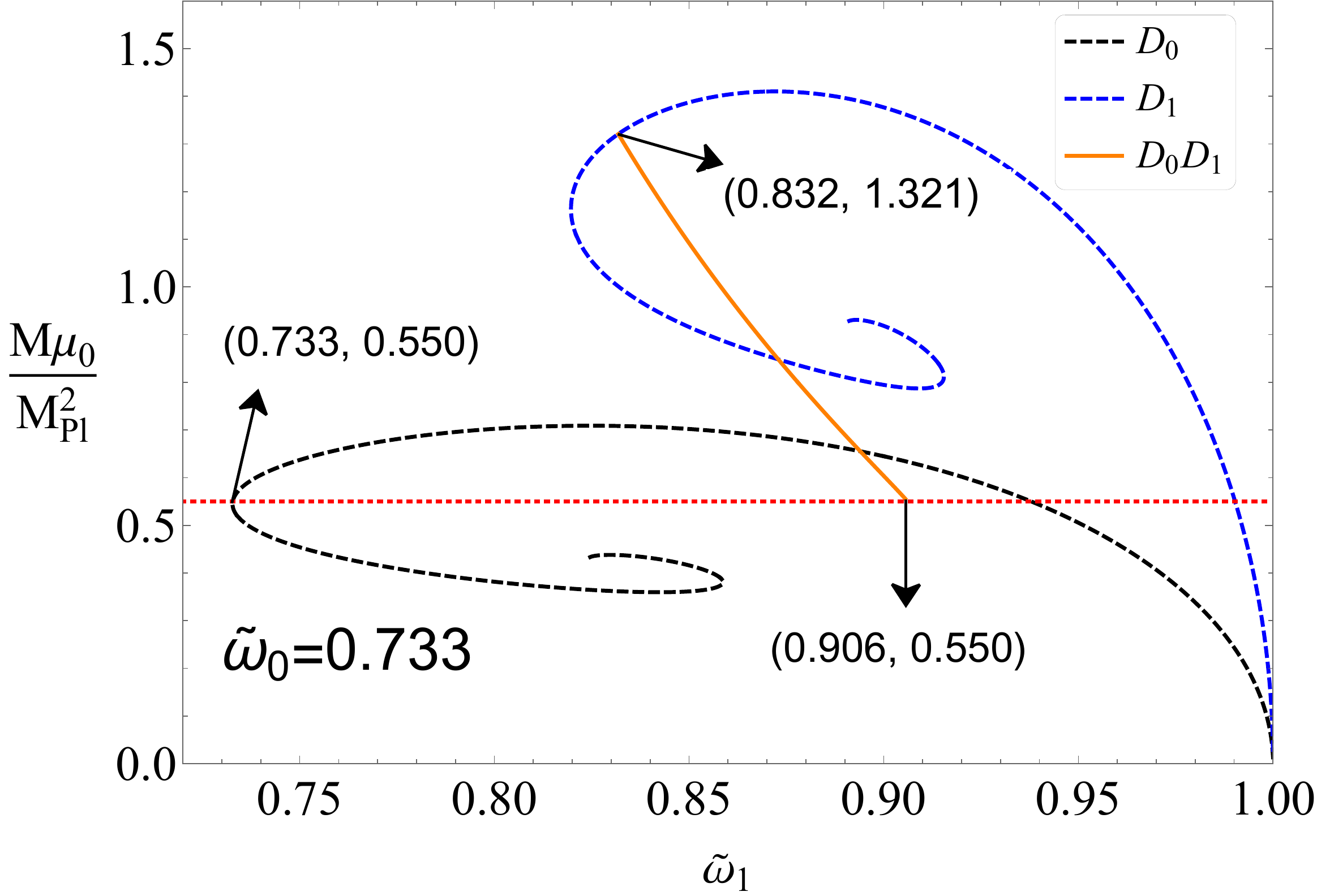}
\end{center}
\caption{The ADM mass $M$ of the MSDSs as a function of the frequency $\tilde{\omega}_1$ for several values of $\tilde{\omega}_0$.}
\label{nsf_single_k-adm}
\end{figure}

\begin{figure}[!htbp]
\begin{center}
    \includegraphics[height=.21\textheight]{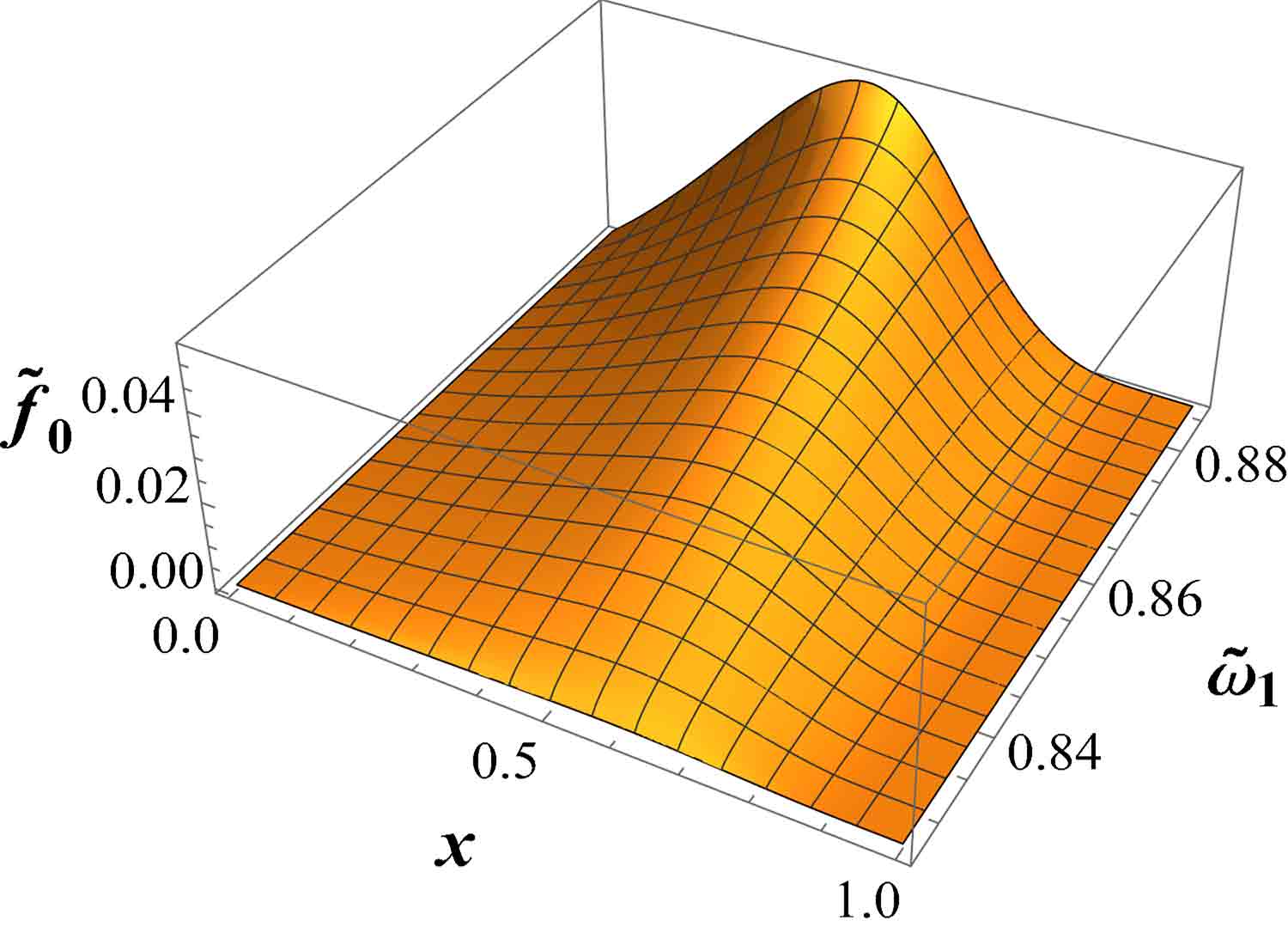}
    \includegraphics[height=.21\textheight]{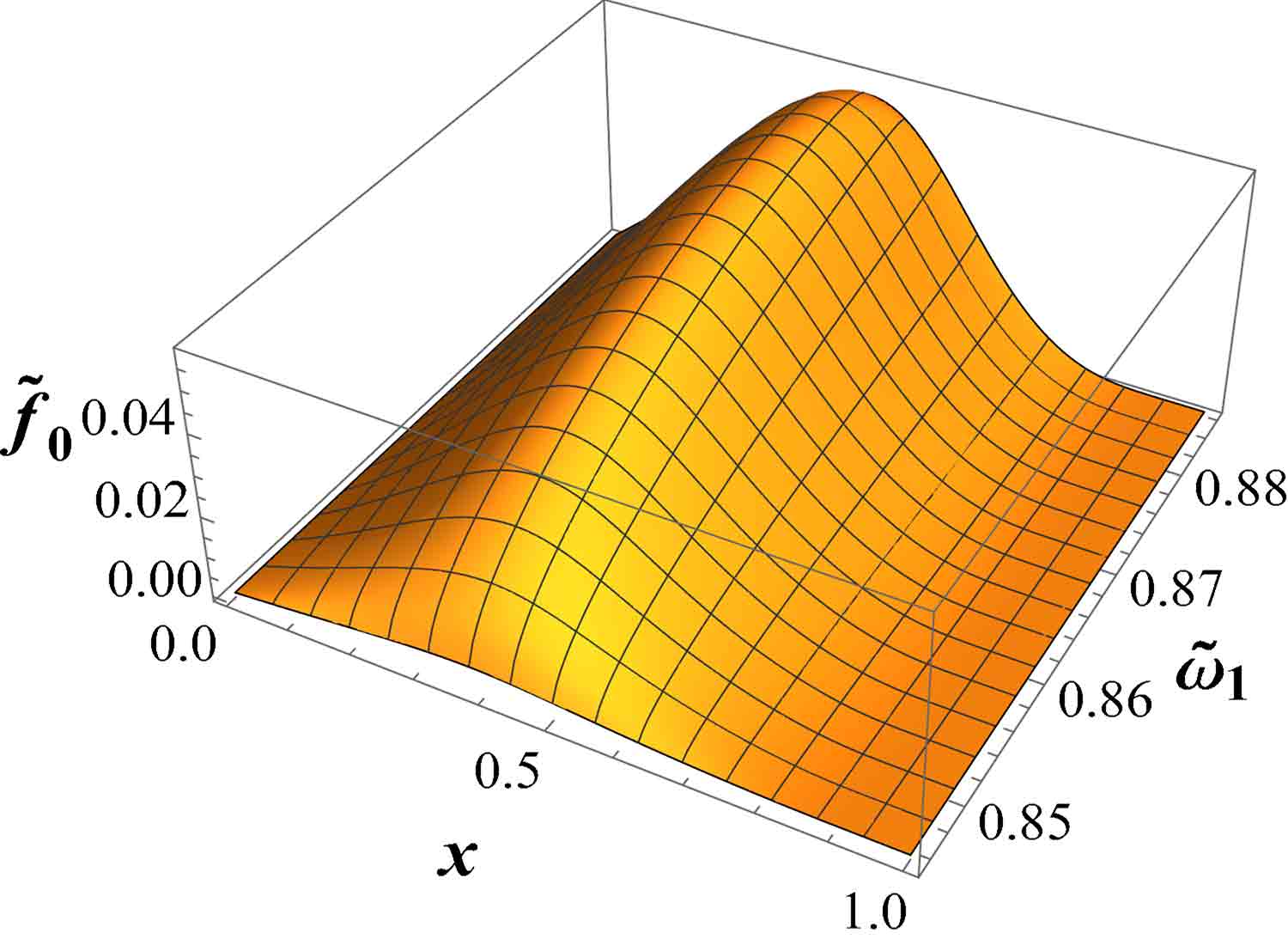}
    \includegraphics[height=.21\textheight]{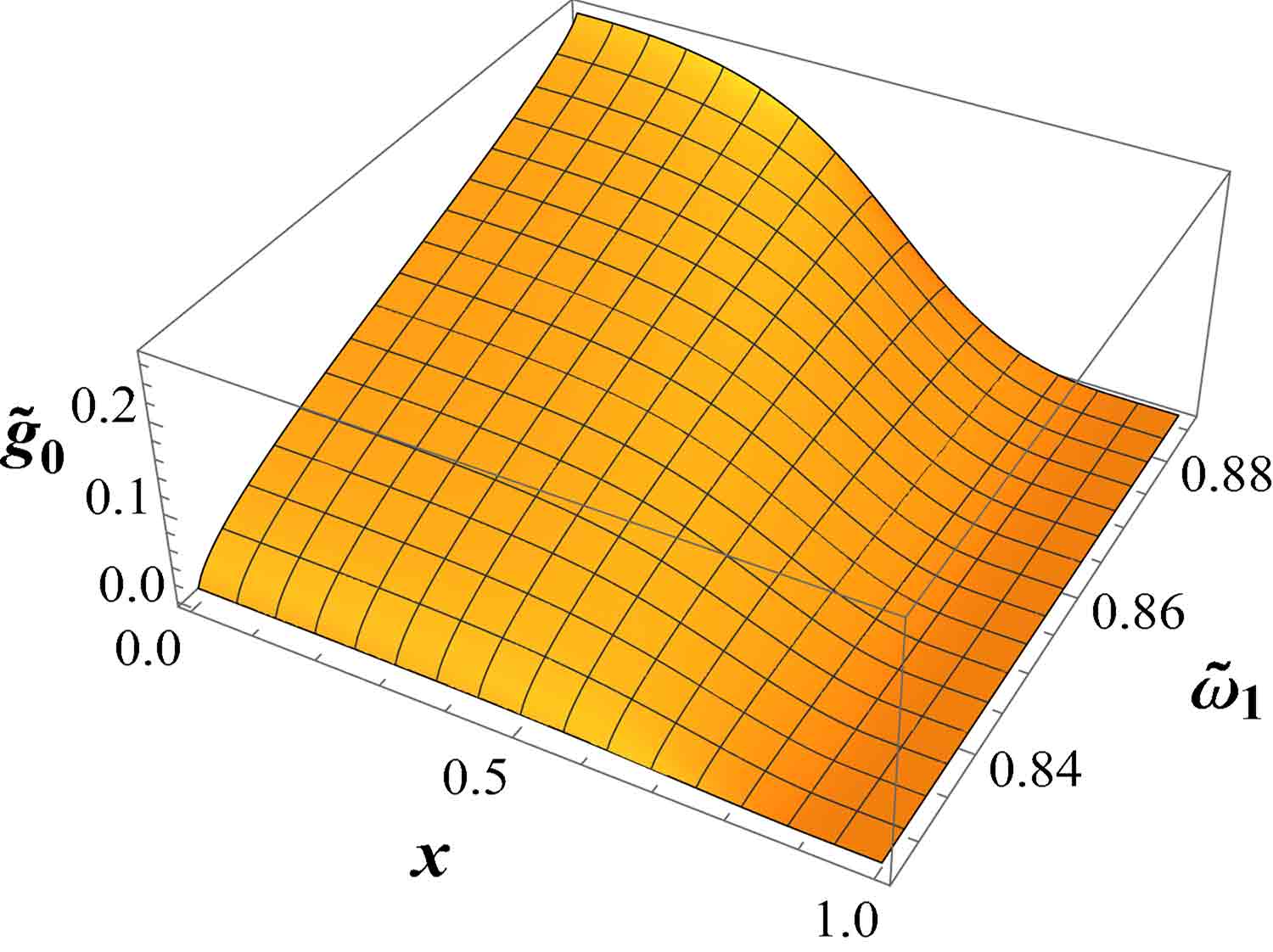}
    \includegraphics[height=.21\textheight]{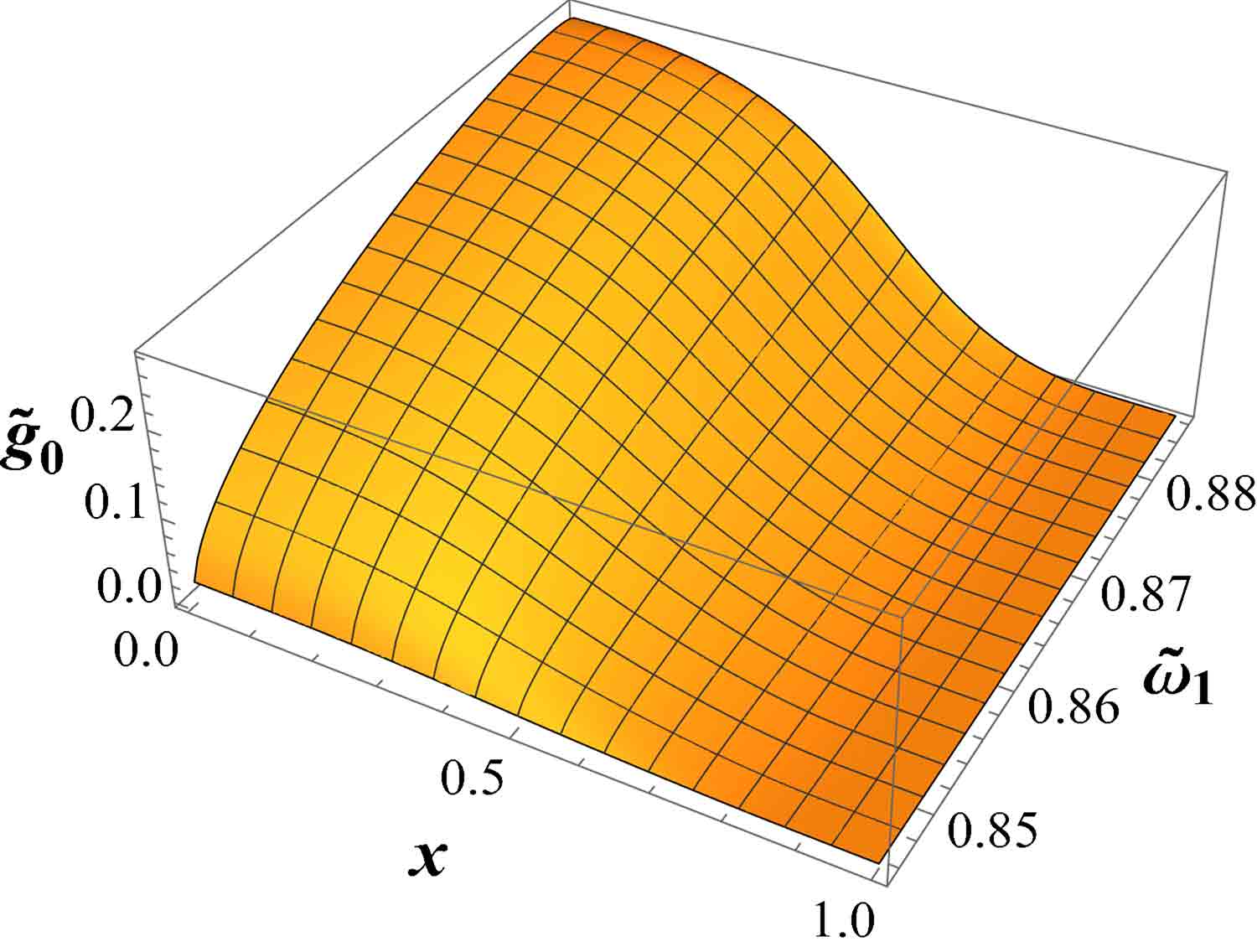}
    \includegraphics[height=.21\textheight]{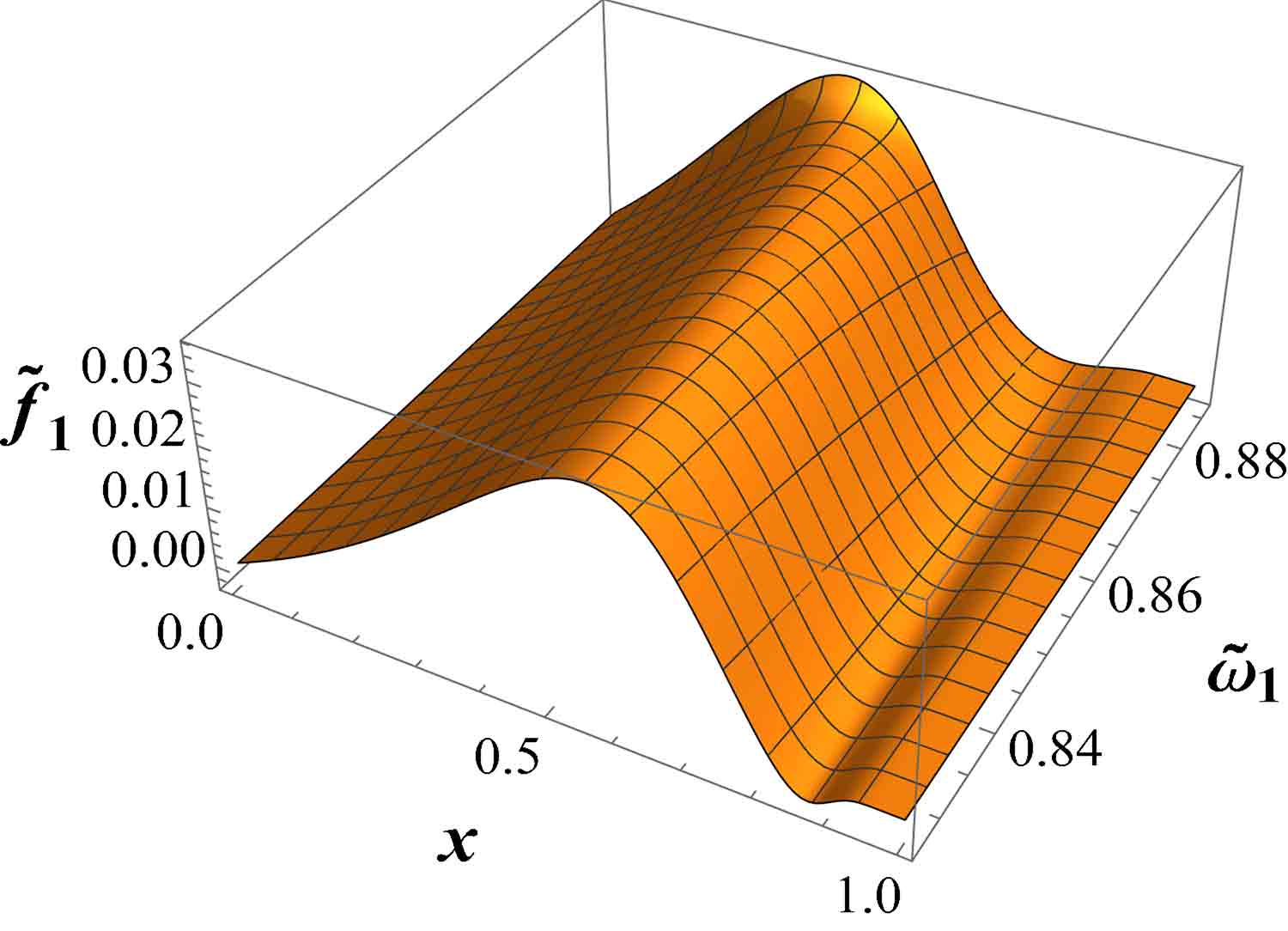}
    \includegraphics[height=.21\textheight]{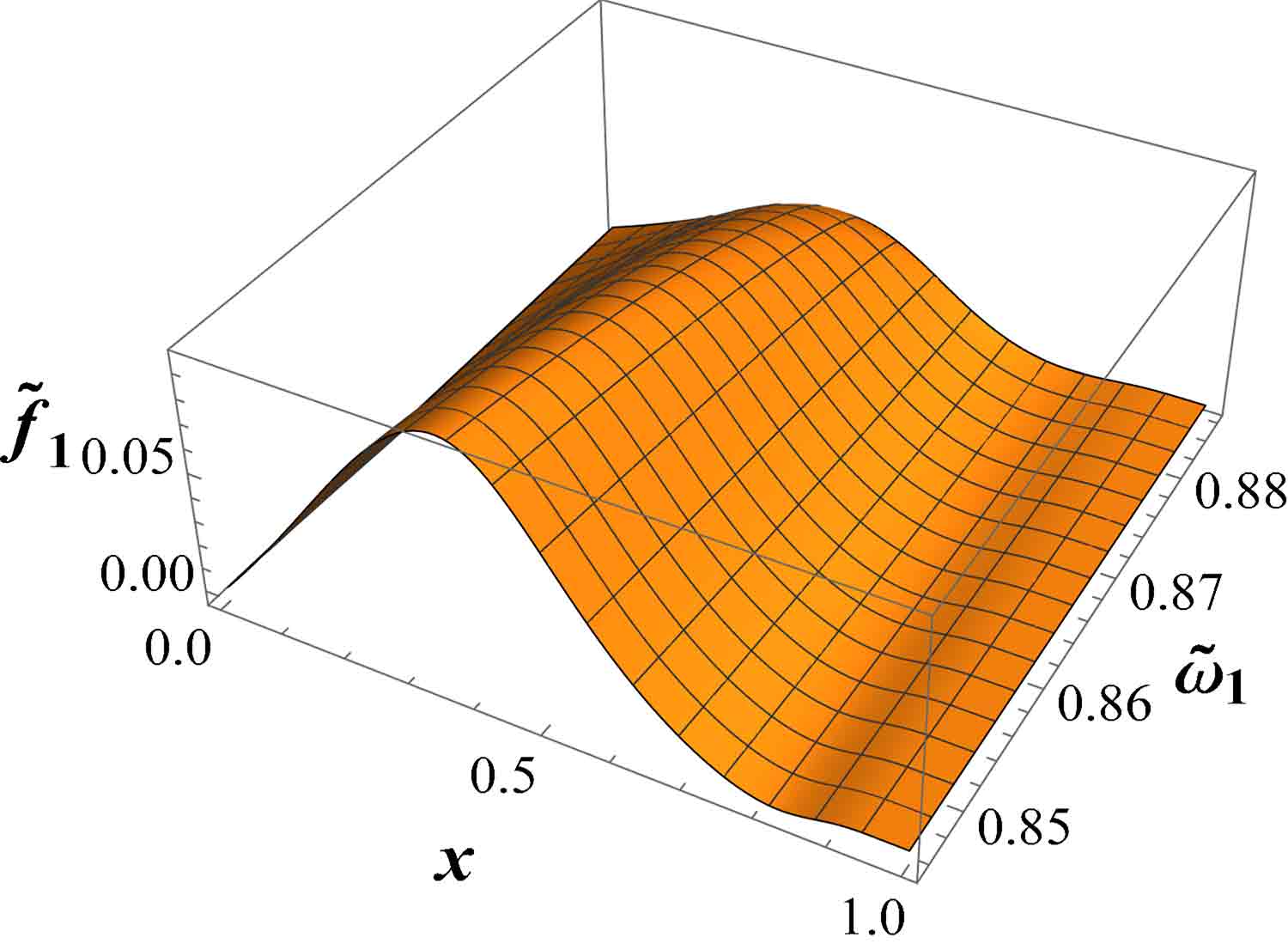}
    \includegraphics[height=.21\textheight]{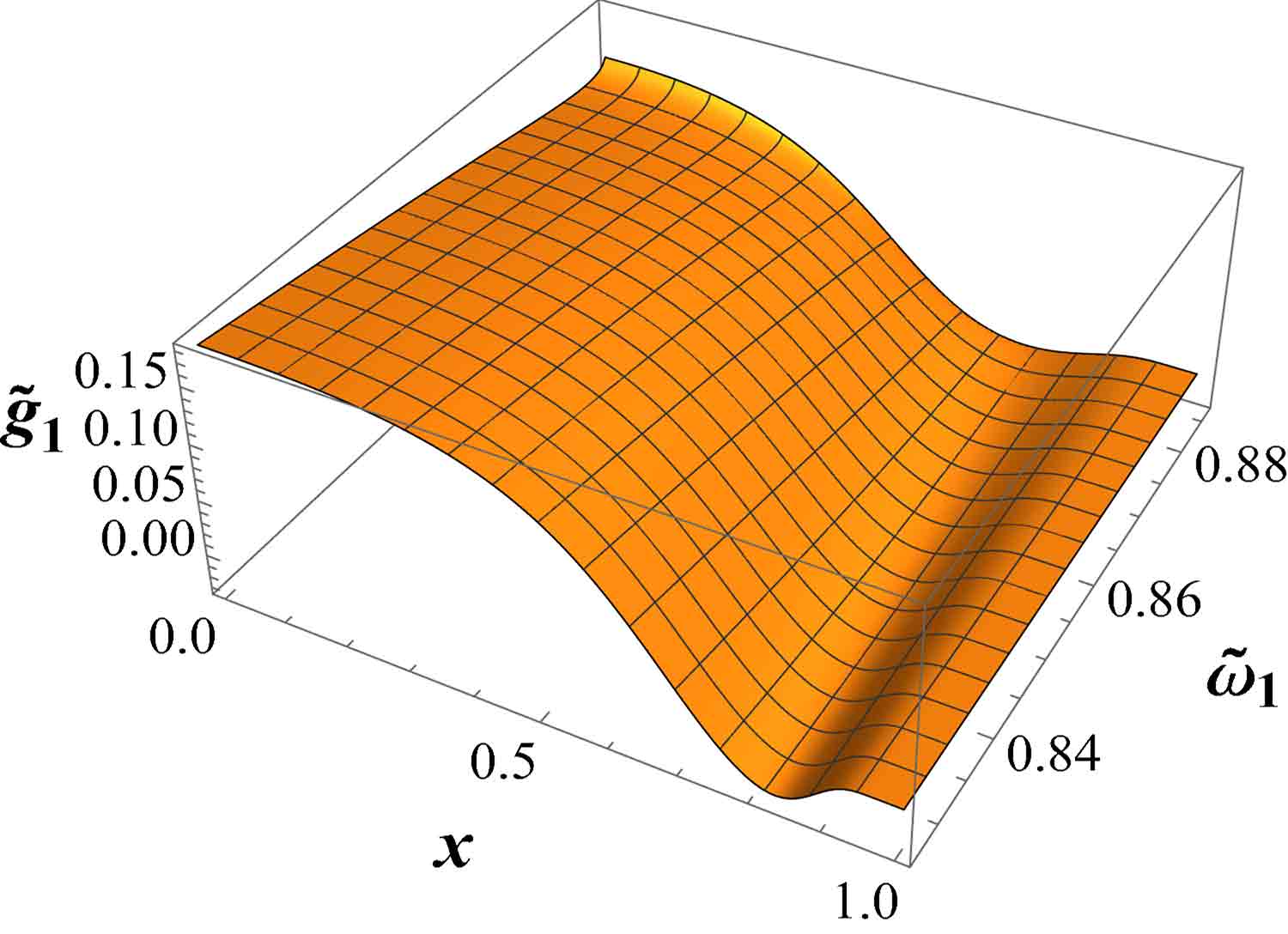}
    \includegraphics[height=.21\textheight]{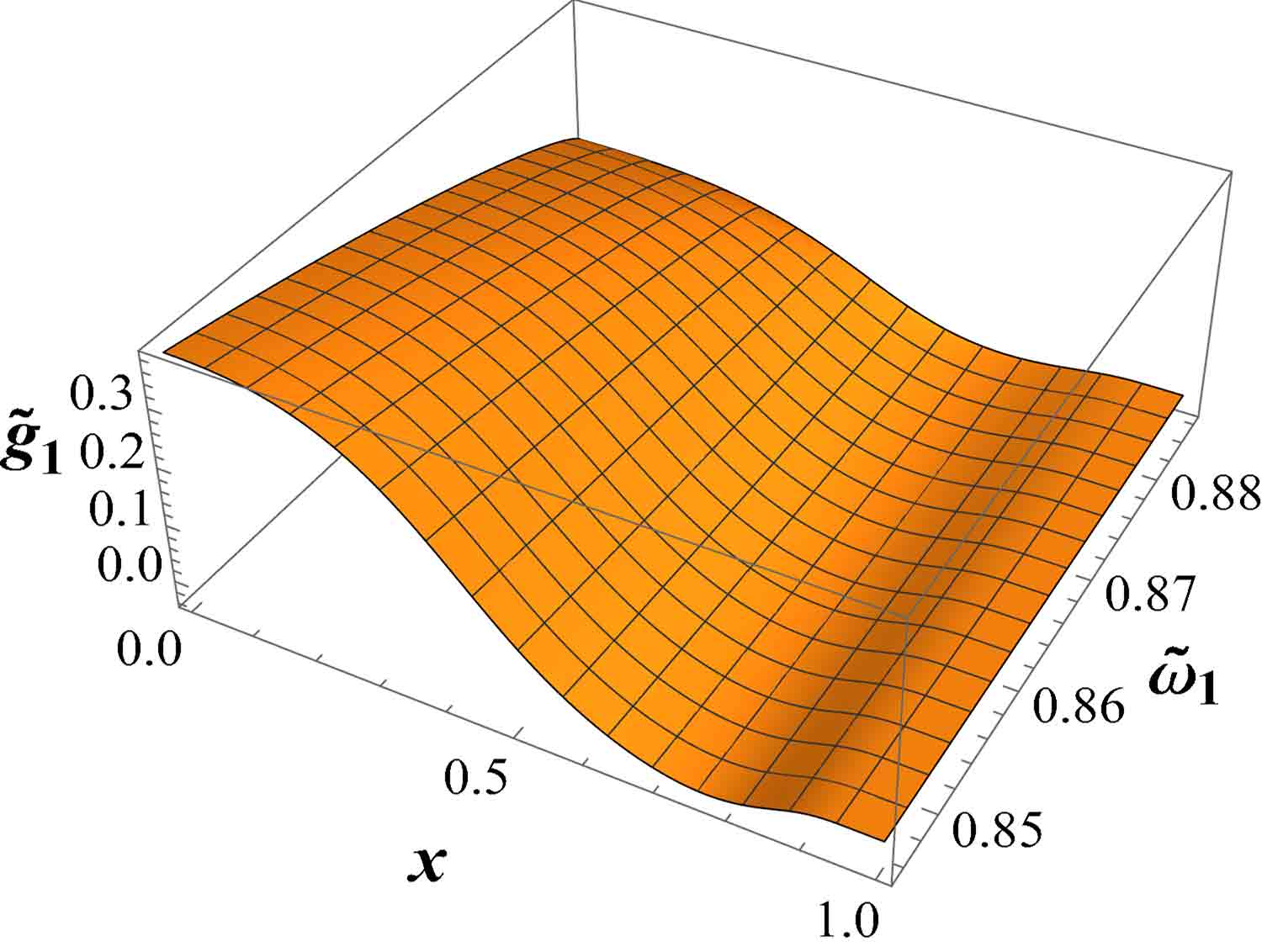}
\end{center}
\caption{The matter functions $\tilde{f}_0$, $\tilde{g}_0$, $\tilde{f}_1$ and $\tilde{g}_1$ on the first (left column) and second branches (right column) of the MSDSs solution function as functions of $x$ and $\tilde{\omega}_1$ for $\tilde{\omega}_0 = 0.721$.}
\label{nsf_double_f_g}
\end{figure}

The relationship between the ADM mass $M$ of nonsynchronized frequency single-branch solutions of the MSDSs and the frequency $\tilde{\omega}_1$ of the excited Dirac field is shown in Fig.~\ref{nsf_single_k-adm}. The black and blue dashed lines represent the ground state and the first excited state Dirac star solutions ($D_0$ and $D_1$), the orange line represents the MSDSs solution, and the red dashed line represents the ADM mass of $D_0$ when the frequency $\tilde{\omega}_0$ of the ground state Dirac field takes the values indicated in each plot.

When the frequency $\tilde{\omega}_0$ of the ground state Dirac field is close to $1$, the range of the frequency $\tilde{\omega}_1$ corresponding to the obtained single-branch solution is very narrow. As the frequency $\tilde{\omega}_0$ gradually decreases, the range of the frequency $\tilde{\omega}_1$ increases gradually.

Additionally, for a fixed frequency $\tilde{\omega}_0$, as the frequency $\tilde{\omega}_1$ increases, the ADM mass of the system decreases gradually. When the ADM mass reaches its maximum value, the frequency $\tilde{\omega}_1$ reaches its minimum value, and the ground state Dirac field disappears, causing the MSDSs to degenerate into $D_1$. Conversely, when the ADM mass reaches its minimum value, this minimum value is the same as the ADM mass of $D_0$ when the frequency $\tilde{\omega}_0$ takes the values indicated in each plot. This is because when the ADM mass of the MSDSs reaches its minimum value, the frequency $\tilde{\omega}_1$ reaches its maximum value, and the excited state Dirac field disappears, resulting in the system degenerating into $D_0$. In other words, the minimum ADM mass of MSDSs depends on the frequency $\tilde{\omega}_0$ of the ground state Dirac field.

\begin{figure}[!htbp]
\begin{center}
    \includegraphics[height=.23\textheight]{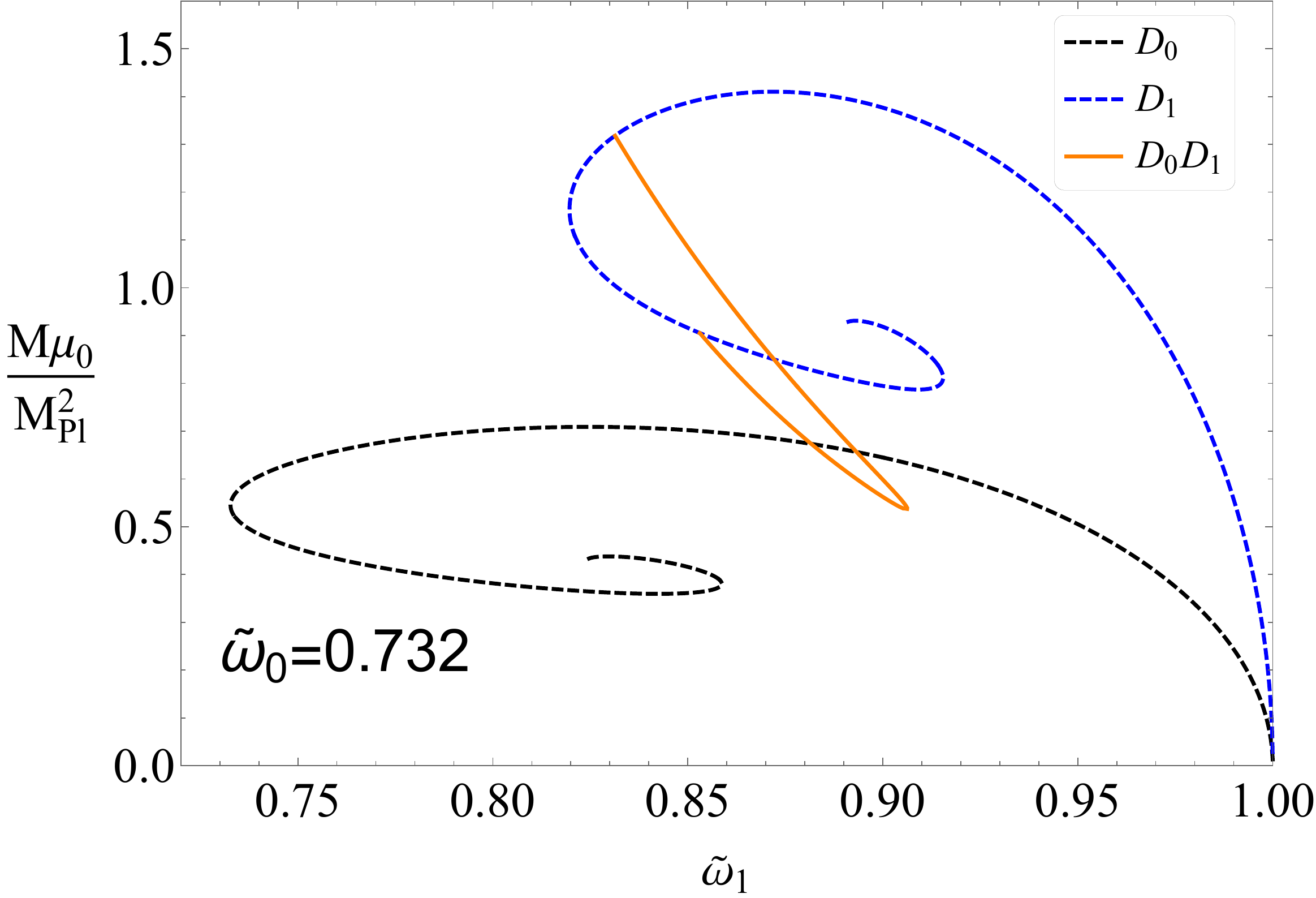}
    \includegraphics[height=.23\textheight]{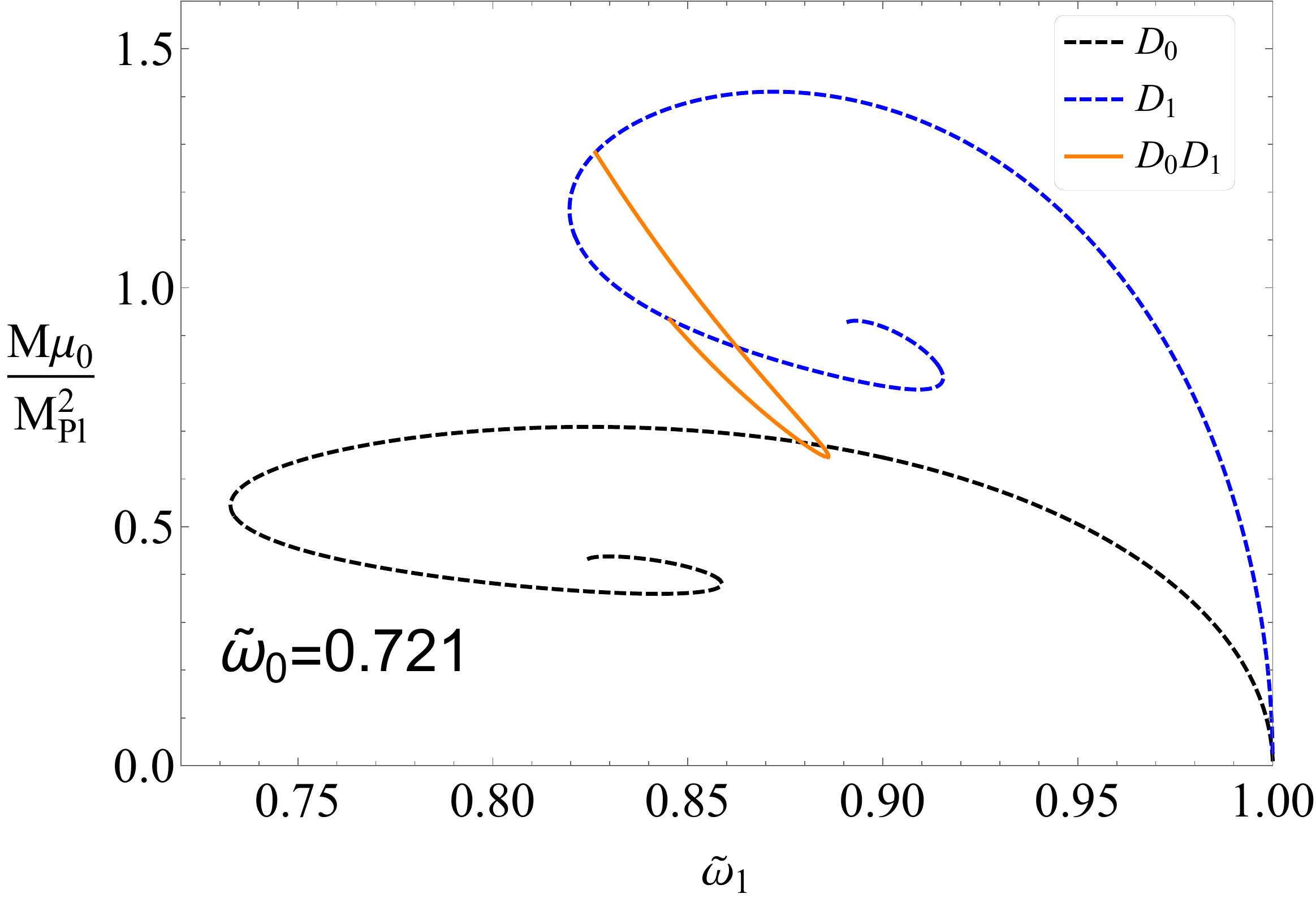}
    \includegraphics[height=.23\textheight]{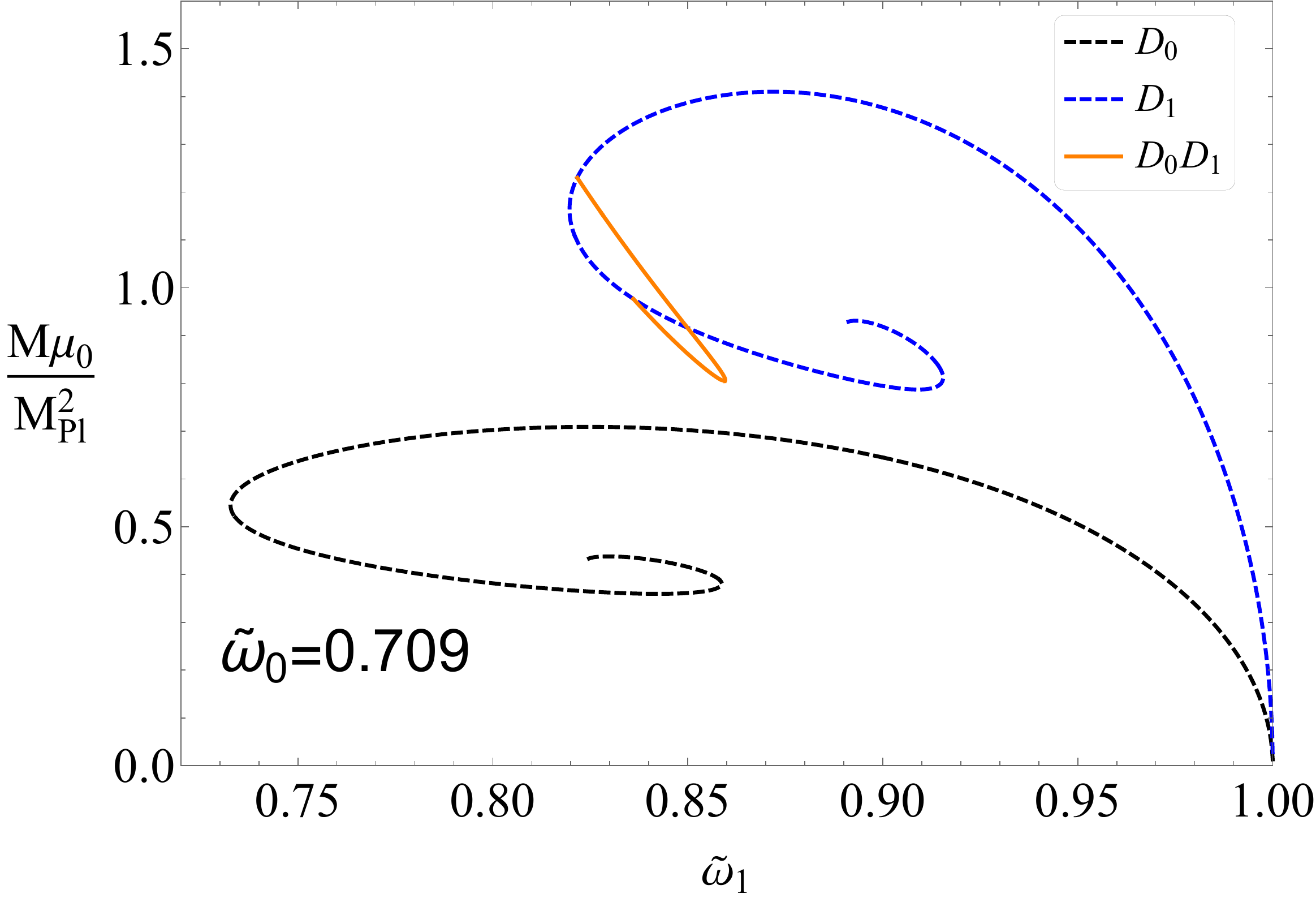}
    \includegraphics[height=.23\textheight]{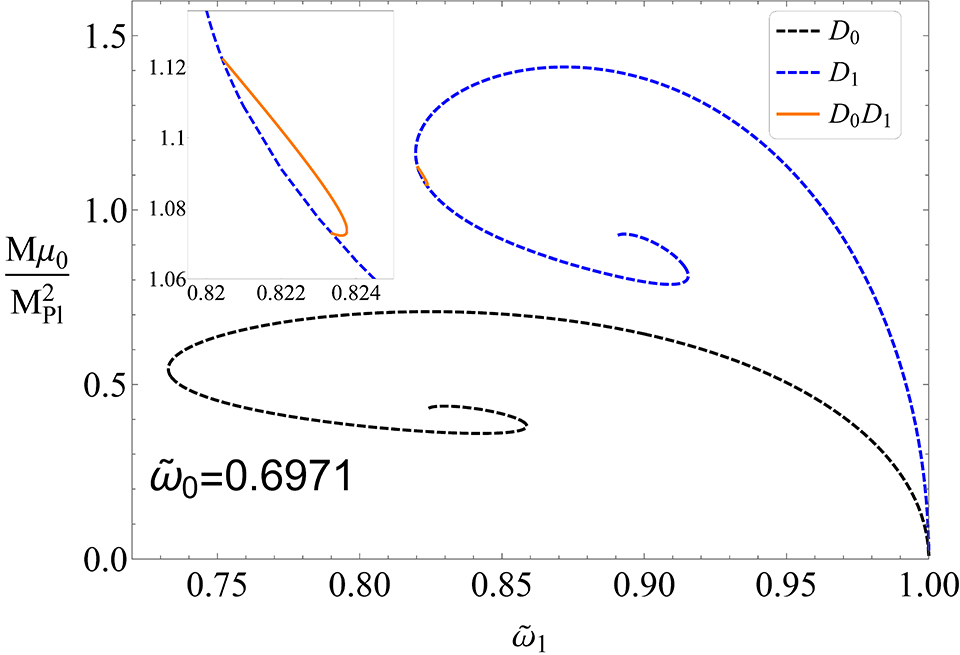}
\end{center}
\caption{The ADM mass $M$ of the MSDSs as a function of the frequency $\tilde{\omega}_1$ for several values of $\tilde{\omega}_0$.}
\label{nsf_double_k-adm}
\end{figure}

\subsubsection{Double-Branch}

Next, we discuss the double-branch solutions of the MSDSs under nonsynchronized frequency. In the single-branch solution discussed earlier, the minimum value of the frequency $\tilde{\omega}_0$ of the ground state Dirac field is $0.733$. Since there is no solution with a frequency less than $0.733$ for $D_0$, the MSDSs cannot degenerate into $D_0$ when the frequency $\tilde{\omega}_0$ is less than $0.733$, and thus the single-branch solution cannot be obtained. Through a series of numerical calculations, we found that when the frequency $\tilde{\omega}_0$ of the ground state field satisfies $0.6971 \le \tilde{\omega}_0 < 0.733$, the double-branch solutions of the MSDSs are obtained.

Fig.~\ref{nsf_double_f_g} shows the relationship between the radial profile of the matter fields that constitute the MSDSs and the nonsynchronized frequency $\tilde{\omega}_1$. The left column shows the field functions on the first branch. As the frequency $\tilde{\omega}_1$ increases, the peak values of the ground state Dirac field functions $\tilde{f}_0$ and $\tilde{g}_0$ gradually increase, while the changes in the excited state field functions $\tilde{f}_1$ and $\tilde{g}_1$ are relatively small. The right column shows the field functions on the second branch. As the frequency $\tilde{\omega}_1$ decreases, the peak values of the ground state Dirac field functions $\tilde{f}_0$ and $\tilde{g}_0$ gradually decrease, while the excited state field functions $\tilde{f}_1$ and $\tilde{g}_1$ gradually increase. It can be observed that when the frequency $\tilde{\omega}_1$ of the first and second branch solutions reaches its minimum value, the ground state Dirac field disappears, while the excited state Dirac field does not disappear for any $\tilde{\omega}_1$.

The relationship between the ADM mass $M$ of the nonsynchronized frequency double-branch solutions of the MSDSs and the frequency $\tilde{\omega}_1$ is shown in Fig.~\ref{nsf_double_k-adm}. The black and blue dashed lines represent the ground state and first excited state solutions of the Dirac star ($D_0$ and $D_1$), and the orange line represents the double-branch solution of the MSDSs ($D_0D_1$). Similar to the case of synchronized frequencies mentioned earlier, a bifurcation occurs when the frequency $\tilde{\omega}_0$ is below the threshold of $0.733$, transforming the single-branch solution into the double-branch solution. As the frequency $\tilde{\omega}_0$ of the ground state Dirac field decreases, the range of existence of the two branches of the double-branch solution gradually decreases with respect to the frequency $\tilde{\omega}_1$. For a fixed ground state field frequency $\tilde{\omega}_0$, when the frequency $\tilde{\omega}_1$ reaches its maximum value, the MSDSs do not degenerate into $D_0$ but transition to another new branch. As the frequency $\tilde{\omega}_1$ decreases further, the ADM mass of the system gradually increases, and eventually, the MSDSs transform into $D_1$. This characteristic change in the system's mass is also reflected in the profile of the field functions presented in Fig.~\ref{nsf_double_f_g}, where the disappearance of the ground state Dirac field functions on the two branches when the frequency $\tilde{\omega}_1$ reaches its minimum indicates the transformation of the MSDSs into $D_1$.

\subsection{Binding energy}

After obtaining various different solutions for the MSDSs, we will analyze the stability of the system from the perspective of binding energy. Consider a MSDS with ADM mass $M$, where the Noether charge for the ground state Dirac field is denoted as $Q_0$, and the Noether charge for the first excited state Dirac field is denoted as $Q_1$. The binding energy $E_B$ of the system can be expressed as:
\begin{equation}
  E_B = M - 2\left( \mu_0 Q_0 + \mu_1 Q_1 \right)\,,
\end{equation}
where the coefficient $2$ outside the parentheses on the right-hand side of the equation arises from the fact that in a spherically symmetric MSDS, both the ground state and the first excited state of the Dirac field have two components.

\begin{figure}[!htbp]
\begin{center}
    \includegraphics[height=.23\textheight]{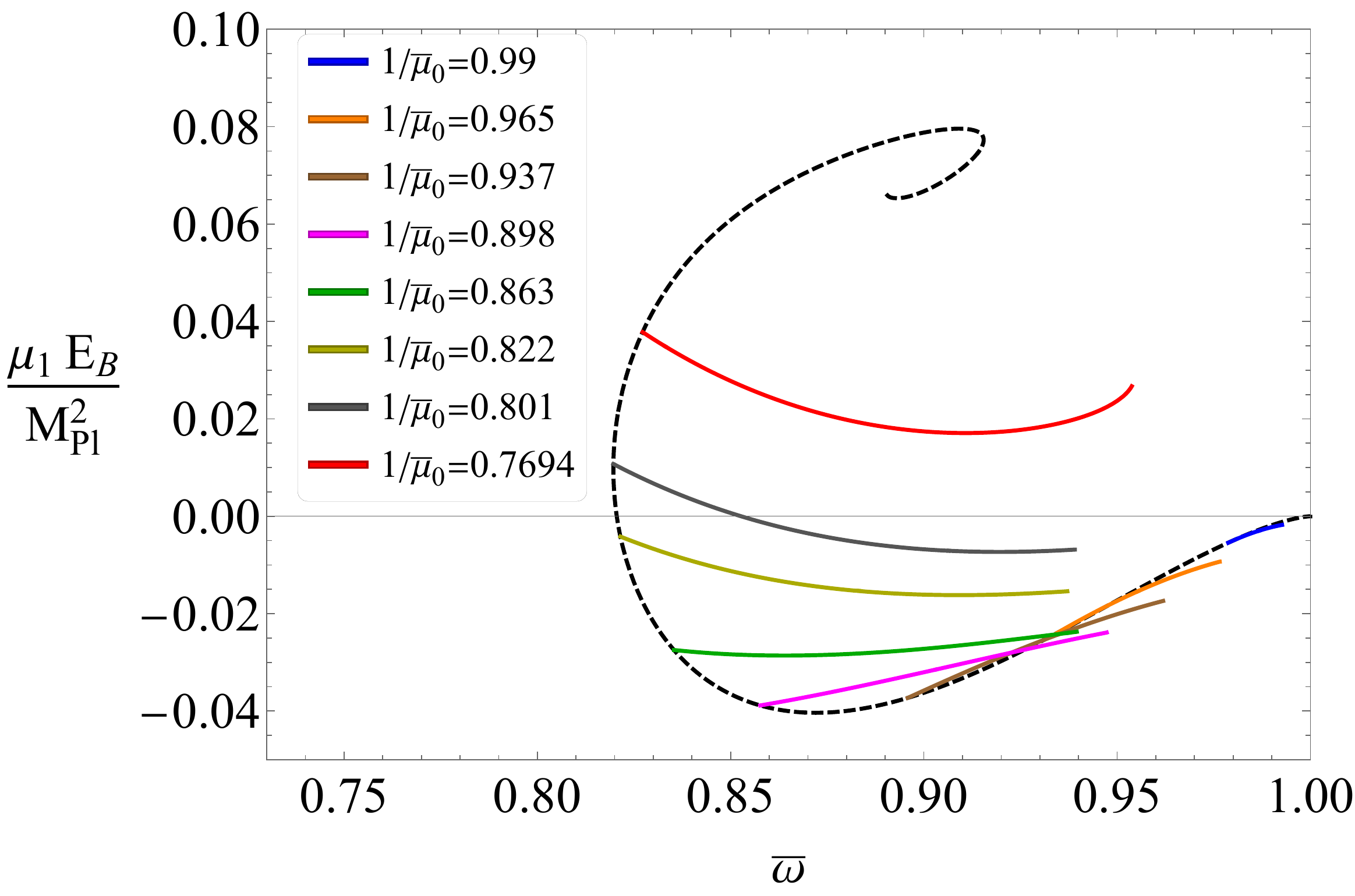}
    \includegraphics[height=.23\textheight]{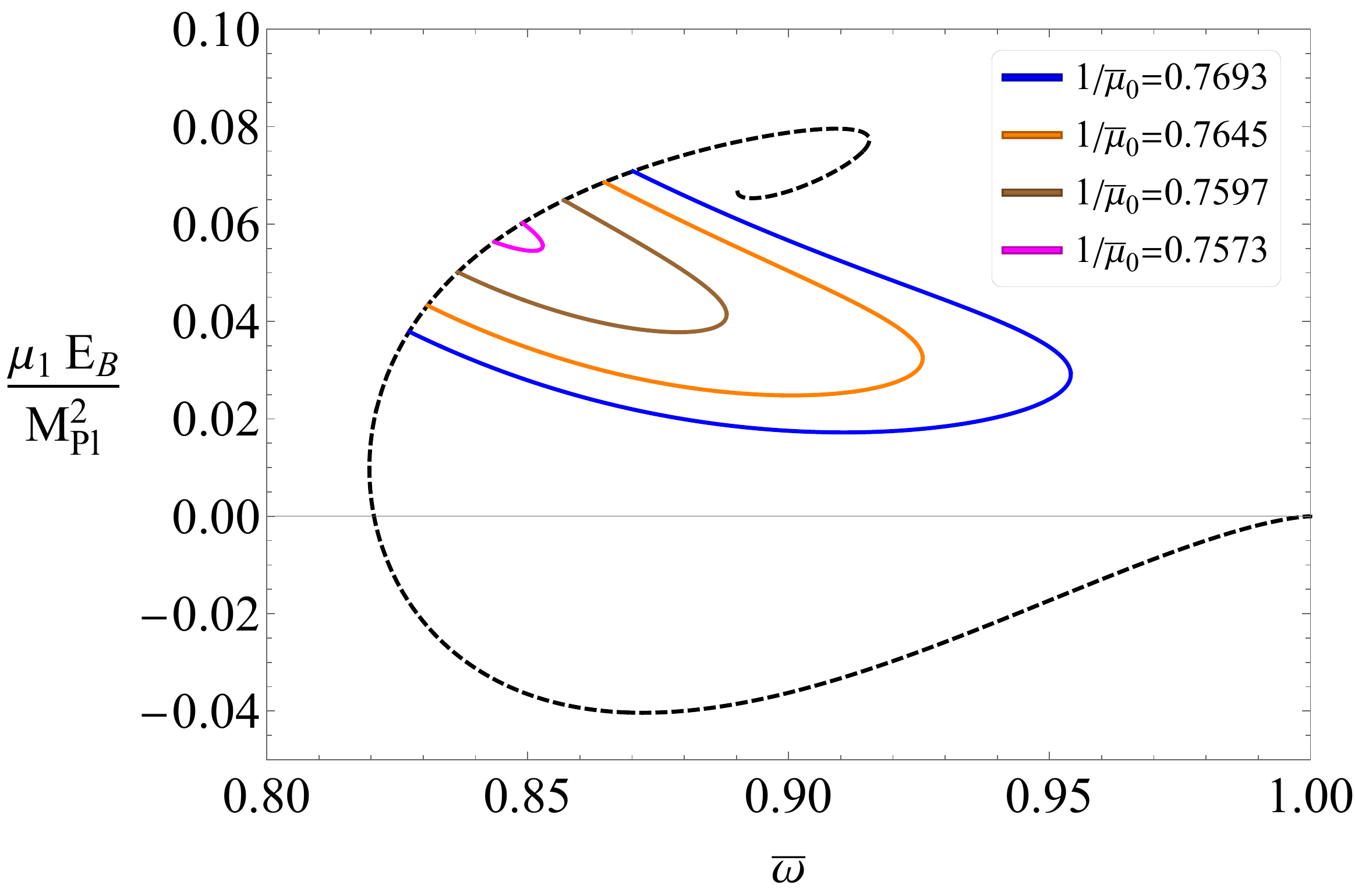}
\end{center}
\caption{The binding energy $E_B$ of the MSDSs as a function of the synchronized frequency $\overline{\omega}$ for several values of $1/\overline{\mu}_0$.}
\label{sf_be}
\end{figure}

\begin{figure}[!htbp]
\begin{center}
    \includegraphics[height=.23\textheight]{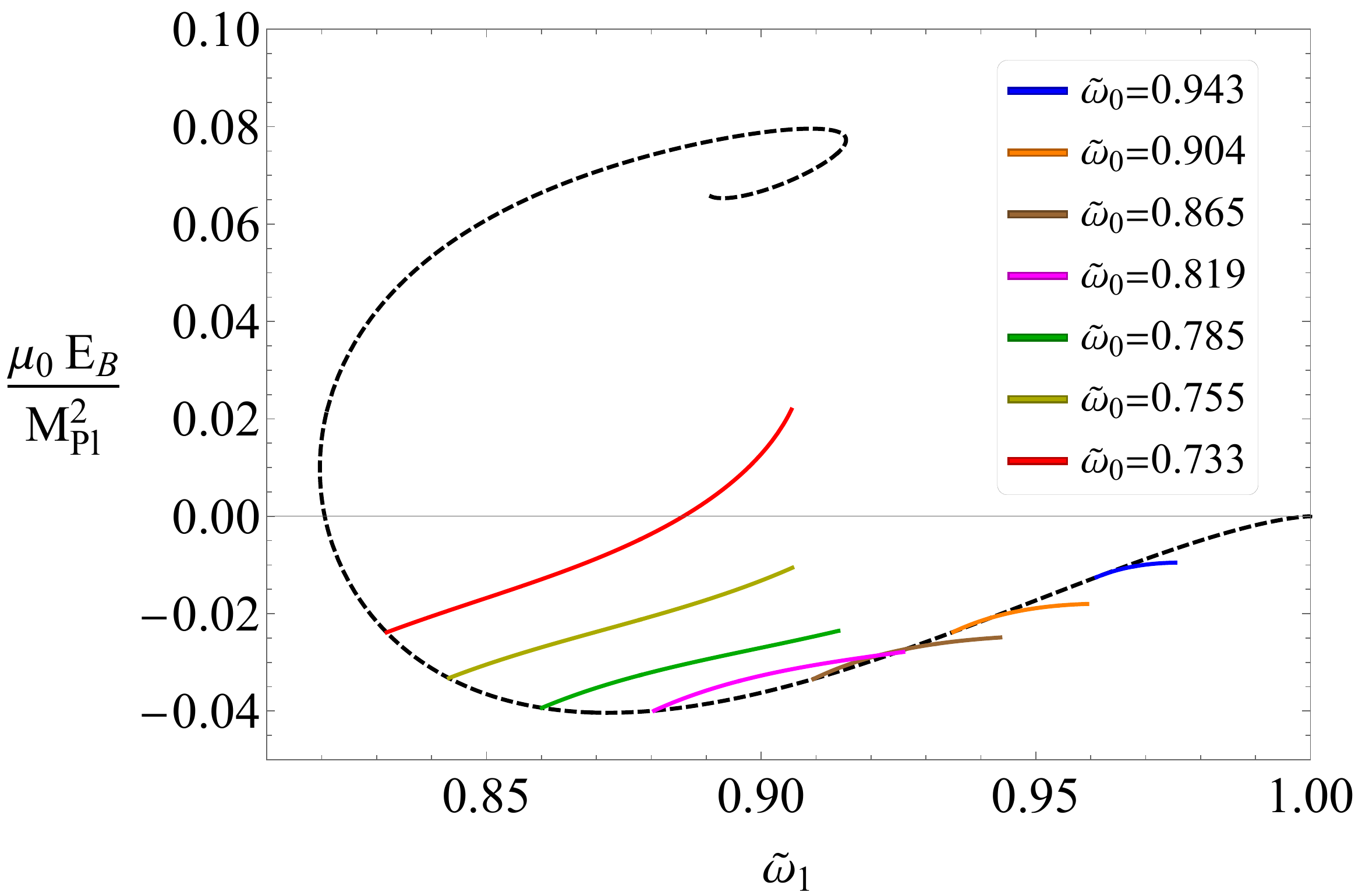}
    \includegraphics[height=.23\textheight]{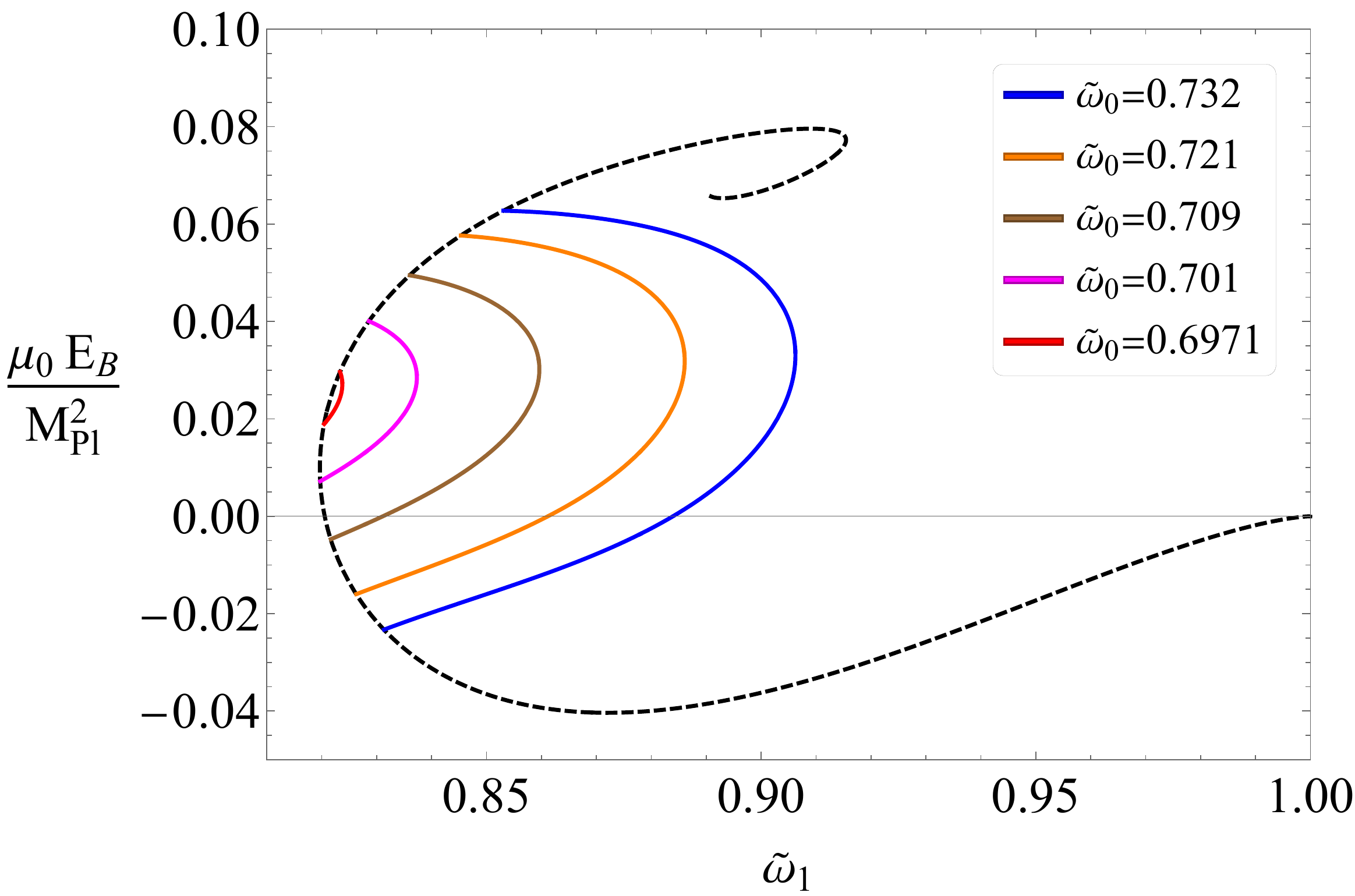}
\end{center}
\caption{The binding energy $E_B$ of the MSDSs as a function of the nonsynchronized frequency $\tilde{\omega}_1$ for several values of $\tilde{\omega}_0$.}
\label{nsf_be}
\end{figure}

We first analyze the binding energy of the MSDSs under synchronized frequency. Fig.~\ref{sf_be} shows the relationship between the binding energy $E_B$ and the synchronized frequency $\overline{\omega}$ of the MSDSs. The left plot represents the single-branch solutions of the MSDSs. When $1/\overline{\mu}_0 > 0.863$, for a fixed value of $\overline{\mu}_0$, the binding energy $E_B$ of the MSDSs monotonically increases with the synchronized frequency $\overline{\omega}$, and the binding energy is always less than zero. When $1/\overline{\mu}_0 \le 0.863$, for a given $\overline{\mu}_0$, the binding energy initially decreases and then increases as the synchronized frequency increases. When $1/\overline{\mu}_0$ becomes sufficiently small, the solutions become unstable (e.g., the red curve). Therefore, when $1/\overline{\mu}_0$ is sufficiently large, i.e., when the masses $\mu_0$ and $\mu_1$ of the ground state and excited state Dirac fields are sufficiently close, the MSDSs are more stable. The right plot represents the double-branch solutions of the MSDSs. As $1/\overline{\mu}_0$ increases, the minimum value of the binding energy gradually decreases, but the binding energies of these double-branch solutions are all greater than zero, indicating that the solutions are unstable.

Next, we consider the binding energy of the MSDSs under nonsynchronized frequency. Fig.~\ref{nsf_be} shows the relationship between the binding energy $E_B$ and the frequency $\tilde{\omega}_1$ for the nonsynchronized frequency solutions of the MSDSs. The left plot represents the single-branch solutions, where for a fixed $\tilde{\omega}_0$, the binding energy $E_B$ monotonically increases with the frequency $\tilde{\omega}_1$ and remains negative when the frequency $\tilde{\omega}_0$ is sufficiently large. However, for small values of $\tilde{\omega}_0$ (e.g., $\tilde{\omega}_0=0.733$), the MSDSs can undergo a transition from a stable solution to an unstable one as the frequency $\tilde{\omega}_1$ increases. The right plot represents the double-branch solutions, where the negative binding energy solutions appear when the frequency $\tilde{\omega}_0$ is sufficiently large (e.g., $\tilde{\omega}_0=0.732$). As the frequency $\tilde{\omega}_0$ decreases, the stable solutions in the double-branch solutions gradually disappear, and all solutions eventually become unstable. Therefore, in the case of nonsynchronized frequency, stable solutions exist within the double-branch solutions.

\subsection{Galactic halos as MSDSs}
The velocity of stars orbiting the central core of a galaxy remains constant over a large range of distances starting from the galactic center. This phenomenon may be attributed to the presence of a dark matter halo in the outer regions of the galaxy. By using boson stars to simulate the dark matter halo, it is possible to obtain results that are consistent with real observational data~\cite{Lee:1995af,Bernal:2009zy,Brito:2023fwr}. In the following, we will analyze the feasibility of simulating the dark matter halo using MSDSs by computing the velocities of test particles orbiting around them. Considering timelike circular geodesics on the equatorial plane, the rotational velocity of the test particles is given by~\cite{Bernal:2009zy,Brito:2023fwr}:
\begin{equation}
 v^2 = r\sqrt{-g_{tt}}\partial_r \sqrt{-g_{tt}}\,,
\end{equation}
Substituting Equ.~(\ref{equ10}) into the expression, we obtain
\begin{equation}
 v = \sqrt{\frac{r\sigma(\sigma N^\prime + 2N \sigma^\prime)}{2}}\,.
\end{equation}
\begin{figure}[!htbp]
  \begin{center}
    \includegraphics[height=.24\textheight]{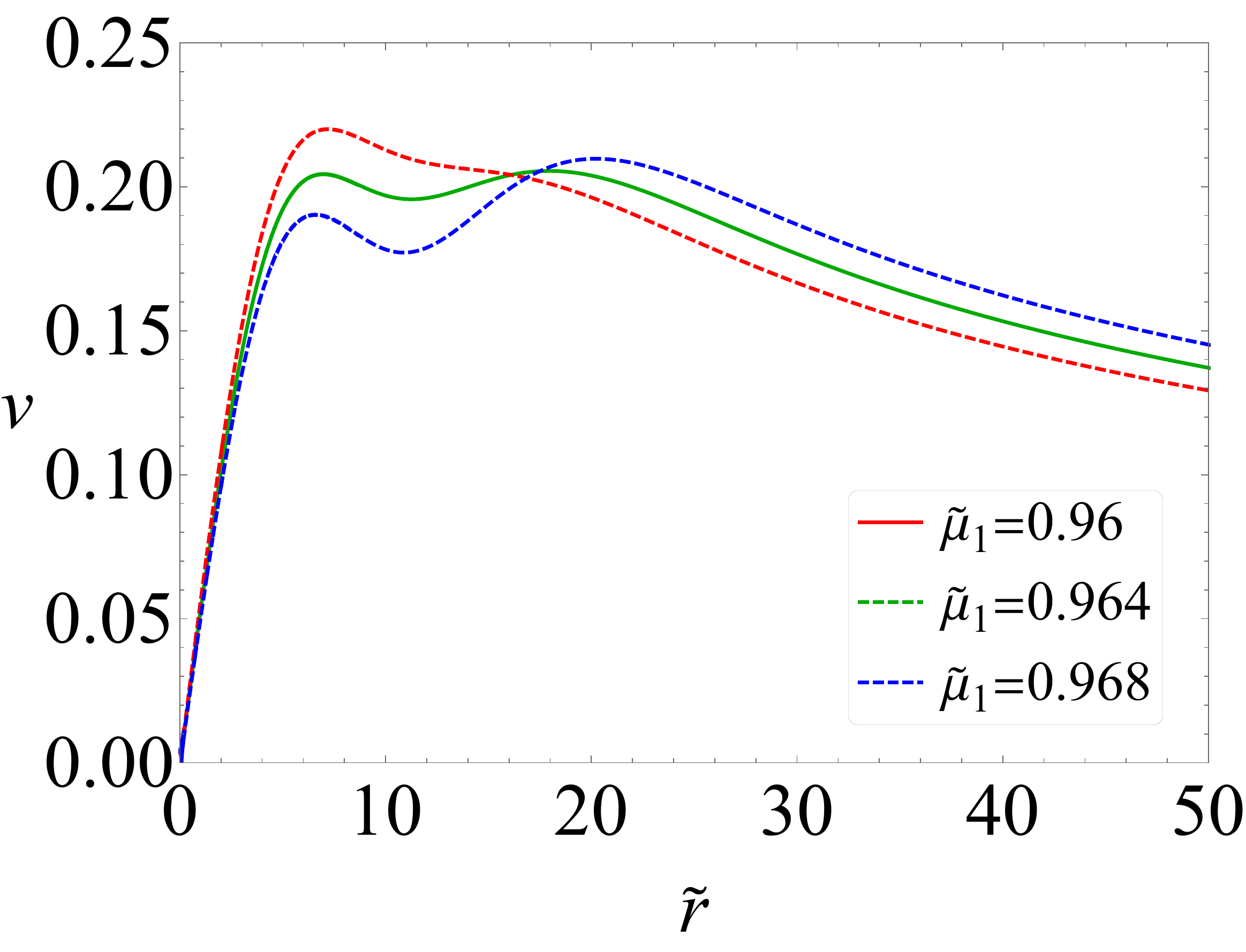}
    \includegraphics[height=.23\textheight]{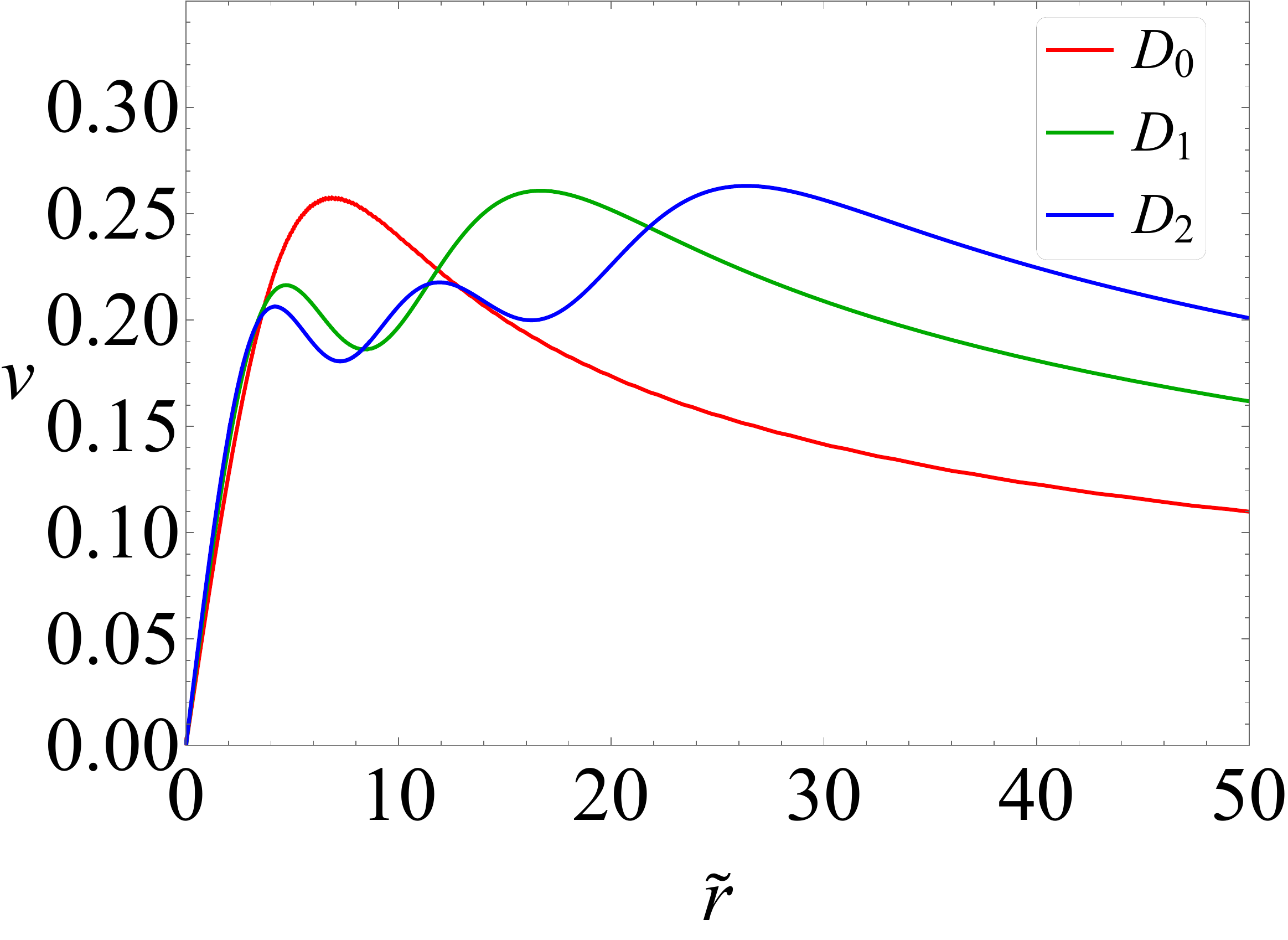}
  \end{center}
  \caption{Left panel: the rotational curves of the MSDSs for several values of $\tilde{\mu}_1$. Right panel: the rotational curves of the Dirac stars at different energy levels.}
  \label{rc}
\end{figure}
Next, we analyze the rotational curves of the MSDSs. As shown in Fig.~\ref{rc}, the left panel displays the rotation curves of the MSDSs' synchronized frequency solutions for different values of the mass $\tilde{\mu}_1$, represented by the red, green, and blue curves. In the right panel, the red, green, and blue curves correspond to the rotation curves of the ground state, first excited state, and second excited state Dirac stars, respectively. All solutions in the figure have a synchronized frequency of $\tilde{\omega} = 0.92$. By examining the left panel, we observe that when the mass of the excited state Dirac field is $\tilde{\mu}_1 = 0.964$, the rotation curve of the MSDSs reaches a peak velocity and then exhibits a relatively flat region with slight oscillations, rather than significantly decreasing. Increasing the mass $\tilde{\mu}_1$ leads to larger oscillation amplitudes, resembling the green curve in the right panel (first excited state Dirac star). Conversely, decreasing the mass $\tilde{\mu}_1$ results in curves resembling the red curve in the right panel (ground state Dirac star).

\section{Conclusion}\label{sec5}
In this paper, we investigate the Einstein-Dirac system and construct spherically symmetric multi-state Dirac stars, wherein two coexisting states of the Dirac field are present. We discuss the field functions, ADM mass, and binding energy of the solutions for the multi-state Dirac star under synchronized and nonsynchronized frequency conditions. Furthermore, we analyze the feasibility of considering the multi-state Dirac star as a candidate for dark matter halos.

In the case of synchronized frequency, we explore different solutions for the MSDSs by varying the ratio of masses between the ground state and excited state Dirac fields ($\mu_0/\mu_1$). Based on the number of solution branches, we classify the obtained numerical results into single-branch solutions and double-branch solutions. For single-branch solutions, the peak values of the ground state and excited state Dirac field functions exhibit monotonic behavior as the synchronized frequency changes. The ADM mass monotonically decreases with increasing synchronized frequency, and at the minimum and maximum synchronized frequencies, the MSDSs degenerate into the first excited state Dirac stars and the ground state Dirac stars, respectively. When $\tilde{\mu}_1$ (or $1/\overline{\mu}_0$) is below the threshold of $0.7694$, the single-branch solution undergoes a transition into a double-branch solution. In the case of double-branch solutions, the excited state Dirac field functions persist while the ground state field functions disappear at the minimum synchronized frequency on both branches, leading to the degeneration of the MSDSs into the first excited state Dirac stars. Moreover, as the mass of the ground state field decreases, the range of synchronized frequency values on the two branches of the double-branch solution gradually diminishes.

Next, for the case of nonsynchronized frequency, we set the ratio of masses between the ground state and excited state Dirac fields as $\mu_0/\mu_1 = 1$ and obtain different solutions by varying the frequency of the ground state Dirac field. Similar to the synchronized frequency case, the nonsynchronized frequency solutions of the MSDSs can also be classified into single-branch solutions and double-branch solutions. In the case of single-branch solutions, the variations of the ground state and excited state field functions with respect to the frequency of the excited state Dirac field exhibit monotonic behavior. The ADM mass decreases as the frequency of the excited state field increases, and its minimum value depends on the frequency of the ground state Dirac field. When $\tilde{\omega}_0$ is below the threshold of $0.733$, the single-branch solution undergoes a transition into a double-branch solution. For the double-branch solutions, as the frequency of the ground state field decreases, the minimum ADM mass of the MSDSs gradually increases, the range of nonsynchronized frequency values on the two branches diminishes, and the MSDSs degenerate into the first excited state Dirac stars when the nonsynchronized frequency becomes sufficiently small.

It is worth noting that the MSDSs solutions exhibit certain similarities between the synchronized and nonsynchronized frequency cases. In the synchronized frequency (nonsynchronized frequency) scenario, when $\tilde{\mu}_1$ ($\tilde{\omega}_0$) falls below the threshold of $0.7694$ ($0.733$), the single-branch solutions undergo an abrupt transition into double-branch solutions. Furthermore, regardless of whether the frequencies are synchronized or not, the single-branch solutions of the MSDSs can always degenerate to the ground state or the first excited state, while the double-branch solutions can only degenerate to the first excited state.

Subsequently, we computed the binding energy of various solutions for the MSDSs. For synchronized frequency solutions, stable solutions only exist in the single-branch solutions, and when $1/\overline{\mu}_0$ is sufficiently small, all stable solutions within the single-branch solutions vanish. For nonsynchronized frequency solutions, stable solutions exist in both  single-branch and double-branch solutions. However, the double-branch solutions become unstable when the frequency of the ground state field reaches a sufficiently low value, whereas the single-branch solutions maintain stable solutions for any frequency of the ground state field. 

Finally, we computed the rotation curves of the MSDSs. It has been observed that MSDSs containing excited state matter fields exhibit a nearly flat region near the velocity peak in their rotation curves, similar to the rotation curves of multi-state boson stars discussed in~\cite{Bernal:2009zy}. In our future work, we plan to construct MSDSs with a greater number of matter field nodes. It is possible that rotational curves of models with a higher number of nodes can provide a closer fit to the observed data~\cite{Brito:2023fwr}.

\section*{Acknowledgements}
This work is supported by National Key Research and Development Program of China (Grant No.~2020YFC2201503) and  the National Natural Science Foundation of China (Grants No.~12275110 and No.~12247101). Parts of computations were performed on the shared memory system at institute of computational physics and complex systems in Lanzhou university.

\providecommand{\href}[2]{#2}\begingroup\raggedright

\end{document}